\documentclass[12pt]{article}

\usepackage[numbers]{natbib}
\usepackage[small,bf]{caption}

\usepackage[hmargin=3.cm,vmargin=3.5cm]{geometry}

\RequirePackage[OT1]{fontenc}

\RequirePackage{amsthm,amsmath}
\RequirePackage[colorlinks,citecolor=blue,urlcolor=blue]{hyperref}
\usepackage{array,amsmath,amsthm, amssymb, float}
\usepackage{epsfig,psfrag,url,verbatim, multirow}
\usepackage{graphics}
\usepackage{graphicx, algorithm}
\usepackage{xargs}
\usepackage{amsfonts}
\usepackage{color}

\usepackage[numbers]{natbib}

\usepackage[parfill]{parskip}
\usepackage{subcaption,graphicx}

\usepackage{xargs}
\usepackage{amsfonts}

\usepackage{color}

\newcommand{\diag}{\mathrm{diag}}

\newcommand{\B}{\boldsymbol}
\newcommand{\M}{\mathbf}

\newcommand{\sbt}{\mathrm{subject\; to }}

\newcommand{\half}{\mbox{$\frac12$}}
\renewcommand{\v}{\mbox{ v}}

\newtheorem{thm}{Theorem}[section]

\newtheorem{prop}{Proposition}
\newtheorem{mydef}{Definition}
\newtheorem{rem}{Remark}

\DeclareMathOperator*{\argmin}{arg\,min}

\newenvironment{myarray}[2][1]
  {\def\arraystretch{#1}\array{#2}}
  {\endarray}
\newcommand{\mmk}{k}

\usepackage{array,graphicx}
\usepackage{booktabs}
\usepackage{pifont}

\newcommand*\rot{\rotatebox{90}}

\begin{document}

\title{Best Subset Selection via a Modern Optimization Lens}
%\author{Dimitris Bertsimas \and Angela King \and Rahul Mazumder}

\date{\sf { \scriptsize{(This is a Revised Version dated May, 2015. First Version Submitted for Publication on June, 2014.})}}

\author{%
Dimitris Bertsimas\thanks{MIT Sloan School of Management and Operations Research Center, Massachusetts Institute of Technology:~\texttt{dbertsim@mit.edu}}%
\and
Angela King\thanks{Operations Research Center, Massachusetts Institute of Technology:~\texttt{aking10@mit.edu}}%
\and
Rahul Mazumder\thanks{MIT Sloan School of Management and Operations Research Center, Massachusetts Institute of Technology:~\texttt{rahulmaz@mit.edu}}%
}

\maketitle

%\begin{frontmatter}

%\title{Best Subset Selection via a Modern Optimization Lens}
%\author{Dimitris Bertsimas \and Angela King \and Rahul Mazumder}

%\begin{aug}
%\author{\fnms{Dimitris} \snm{Bertsimas}\thanksref{m1}\ead[label=e1]{dbertsim@mit.edu}},
%%\and
%\author{\fnms{Angela} \snm{King}\thanksref{m11}\ead[label=e11]{aking10@mit.edu}}
%\and
%\author{\fnms{Rahul} \snm{Mazumder}\thanksref{m2}\ead[label=e2]{rm3184@columbia.edu}}
%
%%\author{\fnms{Third} \snm{Author}\thanksref{t1,m2}
%%\ead[label=e3]{third@somewhere.com}
%%\ead[label=u1,url]{http://www.foo.com}}
%
%%\thankstext{t1}{Some comment}
%%\thankstext{t2}{First supporter of the project}
%%\thankstext{t3}{Second supporter of the project}
%\runauthor{Bertsimas, King and Mazumder}
%
%\affiliation{Massachusetts Institute of Technology\thanksmark{m1}\thanksmark{m11} and Columbia University\thanksmark{m2}}
%
%\address{Dimitris Bertsimas, \\
%MIT Sloan School of Management and \\
%Operations Research Center, \\
%Massachusetts Institute of Technology,\\
%Cambridge, MA. \\
%\printead{e1}}
%%\phantom{E-mail:\ }\printead*{e2}}
%
%\address{Angela King, \\
%Operations Research Center, \\
%Massachusetts Institute of Technology,\\
%Cambridge, MA. \\
%\printead{e11}}
%
%
%\address{Rahul Mazumder,\\
%Department of Statistics,\\
%Columbia University,\\
%New York, NY.\\
%\printead{e2}}
%\end{aug}
%

%%\maketitle

\begin{abstract}
%The belief that mixed integer optimization (MIO) approaches to classical problems in statistics 
%  are not practically relevant was formed in the 1970s and 1980s and it was at the time justified. 
In the last twenty-five years (1990-2014), algorithmic advances in integer optimization combined with hardware improvements have resulted in an astonishing 
 200 billion factor speedup in solving  Mixed Integer Optimization (MIO) problems. 
We present a MIO approach  for solving the classical  best subset selection problem of choosing $\mmk$  out of $p$ features in linear regression given $n$ observations. 
We develop a discrete extension of modern  first order continuous optimization methods to find high quality feasible solutions that we use as warm starts 
to a MIO solver that finds provably optimal solutions.   The resulting algorithm (a) provides a solution with a guarantee on its suboptimality even if 
we terminate the algorithm early, (b) can 
 accommodate side constraints on  the coefficients of the  linear regression and (c)
 extends   to finding best subset solutions for the least absolute deviation loss function.
Using a wide variety of synthetic and real datasets, we demonstrate that our approach solves problems with $n$ in the 1000s and $p$ in the 100s in minutes
to provable optimality,
and finds near optimal  solutions for $n$ in the 100s and $p$ in the 1000s in minutes.   We also establish via numerical experiments that the MIO approach performs better than  {\texttt {Lasso}} and other popularly used sparse learning procedures, in terms of achieving sparse solutions with good predictive power. 
% Overall, we provide  evidence that statistics as a field needs to revisit the belief that 
%MIO  is not practically relevant. 
\end{abstract}

%
%\begin{keyword}[class=MSC]
%\kwd[Primary ]{62J05, 62J07 , 62G35}
%%\kwd{60K35}
%\kwd[; secondary ]{90C11,90C26,90C27}
%\end{keyword}
%
%\begin{keyword}
%%\kwd{sample}
%\kwd{Sparse Linear Regression, Best Subset Selection, $\ell_0$-constrained minimization, {Lasso}, Least Absolute Deviation, Algorithms, Mixed Integer Programming, Global Optimization, Discrete Optimization}
%\end{keyword}
%
%\end{frontmatter}

%%%%%%%%%%%%%%%%%%%%%
%% INTRODUCTION + LIT
%%%%%%%%%%%%%%%%%%%%%

%\textcolor{red}{Applied global changes: \textsc{{\textsc{Gurobi}}}  to {\textsc{Gurobi}}}\\

%\textcolor{red}{Applied global changes: Problem~ to Problem~ ; Section~ to Section~}

\section{Introduction}\label{sec:intro}
We consider the   linear  regression model with response vector $\M{y}_{n \times 1}$, 
model matrix $\M{X}=[\M{x}_{1}, \ldots, \M{x}_{p}] \in  \mathbb{R}^{n \times p}$, regression coefficients $\B\beta \in  \mathbb{R}^{p \times 1}$
and errors $\B\epsilon \in  \mathbb{R}^{n \times 1}$:
%\begin{equation}\label{ls-model}
$$\M{y} = \M{X} \B\beta + \B\epsilon.$$
%\end{equation}
We will assume that the columns of $\M{X}$ have been standardized to have zero means and unit $\ell_{2}$-norm.
In many important classical and modern statistical applications, it is desirable to obtain a parsimonious fit to the data by finding the best $\mmk$-feature 
fit to the response $\M{y}$. 
Especially in the high-dimensional regime  with $p\gg  n,$ in order to conduct statistically meaningful inference, it is desirable to assume that the true regression coefficient $\B\beta$  is sparse or may be well approximated by a sparse vector.  Quite naturally, 
the last few decades have seen a flurry  of activity in estimating sparse linear models with good explanatory power.
Central to this statistical task lies the best subset problem~\cite{miller2002subset} with subset size $\mmk$, which is given by the following optimization problem:
\begin{equation}\label{eq-card-k}
\min_{\B\beta}  \;\; \frac12\| \M{y} - \M{X}\B\beta \|_2^2 \;\;\; \sbt \;\;\; \| \B\beta \|_{0} \leq \mmk, 
\end{equation}
where the $\ell_{0}$ (pseudo)norm of a vector $\B\beta$ counts the number of nonzeros in $\B\beta$ and is given by 
$\| \B\beta \|_{0} = \sum_{i=1}^{p} 1( \beta_{i} \neq 0 ),$ 
 where $1(\cdot)$ denotes the indicator function.
 The cardinality constraint  makes Problem~\eqref{eq-card-k} NP-hard~\cite{natarajan1995sparse}.
Indeed, state-of-the-art algorithms to solve Problem~\eqref{eq-card-k}, as implemented in 
popular statistical packages, like {\texttt {leaps}} in {\texttt {R}}, do not scale to problem sizes larger than $p=30$.
Due to this reason, it is not surprising that the best subset problem has been widely dismissed  as being \emph{intractable} 
by the greater statistical community.  

In this paper we address Problem~\eqref{eq-card-k} using modern optimization methods, specifically mixed integer optimization (MIO) and a discrete extension of 
first order continuous optimization  methods.  Using a wide variety of synthetic and real datasets, we demonstrate
 that our approach 
solves problems with $n$ in the 1000s and $p$ in the 100s in minutes
to provable optimality,
and finds near optimal  solutions for $n$ in the 100s and $p$ in the 1000s in minutes. To the best of our knowledge, this is the first time that MIO  has been demonstrated to be a tractable solution method for Problem~\eqref{eq-card-k}.  
 We note that we use the term tractability not to mean the usual polynomial solvability for problems, but rather the ability to solve problems of realistic size in  times that are appropriate 
 for the applications we consider. 
  
As there is a vast literature on the best subset problem, we next  give a brief and selective overview of related approaches for  the problem.
%%%%%%%%%%%%%%%
% LITERATURE REVIEW
%%%%%%%%%%%%%%%

\subsection*{Brief Context and Background}
To overcome the computational difficulties of the best subset problem,  
computationally tractable convex optimization based methods like  {\texttt {Lasso}}~\cite{Ti96,CDS1998}
 have been proposed as a convex surrogate for Problem~\eqref{eq-card-k}. 
For the linear regression problem, the Lagrangian form of {\texttt {Lasso}} solves
 \begin{equation}\label{lass-lag}
\min_{\B\beta} \half \| \M{y} - \M{X} \B\beta\|_2^2  + \lambda \| \B\beta\|_{1},
\end{equation}
where 
the $\ell_{1}$ penalty on $\B\beta$, i.e., 
$\|\B\beta\|_{1} = \sum_{i} |\beta_{i}|$ 
shrinks the coefficients towards zero and 
naturally produces a sparse solution by setting many coefficients to be exactly zero.
%often sets many coefficients to be exactly zero, by virtue of its sparsity producing nature. 
There has been a substantial amount of impressive work on  
{\texttt {Lasso}}~\cite{LARS,candes2009near,bickel2009simultaneous,ZH08,greenshtein2004persistence,ZY2006,donoho2006,KF2000,MB2006,wainwright2009sharp,tibshirani2011regression} in terms of algorithms and 
understanding of  its theoretical properties---see for example the excellent books or surveys~\cite{buhlmann2011statistics,FHT-09-new,tibshirani2011regression} and the  references therein.

Indeed,  {\texttt {Lasso}} enjoys several attractive statistical properties and has drawn a significant amount of 
attention from the statistics community as well as other closely related fields. Under various conditions on the model matrix $\M{X}$  and
$n, p , \B\beta$ it can be shown that   {\texttt {Lasso}} delivers a sparse model 
with good predictive performance~\cite{buhlmann2011statistics,FHT-09-new}.
In order to perform 
exact variable selection, much stronger assumptions are required~\cite{buhlmann2011statistics}. 
Sufficient conditions under which  {\texttt {Lasso}} gives  a sparse model with  good predictive performance
are the restricted eigenvalue conditions and compatibility conditions~\cite{buhlmann2011statistics}. These involve statements about 
the range of the spectrum of sub-matrices of $\M{X}$ and are difficult  to 
verify, for a given data-matrix $\M{X}$.

An important reason behind the   popularity of {\texttt {Lasso}} is its computational feasibility and scalability to practical sized problems.
Problem~\eqref{lass-lag} is a convex quadratic optimization problem and
there are several   efficient solvers for it, see for example~\cite{nest-07,LARS,FHT2007}. 
%State-of-the art solvers for {\texttt {Lasso}} are capable of solving~\eqref{lass-lag} for a path of values 
%of $\lambda$ and gracefully scale to large problems. 

In spite of its favorable statistical properties,   {\texttt {Lasso}} has several shortcomings. 
In the presence of noise and correlated variables, in order to deliver a model with good predictive accuracy,  
{\texttt {Lasso}} brings in a large number of nonzero coefficients (all of which are shrunk towards zero) including noise variables.
 {\texttt {Lasso}} leads to biased 
regression coefficient estimates, since
%The $\ell_{1}$-norm penalizes both large and small coefficients uniformly; in particular, 
the $\ell_{1}$-norm penalizes the large coefficients  more severely than the smaller coefficients.
In contrast, if the 
best subset selection procedure decides to include a variable in the model, it brings it in without any shrinkage thereby 
draining the effect of its correlated surrogates. 
Upon increasing the degree of regularization, {\texttt {Lasso}} sets more coefficients to zero, but in the process ends up 
leaving out true predictors from the active set. 
Thus, as soon as certain sufficient regularity conditions on the data are violated, 
 {\texttt {Lasso}} becomes suboptimal as (a) a variable selector and (b) in terms of delivering a model with good predictive performance. 

The shortcomings of  {\texttt {Lasso}} are also known in the statistics literature.
In fact, there is 
a significant gap between what can be achieved via best subset selection and {\texttt {Lasso}}: this is supported by empirical (for small problem sizes, i.e., $p\leq 30$) 
and theoretical evidence, see for example, \cite{raskutti2011minimax,zhang2014lower,mhf-09-jasa,greenshtein2006best,zhang2012general,shen2013constrained} and the  references therein. 
Some discussion is also presented herein, in Section~\ref{sec:stat-prop1}.

 To address  the shortcomings, 
non-convex penalized regression is often  used to ``bridge'' the gap between the convex $\ell_{1}$ penalty and the 
combinatorial $\ell_{0}$ penalty~\cite{mhf-09-jasa,FF93,Fan01,zhang2010nearly,ZH08,FJ08,zou2006a, zouli08,tzhang-09N,boyd08-new}. 
Written in Lagrangian form, this gives rise to continuous non-convex 
optimization problems of the form:
\begin{equation}\label{non-convex-1}
\half \| \M{y} - \M{X} \B\beta\|_2^2 + \sum_{i} p(|\beta_{i}| ; \gamma ; \lambda),
\end{equation}
where $p(|\beta| ; \gamma; \lambda)$ is a non-convex function in $\beta$ with $\lambda$ and $\gamma$ denoting the degree of 
regularization and non-convexity, respectively. 
Typical examples of non-convex penalties  include the minimax concave penalty (MCP), the smoothly clipped absolute deviation (SCAD),  and 
 $\ell_{\gamma}$ penalties~(see for example, \cite{FF93,mhf-09-jasa,zouli08,Fan01}).
There is strong statistical evidence indicating the usefulness of estimators obtained as minimizers of non-convex penalized problems~\eqref{non-convex-1} 
over {\texttt {Lasso}}~see for example~\cite{zhang2012general,loh2013regularized,zhang2010nearly,fan2011nonconcave,van2011adaptive,lv2009unified,Lv-2014,fan2013asymptotic}. 
In a recent paper, \cite{Lv-2014} discuss the usefulness of non-convex penalties over convex penalties (like {\texttt {Lasso}}) in identifying important covariates, leading 
to efficient estimation strategies in high dimensions. They describe interesting connections between $\ell_{0}$ regularized least squares and 
least squares with the hard thresholding penalty; and  in the process develop comprehensive global
properties of hard thresholding regularization in terms of various metrics. \cite{fan2013asymptotic} establish asymptotic equivalence of a wide class of regularization
methods in high dimensions with comprehensive sampling properties on both global and computable solutions.

Problem~\eqref{non-convex-1} mainly leads to a family of continuous and non-convex optimization problems. Various effective 
nonlinear optimization based methods~(see for example \cite{zouli08,Fan01,boyd08-new,loh2013regularized,zhang2010nearly,mhf-09-jasa} and the  references therein)
have been proposed in the literature to obtain good local minimizers to Problem~\eqref{non-convex-1}. In particular~\cite{mhf-09-jasa} proposes {\texttt {Sparsenet}}, a 
coordinate-descent procedure to trace out a surface of local minimizers for Problem~\eqref{non-convex-1} for the MCP penalty using effective 
warm start procedures.
None of the existing approaches for solving Problem~\eqref{non-convex-1}, however, come 
with  guarantees of how close the solutions are to the global minimum of Problem~\eqref{non-convex-1}.

%\cite{} proposes very efficient and scalable path-seeking algorithms to obtain high quality local solutions to Problem~\eqref{} across different values of $\lambda, \gamma$ using one-at-a time coordinate descent methods. 

The Lagrangian version of~\eqref{eq-card-k} given by
\begin{equation}\label{L0-lag-1}
\half \| \M{y} - \M{X} \B\beta\|_2^2 + \lambda \sum_{i=1}^{p} 1(\beta_{i} \neq 0), \;\; 
\end{equation}
may be seen as a special case of~\eqref{non-convex-1}.
Note that, due to non-convexity, problems~\eqref{L0-lag-1} and~\eqref{eq-card-k} are \emph{not} equivalent. 
 Problem~\eqref{eq-card-k} allows one to control the exact level of sparsity via the choice of $k$, unlike~\eqref{L0-lag-1} where there is no clear
 correspondence between $\lambda$ and $\mmk$.
 Problem~\eqref{L0-lag-1} is a discrete optimization problem unlike continuous optimization problems~\eqref{non-convex-1} arising 
 from continuous non-convex penalties.
 
 Insightful statistical properties of Problem~\eqref{L0-lag-1} have been explored from a theoretical viewpoint in~\cite{zhang2012general,greenshtein2006best,greenshtein2004persistence,shen2013constrained}.
 %, assuming that  the global solution to Problem~\eqref{L0-lag-1} is computable. 
  \cite{shen2013constrained} points out that~\eqref{eq-card-k}  is preferable over~\eqref{L0-lag-1} in terms of superior statistical properties of the 
  resulting estimator.
The aforementioned papers, however, do not discuss  methods to obtain provably optimal  solutions to problems~\eqref{L0-lag-1} or~\eqref{eq-card-k}, 
  and to the best of our knowledge, computing optimal  solutions to problems~\eqref{L0-lag-1} and~\eqref{eq-card-k} 
 is   deemed as intractable. 

\paragraph{Our Approach}
In this paper, we propose a novel framework via which the best subset selection problem can be solved to   optimality or near optimality 
in problems of practical interest within a reasonable time frame. 
%We thus send out a message to practitioners and researchers to 
%revisit the standard practice of dismissing best subset problems as prohibitively expensive and intractable.  
At the core of our  proposal is a computationally tractable framework
% to compute a globally optimal solution to the best subset Problem~\eqref{eq-card-k}
%bringing to 
that brings to bear the power of modern discrete optimization methods: discrete first order methods motivated by first order methods in 
convex optimization~\cite{nesterov2004introductory} and mixed integer optimization (MIO), see~\cite{bertsimas2005optimization_new}. 
We do not guarantee polynomial time solution times as these do not exist for the best subset problem unless P=NP.  Rather, our view of computational tractability  is the ability of a method to solve problems of practical interest in times that are appropriate for the application addressed.  
An  advantage of our approach  is that it adapts to variants of the best subset 
regression problem of the form:
%\begin{equation}\label{eq-card-form-0-intro}
$$\begin{myarray}{l  l }
\min\limits_{\B\beta}&  \frac12\| \M{y} - \M{X}\B\beta \|_{q}^{q} \\
s.t. & \|\B\beta\|_0 \leq \mmk\\
& \M{A}\B\beta \leq \M{b},
\end{myarray} $$
%\end{equation}
where $\M{A}\B\beta \leq \M{b}$ represents polyhedral constraints and $q\in \{1, 2\}$ refers to a least absolute deviation or 
the least squares loss function on the residuals $\M{r}:= \M{y} - \M{X}\B\beta$.

\paragraph{Existing approaches in the Mathematical Optimization Literature}
In a seminal paper~\cite{FW74}, the authors describe a leaps and bounds procedure for computing global solutions to Problem~\eqref{eq-card-k} 
(for the classical $n > p$ case) which can be achieved with computational effort significantly less than complete enumeration. 
{\texttt {leaps}}, a state-of-the-art {\texttt {R}} package uses this principle to perform best subset selection for problems with $n > p $ and $p\leq 30$.
\cite{bertsimas2009algorithm} proposed a tailored branch-and-bound scheme that can be applied to 
Problem~\eqref{eq-card-k} using ideas from~\cite{FW74} and techniques in 
quadratic optimization, extending and enhancing the proposal of~\cite{bienstock1996computational}. 
The proposal of~\cite{bertsimas2009algorithm} concentrates on obtaining high quality upper bounds for Problem~\eqref{eq-card-k}
and is less scalable than the methods presented in this paper. 

 \paragraph{Contributions}
We summarize our contributions in this paper below:
\begin{enumerate}

\item We use   MIO  to find a  provably optimal solution  for the best subset problem. Our approach
has the appealing characteristic that if we terminate the algorithm early, we obtain a solution with a guarantee on its suboptimality. 
Furthermore, our framework can accommodate side constraints on  $\B\beta$ and also 
 extends   to finding best subset solutions for the least absolute deviation loss function.

\item We introduce a general  algorithmic framework based on a discrete extension of modern  first order continuous optimization methods
that provide near-optimal solutions for 
the best subset problem.   
The MIO algorithm  significantly benefits from  solutions obtained by the first order methods  and 
problem specific information that can be computed in a data-driven fashion.

\item We report computational results with both synthetic and real-world datasets that show that our proposed framework 
can deliver provably  optimal solutions for problems of size $n$ in the 1000s and $p$ in the 100s in minutes. 
 For high-dimensional  problems with $n\in \{50, 100\}$ and    $p \in \{ 1000 , 2000\}$, with the aid of warm starts and further problem-specific information, our approach
 finds near optimal solutions in minutes but takes hours to prove optimality.

\item We investigate the statistical properties of best subset selection procedures for practical problem sizes, 
which to the best of our knowledge, have remained largely unexplored to date.  We demonstrate the favorable 
predictive performance and sparsity-inducing properties of the best subset selection procedure over its competitors
%{\texttt {Lasso}},  {\texttt {Sparsenet}} and stepwise regression 
in a wide variety of real and synthetic examples for the least squares and absolute deviation loss functions.

\end{enumerate}

The structure of the paper is as follows. 
In Section \ref{sec:MIO},  we present a brief overview of MIO, including  a summary of  the  computational  advances it has enjoyed  in the last twenty-five years. 
We present the  proposed MIO formulations for the best subset problem as well  as some connections with the compressed sensing literature
for estimating parameters and providing lower bounds  for the MIO formulations that improve their computational performance. 
In Section \ref{sec:DFOM}, we develop a discrete extension of   first order methods  in convex optimization
 to obtain near optimal solutions for  the best subset problem and establish its convergence properties---the proposed algorithm and its properties may be of independent interest.
Section~\ref{sec:stat-prop1} briefly reviews some of the statistical properties of the best-subset solution, highlighting the performance gaps in prediction error, 
over regular {\texttt{Lasso}}-type 
estimators. 
In Section~\ref{sec:computation-ls}, we   perform a variety of computational tests on synthetic  and real datasets
 to assess the algorithmic and statistical performances of  our  approach for the least squares loss function for 
both the classical overdetermined  case   $n > p$, and the   high-dimensional   case  $ p\gg  n$.
In Section \ref{sec:cLAD}, we report computational results for the least absolute deviation loss function. 
In Section \ref{sec:conc}, we include our concluding remarks. Due to space limitations, some of the material has been 
relegated to the Appendix.

\section{Mixed Integer Optimization Formulations}\label{sec:MIO}
In this section, we present a brief overview of MIO, including   the simply astonishing  advances it has enjoyed  in the last twenty-five years. 
We then   present the  proposed MIO formulations for the best subset problem as well as some connections with the compressed sensing literature
for estimating parameters. We also present completely data driven methods  to estimate parameters in the MIO formulations that improve their computational performance. 

\subsection{Brief Background on MIO}\label{sec:background-mio}
The general form  of a  Mixed Integer Quadratic Optimization (MIQO) problem is as follows:
%The generic MIQP framework concerns the following optimization problem:
%\begin{equation}\label{miqp_general}
$$\begin{myarray}[1.5]{l c cl l}
%\mini & \begin{pmatrix} \B\theta \\ \B\alpha \end{pmatrix}^T  \M{Q} \begin{pmatrix} \B\theta \\ \B\alpha \end{pmatrix}  + \B\theta'\B{b}  + \B\alpha'\B{d}\\
\min & \B\alpha^T  \M{Q} \B\alpha +  \B\alpha^{T}  \M{a} \\
s.t. & \;\; \M{A} \B\alpha  \B{\leq} \M{b}\\
& \;\; \alpha_i \in \{ 0 , 1 \},\quad\forall i\in {\mathcal I}\\
& \;\; \alpha_j \in \mathbb{R}_+,\quad\forall j \notin {\mathcal I},
\end{myarray}$$
%\end{equation}
where $\M{a}\in \mathbb{R}^{m}, \M{A} \in  \mathbb{R}^{k \times m}, \M{b} \in  \mathbb{R}^{k}$ and $\M{Q}\in  \mathbb{R}^{m \times m}$ (positive semidefinite) are the given parameters of the  problem; 
$\mathbb{R}_+$ denotes the non-negative reals, the symbol $\B{\leq}$ denotes  element-wise inequalities and
we optimize over $\B\alpha \in  \mathbb{R}^{m}$ containing both discrete ($\alpha_i, i \in {\mathcal I}$)  and continuous ($\alpha_i, i \notin {\mathcal I}$) variables, with ${\mathcal I} \subset \{1, \ldots, m\}$. For background on MIO see \cite{bertsimas2005optimization_new}.  Subclasses of MIQO  problems include convex quadratic optimization problems (${\mathcal I} = \emptyset$), mixed integer  
 ($\M{Q} = \M{0}_{m \times m}$)  and linear optimization problems
( $ {\mathcal I} = \emptyset, \M{Q} = \M{0}_{m \times m}$). 
Modern integer optimization solvers such as {\textsc{Gurobi}} and {\textsc{Cplex}} are able to tackle  MIQO  problems. 

 In the last twenty-five  years (1991-2014)  the computational power of MIO  solvers has increased at an astonishing rate. In \cite{bixby}, to measure the speedup of MIO solvers, the same set of MIO problems were tested on the same computers using twelve consecutive versions of {\textsc{Cplex}} and version-on-version speedups were reported. The versions tested ranged from {\textsc{Cplex}} 1.2, released in 1991 to {\textsc{Cplex}} 11, released in 2007. Each version released in these years produced a speed improvement on the previous version, leading to a total speedup factor of more than 29,000 between the first and last version tested (see \cite{bixby}, \cite{nem} for details). {\textsc{Gurobi}} 1.0, a MIO solver which was first released in 2009, was measured to have similar performance to {\textsc{Cplex}} 11. Version-on-version speed comparisons of successive {\textsc{Gurobi}} releases have shown a speedup factor of more than 20 between {\textsc{Gurobi}} 5.5, released in 2013, and {\textsc{Gurobi}} 1.0 (\cite{bixby}, \cite{nem}). The combined machine-independent speedup factor in MIO solvers between 1991 and 2013 is 580,000. 
 This impressive  speedup factor is due to incorporating both theoretical and practical advances   into MIO solvers. Cutting plane theory, disjunctive programming for branching rules, improved heuristic methods, techniques for preprocessing MIOs, using linear optimization  as a black box to be called by MIO solvers, and improved linear optimization  methods have all contributed greatly to the speed improvements in MIO solvers~\cite{bixby}. 

\begin{figure}[H]
\centering
\includegraphics[height = .2\textheight, width = .4\textwidth, trim = 5mm 6mm 3mm 1mm, clip]{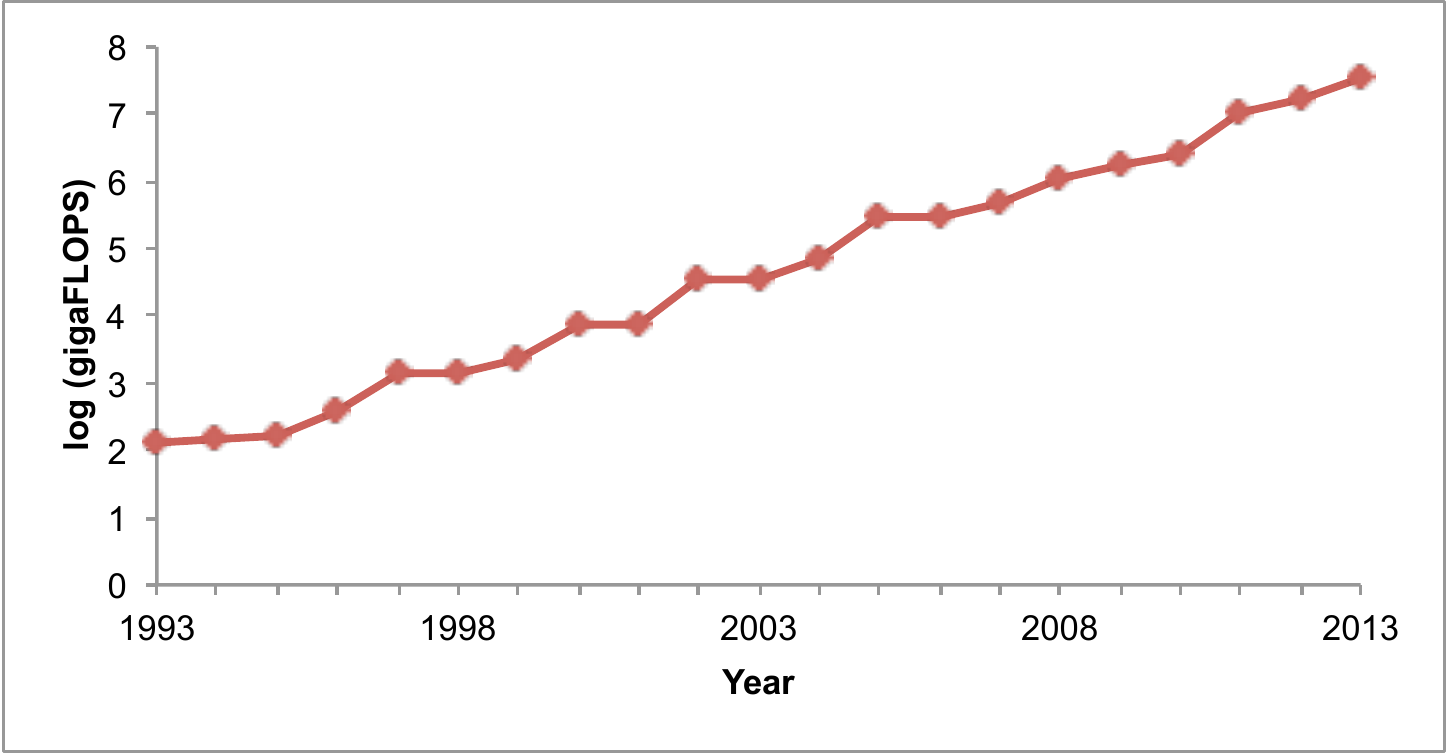}
\caption{Log of Peak Supercomputer Speed from 1993--2013.}
\label{super}
\end{figure}

In addition, the past twenty   years have also brought dramatic improvements in hardware. 
Figure~\ref{super}  shows the exponentially increasing speed of supercomputers over the past twenty years, measured in billion floating point operations per second \cite{supercomputer}. The hardware speedup from 1993 to 2013 is approximately $10^{5.5}\sim 320,000$. 
When both hardware and software improvements are considered, the overall speedup is approximately 200 billion!
Note that the speedup factors cited here refer to mixed integer linear optimization problems, not MIQO problems.   The speedup factors for MIQO problems are similar.
  MIO solvers provide both feasible solutions as well as   lower bounds to the optimal value. As the MIO solver progresses towards the optimal solution, the lower bounds improve and provide an
increasingly better  guarantee of suboptimality, which is especially useful
if the MIO solver  is stopped before reaching the global optimum. In contrast, heuristic methods do not provide such a certificate of suboptimality.

The belief that MIO approaches to problems in statistics are   not practically relevant was formed in the 1970s and 1980s and it was at the time justified. 
Given the astonishing   speedup of MIO solvers and computer hardware   in the last twenty-five years,
the mindset of MIO  as theoretically elegant but practically irrelevant  is  no longer justified. In this paper, we provide   empirical evidence of this 
fact in the context of the best subset selection problem.

\subsection{MIO Formulations for the Best Subset Selection Problem}\label{sec:formulations}
%\paragraph{Big-M Formulation}
We first present a simple reformulation to Problem~\eqref{eq-card-k} as a MIO (in fact a MIQO) problem:
\begin{equation}\label{eq-card-form1}
\begin{myarray}[1.5]{l  l }
Z_1=\min\limits_{\B\beta, \M{z} }&  \;\; \frac12\| \M{y} - \M{X}\B\beta \|_2^2 \\
s.t. & \;\;  -{\mathcal M}_{U}z_{i} \leq \beta_{i} \leq {\mathcal M}_{U}  z_{i}, i = 1, \ldots, p\\
& \;\; z_{i} \in \{ 0 , 1 \}, i = 1, \ldots, p\\
& \;\; \sum\limits_{i=1}^{p} z_{i} \leq \mmk ,
\end{myarray} 
\end{equation}
%%@@
where $\M{z} \in \{ 0 , 1\}^{p}$ is a binary variable and 
${\mathcal M}_{U} $ is a constant such that if $\widehat{\B\beta}$ is a minimizer of Problem~\eqref{eq-card-form1}, then
${\mathcal M}_{U} \geq \|\widehat{\B\beta}\|_\infty$. 
%Such formulations are called ``big-M" formulations and are common in 
%optimization problems. 
If $z_{i} = 1$, then $|\beta_{i}| \leq {\mathcal M}_{U} $ and if $z_{i} = 0$, then $\beta_{i} =0$.
Thus, $\sum_{i=1}^p z_{i}$ is an indicator of the number of non-zeros in $\B\beta$. 

Provided that ${\mathcal M}_{U} $ is chosen to be sufficently large with
${\mathcal M}_{U}  \geq \|\widehat{\B\beta}\|_\infty$, a solution to Problem~\eqref{eq-card-form1} will be a solution to Problem~\eqref{eq-card-k}.
Of course, ${\mathcal M}_{U} $ is not known a priori, and 
a small value of ${\mathcal M}_{U} $ may lead to a solution different from~\eqref{eq-card-k}.
The choice of ${\mathcal M}_{U} $ affects the strength of the formulation and is critical for obtaining good lower bounds in practice.
In Section~\ref{sec:enhanced_form-1} we describe how to find appropriate values for   ${\mathcal M}_{U}$.
Note that there are other MIO formulations, presented herein (See Problem~\eqref{eq-card-form2}) that do not rely on a-priori
specifications of ${\mathcal M}_{U}$. However, we will stick to formulation~\eqref{eq-card-form1} for the time being, since it provides some interesting
connections to the {\texttt{Lasso}}.

Formulation~\eqref{eq-card-form1} leads to interesting insights, especially via the structure of the convex hull of its constraints, as illustrated next:
%$$
%\begin{myarray}[1.2]{l l }
%& \text{Conv} \left( \left\{ \B\beta : |\beta_{i}| \leq {\mathcal M}_{U}  z_{i}, z_{i} \in \{ 0 , 1 \}, i = 1, \ldots, p, \sum\limits_{i=1}^{p} z_{i} \leq {\mmk} \right\} \right)  \\
%=& \{ \B\beta : \|\B\beta\|_{\infty} \leq {\mathcal M}_{U}  ,  \| \B\beta\|_1 \leq {\mathcal M}_{U}  {\mmk} \}  \\
%\subseteq & \{ \B\beta : \| \B\beta\|_1 \leq {\mathcal M}_{U}  {\mmk} \}  .
%\end{myarray}
%$$
$$
\begin{myarray}[1.2]{l l }
& \text{Conv} \left( \left\{ \B\beta : |\beta_{i}| \leq {\mathcal M}_{U}  z_{i}, z_{i} \in \{ 0 , 1 \}, i = 1, \ldots, p, \sum\limits_{i=1}^{p} z_{i} \leq {\mmk} \right\} \right)  \\
=& \{ \B\beta : \|\B\beta\|_{\infty} \leq {\mathcal M}_{U}  ,  \| \B\beta\|_1 \leq {\mathcal M}_{U}  {\mmk} \}  \subseteq  \{ \B\beta : \| \B\beta\|_1 \leq {\mathcal M}_{U}  {\mmk} \}  .
\end{myarray}
$$
Thus,  the minimum of Problem~\eqref{eq-card-form1} is lower-bounded by the optimum objective value of both the following convex optimization problems:
\begin{align}
Z_{2}:=& \min\limits_{\B\beta}  \;\; \frac12\| \M{y} - \M{X}\B\beta \|_2^2 \;\; &\sbt &\;\; \|\B\beta\|_{\infty} \leq {\mathcal M}_{U}  ,   \| \B\beta \|_1 \leq {\mathcal M}_{U} {\mmk} \label{eq-relax-1}\\
Z_{3}:=& \min\limits_{\B\beta}  \;\; \frac12\| \M{y} - \M{X}\B\beta \|_2^2 \;\; &\sbt& \;\;  \| \B\beta \|_1 \leq {\mathcal M}_{U}  {\mmk} \label{eq-relax-2} ,
\end{align}
where~\eqref{eq-relax-2} is the familiar {\texttt {Lasso}} in constrained form. This is a weaker relaxation than formulation~\eqref{eq-relax-1}, which in addition 
to the $\ell_{1}$ constraint on $\B\beta$, has box-constraints controlling the values  of the $\beta_{i}$'s. 
It is easy to see  that the following ordering exists: $ Z_{3} \leq Z_2\leq Z_1,$
with the inequalities being strict in most instances. 

In terms of approximating the optimal solution to Problem~\eqref{eq-card-form1},
the MIO solver begins by first solving a continuous relaxation of Problem~\eqref{eq-card-form1}. 
The {\texttt {Lasso}} formulation~\eqref{eq-relax-2} is weaker than this root node relaxation.
Additionally, MIO is typically able to significantly improve the quality of the root node solution as the MIO solver progresses toward the optimal solution.

%\textcolor{red}{comments on above paragraph: we should not say MIO "solver"(?), I do not think {\textsc{Gurobi}} solves this relaxation at the root-node. We should make it more specific, like:
%"MIO branch and bound procedure begins by solving..."}

\begin{figure}[H]
\centering
\scalebox{.99}[.8]{\begin{tabular}{ c c }
 \small{ \sf  $\mmk = 6$}  &  \small{$\mmk = 7$}   \\
\includegraphics[height = .3\textheight, width = .45\textwidth, trim = 0mm 12mm 3mm 14mm, clip]{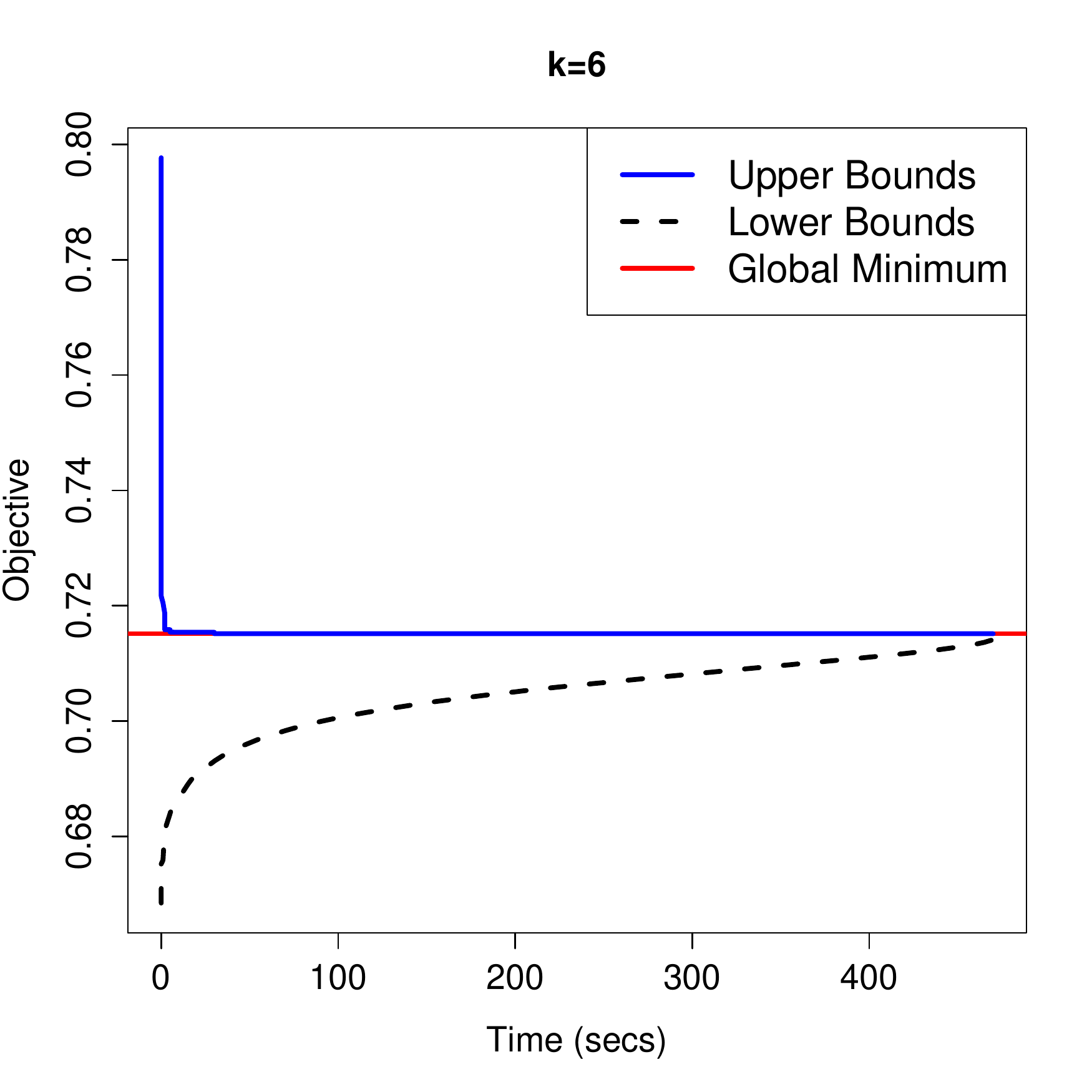}&
\includegraphics[height = .3\textheight,width = .45\textwidth,trim = 6mm 12mm 3mm 14mm, clip]{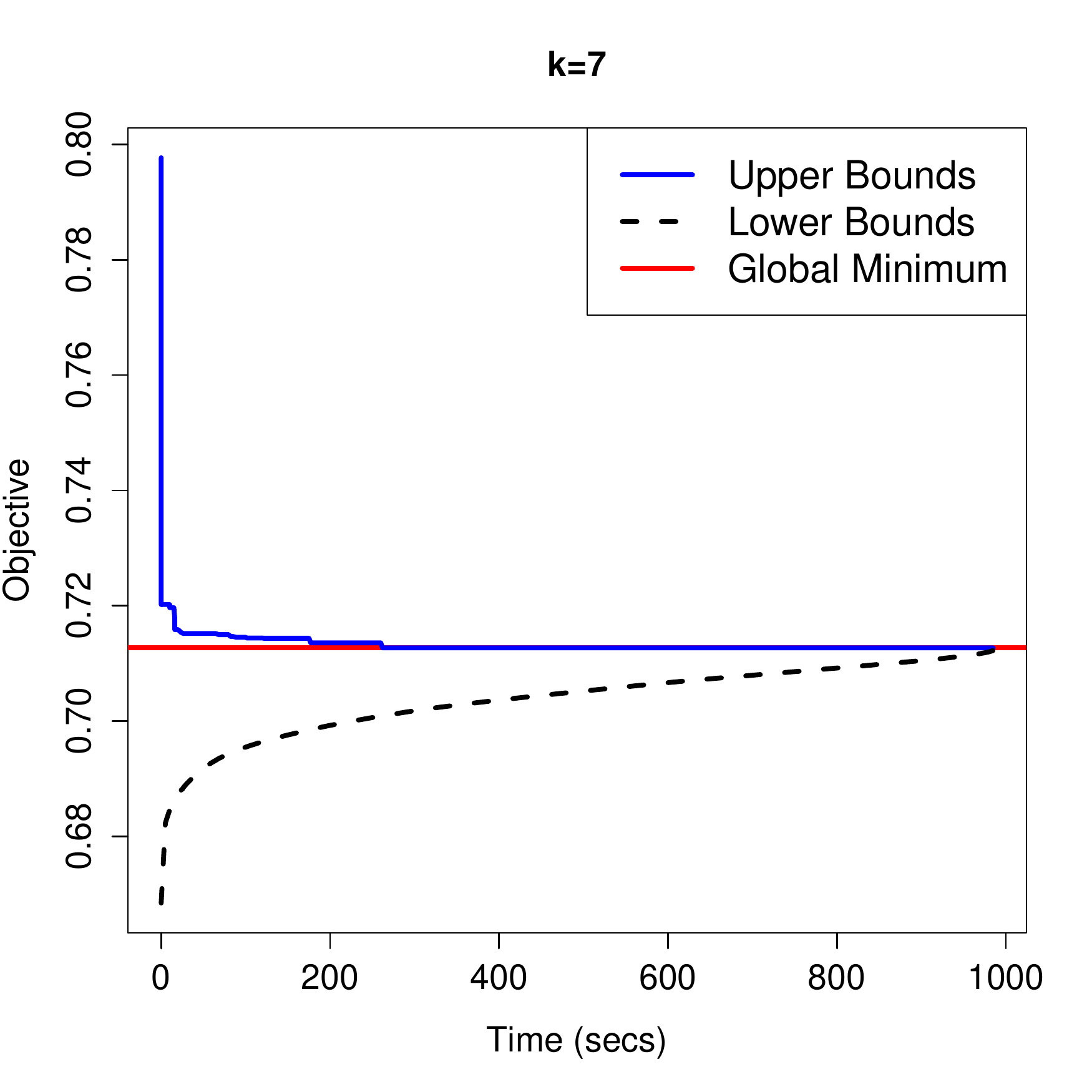} \\
\includegraphics[height = .3\textheight,width = .45\textwidth,trim = 0mm 12mm 3mm 14mm, clip]{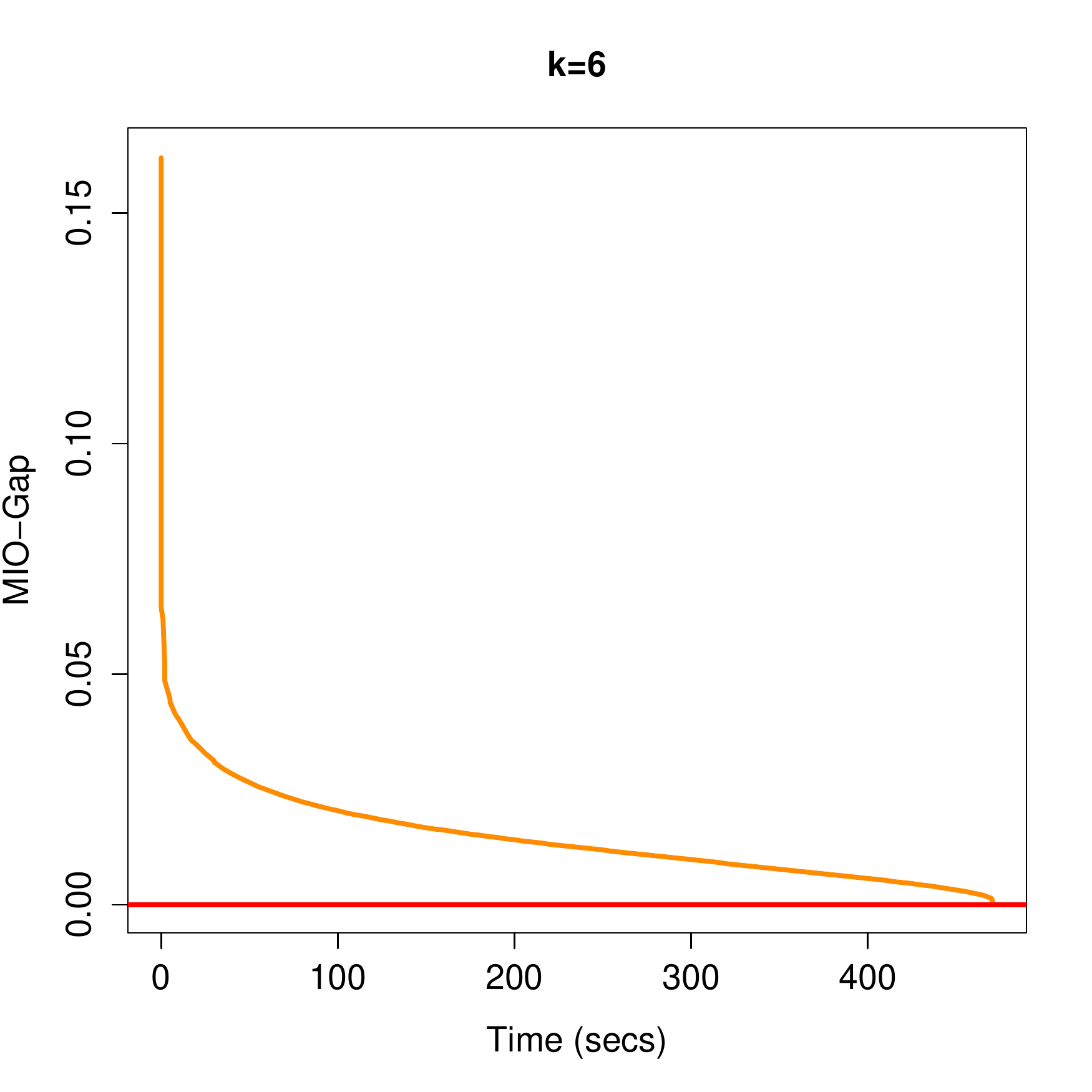}&
\includegraphics[height = .3\textheight,width = .45\textwidth,trim = 6mm 12mm 3mm 14mm, clip]{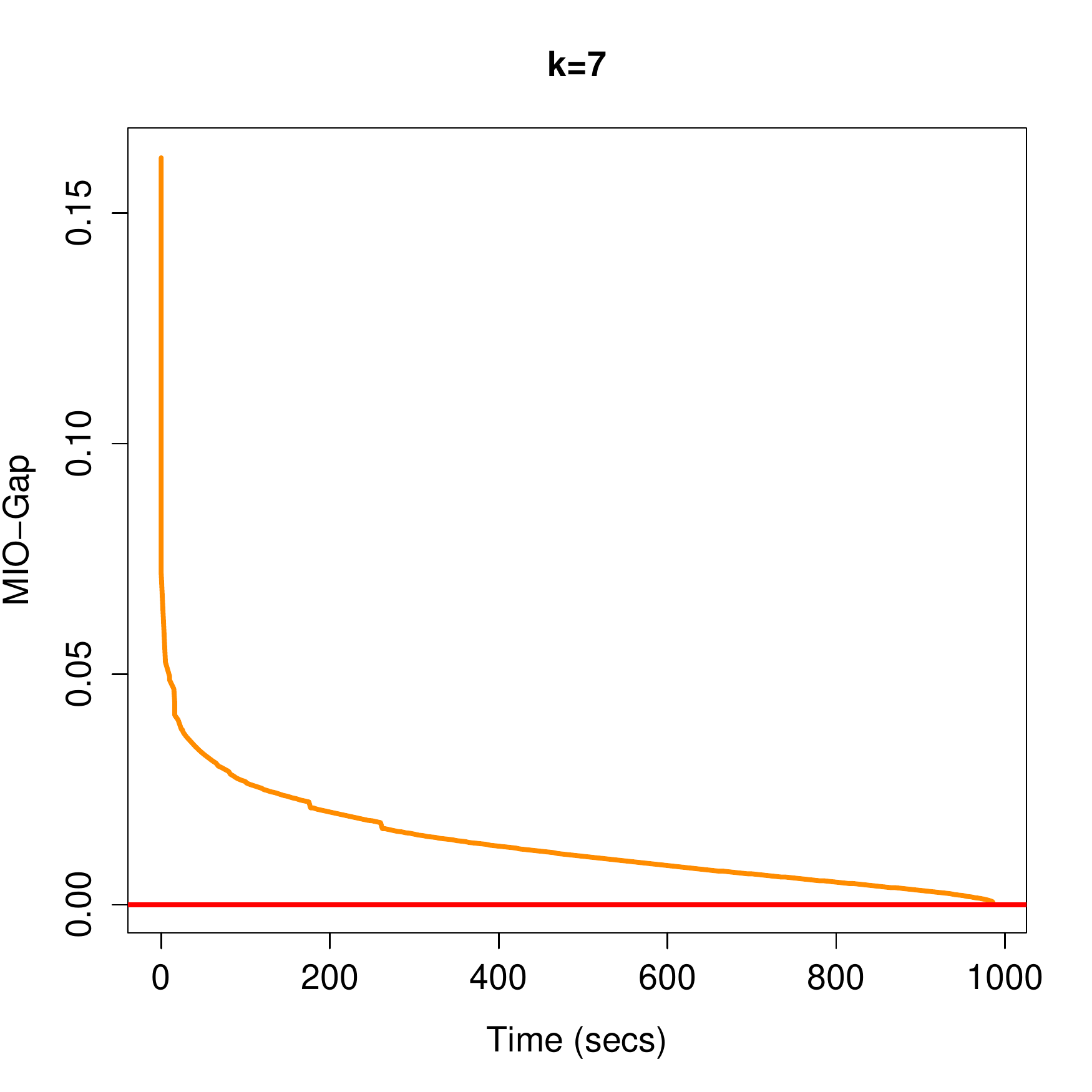}\\
 {\small \sf  Time (secs) } &  {\small \sf  Time (secs)   }\\
 \end{tabular}}
\caption{{ \small {The typical evolution of the MIO formulation~\eqref{eq-card-form2} for the diabetes
dataset with $n = 350 , p = 64$ with $\mmk = 6$ (left panel) and 
$\mmk = 7$ (right panel).  The top panel shows the evolution of upper and lower bounds with time. The lower panel shows the evolution of the corresponding 
MIO gap with time. Optimal solutions for both the problems are found in a few seconds  in both examples,  but it takes 10-20 minutes to certify  optimality via the lower bounds.  
Note that the time taken for the MIO to certify convergence to the global optimum increases with increasing $\mmk$.} }  }\label{fig-diab-data1}
\end{figure}

To motivate the reader we provide an example of the evolution (see Figure~\ref{fig-diab-data1}) 
of the MIO formulation~\eqref{eq-card-form2} for the Diabetes dataset~\cite{LARS}, 
with $n = 350 , p = 64$ (for further details on the dataset see Section~\ref{sec:computation-ls}).

%From the viewpoint of MIO in terms of approximating the minimum solution to Problem~\eqref{eq-card-form1}
% the {\texttt {Lasso}} formulation~\eqref{eq-relax-2}  is weaker than the root-node relaxation of Problem~\eqref{eq-card-form1} --- 
%typically branch-and-bound and branch and cut methods (as a part of the MIO algorithm) significantly improve the quality of the root node solution 
%with subsequent iterations in the MIO algorithm. 

%\begin{equation}\label{eq-card-form1-new}
%\begin{myarray}[1.5]{l  l }
%\mini\limits_{\B\beta, \M{z} }&  \;\; \frac12\| \M{y} - \M{X}\B\beta \|^q_q \\
%\sbt & \| \B\beta \|_1 \leq M {\mmk},
%\end{myarray} 
%\end{equation}
%the constraint $\| \B\beta\|_1 \leq Mk$ is the familiar {Lasso} constraint, which as we know, has a two-fold effect:
%\begin{itemize}
%\item induces sparsity in the regression coefficients, and 
%\item shrinks the regression coefficients to control variance and thus predictive accuracy
%\end{itemize}  

Since formulation~\eqref{eq-card-form1} is sensitive to the choice of   ${\mathcal M}_{U} $, we consider an alternative   MIO formulation based on 
Specially Ordered Sets~\cite{bertsimas2005optimization_new} as described next.
\paragraph{Formulations via Specially Ordered Sets}
%We first present a simple reformulation to Problem~\eqref{eq-card-k} as a MIQ:
%Here we propose another MIO formulation to Problem~\eqref{eq-card-form-0}:
Any feasible solution to formulation~\eqref{eq-card-form1} will have
$(1 - z_{i}) \beta_{i} = 0$ for every $i \in \{1, \ldots, p \}$. This constraint 
 can be modeled via integer optimization using Specially Ordered Sets of Type 1~\cite{bertsimas2005optimization_new} (SOS-1). In an SOS-1 constraint, at most one variable in the set can take a nonzero value, that is
$$
 (1 - z_{i}) \beta_{i} = 0 \;\;  \iff  \;\; (\beta_{i}, 1 - z_{i}) : \text{SOS-1}, 
$$
for every $i = 1, \ldots, p.$ 
This leads to the following formulation of~\eqref{eq-card-k}:
\begin{equation}\label{eq-card-form2}
\begin{myarray}[1.5]{l  l }
\min\limits_{\B\beta, \M{z} }&  \;\;\frac{1}{2} \; \| \M{y} - \M{X}\B\beta \|_2^2 \\
s.t. & \;\; (\beta_{i}, 1 - z_{i}) : \text{SOS-1,}\;\; i = 1, \ldots, p\\
& \;\; z_{i} \in \{ 0 , 1 \}, i = 1, \ldots, p\\
& \;\; \sum\limits_{i=1}^{p} z_{i} \leq \mmk.
\end{myarray} 
\end{equation}
We note that Problem~\eqref{eq-card-form2} can in principle be used to obtain global solutions to Problem~\eqref{eq-card-k} --- Problem~\eqref{eq-card-form2} unlike Problem~\eqref{eq-card-form1} does not require any specification of 
the parameter ${\mathcal M}_{U}$.

%In presence of additional constraints on the amplitudes of $\B\beta$ and $\ell_{1}$-norm of $\B\beta$, we have 
%the following version of the best subset problem, with additional linear constraints:
%\begin{equation}\label{prob-bs-ngep-1}
%\mini\limits_{\B\beta}\;\; \frac12\| \M{y} - \M{X}\B\beta \|_2^2 \;\; \sbt \;\; \|\B\beta\|_{\infty} \leq {\mathcal M}_{U} ,   \| \B\beta \|_1 \leq {\mathcal M}_{\ell} 
%\end{equation}

We now proceed to present a more structured representation of Problem~\eqref{eq-card-form2}. Note that objective in this problem is a convex quadratic function in the continuous variable $\B\beta$, which can be formulated explicitly as:
\begin{equation}\label{eq-card-form2-0}
\begin{myarray}[1.5]{l  l }
\min\limits_{\B\beta, \M{z} }&  \;\;\frac{1}{2}\;  \B\beta^{T} \M{X}^{T}\M{X} \B\beta - \langle \M{X}'\M{y}, \B\beta  \rangle  + \frac{1}{2}\; \| \M{y}\|_2^2\\
s.t.  & \;\; (\beta_{i}, 1 - z_{i}) : \text{SOS-1,}\;\; i = 1, \ldots, p\\
& \;\; z_{i} \in \{ 0 , 1 \}, i = 1, \ldots, p\\
& \;\; \sum\limits_{i=1}^{p} z_{i} \leq \mmk \\
&  - {\mathcal M}_{U}  \leq \beta_{i} \leq {\mathcal M}_{U}  , i = 1, \ldots, p\\
&  \| \B\beta\|_{1} \leq {\mathcal M}_{\ell}  .
\end{myarray} 
\end{equation}

We also provide problem-dependent constants ${\mathcal M}_{U}$ and ${\mathcal M}_{\ell}  \in [0, \infty]$. ${\mathcal M}_{U} $
provides an upper bound on the absolute value of the regression coefficients and ${\mathcal M}_{\ell} $  provides an upper bound on the $\ell_{1}$-norm of $\B\beta$. Adding these bounds typically leads to improved performance of the MIO, especially in 
delivering lower bound certificates. In Section~\ref{sec:enhanced_form-1}, we describe several approaches  to compute these parameters from the data.

 We also  consider another formulation for~\eqref{eq-card-form2-0}:
\begin{equation}
\label{eq-card-form2-1}
\begin{myarray}[1.5]{l  l }
\min\limits_{\B\beta, \M{z} , \B\zeta }&  \;\;  \frac{1}{2}\; \B\zeta^{T}\B\zeta - \langle  \M{X}'\M{y}, \B\beta   \rangle  +  \frac{1}{2}\; \| \M{y}\|_2^2  \\
s.t. \;\; & \B\zeta = \M{X} \B\beta \\
& (\beta_{i}, 1 - z_{i}) : \text{SOS-1,}\;\; i = 1, \ldots, p\\
& \;\; z_{i} \in \{ 0 , 1 \}, i = 1, \ldots, p\\
& \;\; \sum\limits_{i=1}^{p} z_{i} \leq \mmk \\
%\label{eq-card-form2-1}
%\begin{myarray}[1.5]{l  l }
&  - {\mathcal M}_{U}  \leq \beta_{i} \leq {\mathcal M}_{U}  , i = 1, \ldots, p\\
&  \| \B\beta\|_{1} \leq {\mathcal M}_{\ell} \\
&  - {\mathcal M}^{\zeta}_{U} \leq \zeta_{i} \leq {\mathcal M}^{\zeta}_{U} , i = 1, \ldots, n \\
&  \| \B\zeta\|_{1} \leq {\mathcal M}^{\zeta}_{\ell},
\end{myarray} 
\end{equation}

where the optimization variables are $\B\beta \in  \mathbb{R}^{p}, \B\zeta \in  \mathbb{R}^{n}$, $\M{z} \in  \{0,1\}^p$
and $ {\mathcal M}_{U}  ,{\mathcal M}_{\ell} ,{\mathcal M}^{\zeta}_{U}, {\mathcal M}^{\zeta}_{\ell} \in [ 0 , \infty] $ are problem specific parameters.
Note that the objective function in formulation~\eqref{eq-card-form2-1} involves a quadratic form in $n$ variables and a linear  function in $p$ variables.
Problem~\eqref{eq-card-form2-1} is equivalent to the following variant of the best subset problem:
\begin{equation}\label{prob-bs-nLep-1}
\begin{myarray}[1.5]{l  l }
\min\limits_{\B\beta}& \;\; \frac12\| \M{y} - \M{X}\B\beta \|_2^2 \\
s.t.  & \;\; \|\B\beta\|_{0} \leq k~~~~~~~~~~~ \\
& \;\; \|\B\beta\|_{\infty} \leq {\mathcal M}_{U} ,   \| \B\beta \|_1 \leq {\mathcal M}_{\ell} \\
 &  \;\; \|\M{X}\B\beta\|_{\infty}\leq {\mathcal M}^{\zeta}_{U}, \|\M{X}\B\beta\|_{1} \leq {\mathcal M}^{\zeta}_{\ell}.
 \end{myarray}
\end{equation}

Formulations~\eqref{eq-card-form2-0} and~\eqref{eq-card-form2-1} 
%are \emph{not} equivalent as MIO formulations; they
differ in the size of the quadratic forms that are involved. The current state-of-the-art MIO solvers are better-equipped to handle mixed integer linear optimization problems
 than MIQO problems.
%formulations~\eqref{eq-card-form2-0} and~\eqref{eq-card-form2-1} differ in the size of the quadratic forms that are involved --- this gives some intuition as to why 
%the two formulations~\eqref{eq-card-form2-0} and~\eqref{eq-card-form2-1} are different. 
Formulation  \eqref{eq-card-form2-0} has fewer variables but a quadratic form in $p$ variables---we find this formulation more useful in the $n > p$ regime, with $p$ in the $100$s.
Formulation \eqref{eq-card-form2-1} on the other hand has more variables, but involves a quadratic form in $n$ variables---this formulation is more useful for high-dimensional   problems
 $ p\gg  n $, with $n$ in the $100$s  and $p$ in the $1000$s.

As we said earlier, the bounds on $\B\beta$ and $\B\zeta$ are not required,
%(i.e., can be set to be arbitrarily large) 
but if these constraints are provided, they improve the strength of the MIO formulation. In other words, formulations with 
tightly specified bounds provide better lower bounds to the global optimization problem in a specified amount of time, when compared to a MIO formulation
with loose bound specifications.  
We  next show how these bounds 
can be computed  from given data.

%\subsection{Computing explicit bounds on $\widehat{\B\beta}$, $\widehat{\B\zeta}$}
\subsection{Specification of Parameters}\label{sec:enhanced_form-1}
In this section,  we   obtain estimates for  the quantities ${\mathcal M}_{U} ,{\mathcal M}_{\ell},{\mathcal M}^{\zeta}_{U}, {\mathcal M}^{\zeta}_{\ell}$ 
such that an optimal solution to   
 Problem~\eqref{prob-bs-nLep-1} is also an optimal   solution to Problem~\eqref{eq-card-k}, and vice-versa.

\subsection*{Coherence and Restricted Eigenvalues  of a Model Matrix}
Given a model matrix $\M{X}$,  \cite{tropp2006just} introduced the cumulative coherence function
$$ \mu[k] :=  \max_{|I | = k} \;\; \max_{j \notin I} \sum_{i \in I } | \langle \M{X}_{j} , \M{X}_{i} \rangle  |,$$
where, $\M{X}_{j}$, $j=1,\ldots ,p$ represent the columns of $\M{X}$, i.e., features.

For $k=1$, we obtain the notion of coherence 
 introduced in ~\cite{donoho2001uncertainty,donoho2003optimally} as a measure of the maximal pairwise correlation in absolute value of the columns of $\M{X}$:
%$$\mu := \mu[1] = \max_{i \neq j} |\langle \M{X}_{i}, \M{X}_{j} \rangle|.$$
$\mu := \mu[1] = \max_{i \neq j} |\langle \M{X}_{i}, \M{X}_{j} \rangle|.$

\cite{candes2006near,candes2008restricted}  (see also~\cite{buhlmann2011statistics} and references therein)  introduced the notion 
that a matrix $\M{X}$  
satisfies a restricted eigenvalue condition if 
% $$\M{X}_{I}'\M{X}_{I} \succeq \eta_{k} \cdot \M{I}~~{\rm for~every} ~I  \subset \{ 1, \ldots, p \}: ~| I |\leq k.,~{\rm i.e.,}$$
\begin{equation}
\label{eq:ab1}
\lambda_{\min}( \M{X}_{I}'\M{X}_{I} )\geq \eta_k ~~{\rm for~every} ~I  \subset \{ 1, \ldots, p \}: ~| I |\leq k,
 \end{equation}
 where $\lambda_{\min}( \M{X}_{I}'\M{X}_{I} )$ denotes the smallest eigenvalue of the matrix $ \M{X}_{I}'\M{X}_{I} $. 
An inequality linking  $\mu [k]$ and $\eta_k$ is as follows. 
\begin{prop}\label{prop-cumul-coherence-1}  The following bounds hold :
\begin{enumerate}
\item[{\bf (a)}]  \cite{tropp2006just}:~~~$\mu[k] \leq \mu \cdot k .$
\item[{\bf (b)}]  \cite{donoho2003optimally} :~~~$\eta_k\geq 1 - \mu[k-1] \geq 1-  \mu \cdot (k-1) .$
\end{enumerate}
\end{prop}
The computations of $\mu [k]$ and $\eta_k$ for general $k $ are difficult, while $\mu$ is simple to compute. 
Proposition \ref{prop-cumul-coherence-1}  provides bounds for $\mu [k]$ and $\eta_k$  in terms of the coherence $\mu$. 

\subsection*{Operator Norms of Submatrices}
The  $(p,q)$ operator norm of matrix $\M{A}$ 
 is 
$$ \|\M{A}\|_{p,q} := \max_{\|\M{u} \|_{q} = 1} \| \M{A u} \|_{p}.$$ 
We will use extensively here the   $(1,1)$ operator norm.
We   assume that  each column vector of $\M{X}$
 has unit $\ell_{2}$-norm.  The results derived in the next proposition borrow and enhance techniques developed by~\cite{tropp2006just} in the context of 
analyzing  the $\ell_{1}$---$\ell_{0}$ equivalence  in compressed sensing.

\begin{prop}\label{prop-oper-norm-inverse-1}
For any $I  \subset \{ 1, \ldots, p\}$ with $| I |=k$  we have :
\begin{enumerate}
\item[{\bf (a)}]  
    $ \|\M{X}'_{I}\M{X}_{I} -   \M{I}\|_{1,1} \leq   \mu [k-1].$
    % \leq  \mu \cdot (k-1).$
 \item[{\bf (b)}]  If  the matrix  $\M{X}'_{I} \M{X}_{I}$ is invertible and  $\|\M{X}'_{I}\M{X}_{I} -   \M{I}\|_{1,1} < 1$, then 
$ \| (\M{X}'_{I} \M{X}_{I})^{-1}\|_{1,1} \leq \frac{1}{1 - \mu [k-1]}.$
%\begin{equation}
%\label{eq:aa1}
%\| (\M{X}'_{I} \M{X}_{I})^{-1}\|_{1,1} \leq \frac{1}{1 - \mu [k-1]}.
%\end{equation}
% \item[{\bf (c)}]   If $\mu[k-1] < 1$, then the matrix $\M{X}'_{I} \M{X}_{I}$ is invertible for every $I$ with $| I |=k$. 
 \end{enumerate}
\begin{proof}
See Section~\ref{proof-prop-oper-norm-1}.
\end{proof}
 \end{prop}

We note that Part (b)   also appears in~\cite{tropp2006just} for 
the operator norm $\|  (\M{X}'_{I} \M{X}_{I})^{-1} \|_{\infty,\infty}$.

Given a set $I\subset \{1,\ldots, p\}$  with $| I |=k$ we let $\widehat{\B\beta}_{I}$ denote the 
least squares regression coefficients obtained by regressing $\M{y}$ on $\M{X}_{I}$, i.e., 
$\widehat{\B\beta}_{I} = (\M{X}'_{I} \M{X}_{I})^{-1}\M{X}'_{I}\M{y}$. 
  If we append $\widehat{\B\beta}_{I}$ with zeros in the remaining coordinates we   obtain 
$\widehat{\B\beta}$ as follows: $\widehat{\B\beta} \in \argmin_{\B\beta: \beta_{i} = 0, i \notin I} \| \M{y} - \M{X} \B\beta\|_{2}^2.$
%$$
%\widehat{\B\beta} \in \argmin_{\B\beta: \beta_{i} = 0, i \notin I} \| \M{y} - \M{X} \B\beta\|_{2}^2.
%$$
Note that $\widehat{\B\beta}$ depends  on $I$ but we will suppress the dependence on $I$ for notational convenience.

\subsubsection{Specification of Parameters in terms of Coherence and Restricted Strong Convexity}\label{subsub-nu-eta-1}
Recall that $\M{X}_{j}$, $j=1,\ldots ,p$ represent the columns of $\M{X}$;
and we will use $\M{x}_{i},~i=1,\ldots ,n$ to denote the rows of $\M{X}$. 
As discussed above  $\|\M{X}_{j}\|=1$. 
We   
order the   correlations $| \langle \M{X}_{j} , \M{y} \rangle|$: 
\begin{equation}\label{defn-order-1-ips}
| \langle \M{X}_{(1)} , \M{y} \rangle | \geq | \langle \M{X}_{(2)} , \M{y} \rangle | \ldots \geq | \langle \M{X}_{(p)} , \M{y} \rangle |.
\end{equation}
We finally denote by 
 $\| \M{x}_{i}\|_{1:k}$    the sum of the top $k$ absolute values of the entries of 
$x_{ij}, j \in \{1, 2, \ldots, p \}$. 
%In other words, $ \langle \M{x}_{(i)} , \M{y} \rangle |$ denotes the $i$th largest absolute inner-product.

 \begin{thm}\label{prop-beta-bound1}
 For any $k \geq 1$ such that $\mu[k - 1] < 1$  any optimal solution  $\widehat{\B\beta}$ to  
   \eqref{eq-card-k}
satisfies:
\begin{eqnarray} 
{\bf (a)}~~~~ & \|\widehat{\B\beta}\|_{1} \leq &  \frac{1}{ 1 - \mu[k - 1] } \sum_{j= 1}^{k} | \langle \M{X}_{(j)}, \M{y}  \rangle | \label{l1-norm-bd-beta} .\\
{\bf (b)} ~~~~&  \|\widehat{\B\beta} \|_{\infty}  \leq& \min \left \{
\frac{1}{\eta_k}\sqrt{\sum_{j=1}^{k} | \langle \M{X}_{(j)}, \M{y} \rangle |^2}, \frac{1}{\sqrt{\eta_k}}\|\M{y}\|_2   \right \}  \label{upper-bd-linbet011}.\\
{\bf (c)} ~~~~&
\|\M{X}\widehat{\B\beta}\|_{1} \leq& \min \left\{ \sum_{i=1}^{n} \|\M{x}_{i}\|_{\infty}\|\widehat{\B\beta}\|_{1} , \sqrt{k} \|\M{y}\|_2 \right\} .  \label{l1-norm-bd-xbeta} \\
{\bf (d)} ~~~~&
\|\M{X}\widehat{\B\beta} \|_{\infty}  \leq&    \left( \max_{i = 1, \ldots, n }  \| \M{x}_{i}\|_{1:k} \right) \|\widehat{\B\beta}\|_{\infty} \label{upper-bd-linxbet011}.
\end{eqnarray}
\begin{proof}
For proof see Section~\ref{proof-prop-beta-bound1}.
\end{proof}
\end{thm} 

%\textcolor{red}{Modified by RM - starts}

We note that in the above theorem, the upper bound in Part (a) becomes infinite as soon as $\mu[k - 1] \geq 1$. 
In such a case, we can use purely data-driven bounds by using convex optimization techniques, as described in Section~\ref{subsub-opt-cvx-1}. 

The interesting message conveyed by Theorem~\ref{prop-beta-bound1} is that the upper bounds on 
$ \|\widehat{\B\beta}\|_{1},$ $\|\widehat{\B\beta} \|_{\infty}$,
$\|\M{X}\widehat{\B\beta}\|_{1}$ and
$\|\M{X}\widehat{\B\beta} \|_{\infty}$, corresponding to the Problem~\eqref{prob-bs-nLep-1}
 can all be obtained in terms of $\eta_{k}$ and $\mu[k-1]$, quantities of fundamental interest
 appearing in the analysis of $\ell_{1}$ regularization methods and understanding how close they are to $\ell_{0}$ solutions~\cite{tropp2006just,donoho2001uncertainty,donoho2003optimally,candes2006near,candes2008restricted}.
 On a different note,  Theorem~\ref{prop-beta-bound1} arises from a purely computational motivation and quite curiously, involves the same 
 quantities: cumulative coherence and restricted eigenvalues. 

Note that the quantities~$\mu[k-1], \eta_{k}$ are difficult to compute exactly, but they can be approximated by Proposition~\ref{prop-cumul-coherence-1} which 
provides bounds commonly used in the compressed sensing literature.  
Of course, approximations to these quantities can also be obtained by using subsampling schemes. 
%However, precisely quantifying the quality of approximation becomes difficult in such cases. 

\subsubsection{Specification of Parameters via Convex Quadratic Optimization}\label{subsub-opt-cvx-1}
We provide an alternative purely data-driven way to compute the upper bounds to the parameters
 by solving several simple convex quadratic optimization problems.

\subsubsection*{Bounds on $\hat{\beta}_{i}$'s} For the case $n > p$,
upper and lower bounds on $\hat{\beta}_{i}$ can be obtained by solving the following pair of convex optimization problems: 
\begin{equation}\label{ubs-data-1}
\begin{myarray}[1.5]{l l   l }
u^+_{i} :=&  \max\limits_{\B\beta}  \;\; \beta_{i} & \\
s.t. & \frac12\| \M{y} - \M{X} \B\beta \|^2_{2} \leq \text{UB},&
\end{myarray} \qquad \qquad
\begin{myarray}[1.5]{l  l l }
u^-_{i} :=&  \min\limits_{\B\beta}  \;\; \beta_{i} & \\
s.t. & \frac12\| \M{y} - \M{X} \B\beta \|^2_{2} \leq \text{UB},&
\end{myarray}
\end{equation}
for $i = 1, \ldots, p$. Above, $\text{UB}$ is an upper bound to the minimum of the $k$-subset least squares problem~\eqref{eq-card-k}.
$u_{i}^+$ is an upper bound to $\hat{\beta}_{i}$, since the cardinality constraint $\|\B\beta\|_{0} \leq k$ does not appear in the optimization problem. 
Similarly, $u^{-}_{i}$ is a lower bound to $\hat{\beta}_{i}$. 
The quantity ${\mathcal M}^i_{U} = \max \{ |u^+_{i}| , |  u^{-}_{i} |  \}$ 
serves as an upper bound
to $| \hat{\beta}_{i} |$.
A reasonable choice for $\text{UB}$ is obtained by using the discrete first order methods (Algorithms~1 and 2 as described in Section~\ref{sec:DFOM}) in combination with the 
MIO formulation~\eqref{eq-card-form2} (for a predefined amount of time).
Having obtained ${\mathcal M}^i_{U}$ as described above, we can obtain an upper bound to $\|\widehat{\B\beta}\|_{\infty}$
and $\|\widehat{\B\beta}\|_{1}$ as follows:
%as ${\mathcal M}_{U} =  \max_{i=1, \ldots, p} \;\; {\mathcal M}^i_{U}$.
%This leads to the following upper bound on $\|\widehat{\B\beta}\|_{1}$:
%\begin{equation}\label{bound-l1-1}
%{\mathcal M}_{U} =  \max_{i=1, \ldots, p} \;\; {\mathcal M}^i_{U} \quad \text{and} \quad \|\widehat{\B\beta}\|_{1}  \leq  \sum_{i=1}^{k}  {\mathcal M}^{(i)}_{U} \leq  k{\mathcal M}_{U} ,
% \end{equation}
${\mathcal M}_{U} =  \max_{i}  {\mathcal M}^i_{U}$ and $ \|\widehat{\B\beta}\|_{1}  \leq  \sum_{i=1}^{k}  {\mathcal M}^{(i)}_{U}$
where, ${\mathcal M}^{(1)}_{U} \geq {\mathcal M}^{(2)}_{U} \geq \ldots  \geq {\mathcal M}^{(p)}_{U}$.

Similarly, bounds corresponding to Parts (c) and (d) in Theorem~\ref{prop-beta-bound1} can be obtained by 
using the upper bounds on $\|\widehat{\B\beta}\|_{\infty}, \|\widehat{\B\beta}\|_{1}$ as described above.

Note that the quantities $u_{i}^+$ and $u_{i}^-$ are finite when the level sets of the least squares 
loss function are finite. In particular, the bounds are loose when $p > n$. In the following we describe methods to obtain non-trivial bounds on $\langle \M{x}_{i}, \B\beta \rangle$, for 
$i = 1, \ldots, n$ that apply for arbitrary $n,p$.

\subsubsection*{Bounds on $\langle \M{x}_{i}, \widehat{\B\beta} \rangle$'s} We now provide a generic method to obtain upper and lower bounds on the quantities $\langle \M{x}_{i}, \widehat{\B\beta} \rangle$:
\begin{equation}\label{ubs-data-2}
\begin{myarray}[1.5]{l l   l }
v^+_{i} :=&  \max\limits_{\B\beta}  \;\; \langle \M{x}_{i}, \B\beta \rangle & \\
s.t. & \frac12\| \M{y} - \M{X} \B\beta \|^2_{2} \leq \text{UB},&
\end{myarray} \qquad \qquad
\begin{myarray}[1.5]{l  l l }
v^-_{i} :=&  \min\limits_{\B\beta}  \;\;\langle \M{x}_{i}, \B\beta \rangle & \\
s.t. & \frac12\| \M{y} - \M{X} \B\beta \|^2_{2} \leq \text{UB},&
\end{myarray}
\end{equation}
for $i = 1, \ldots, n$. Note that the bounds obtained from~\eqref{ubs-data-2} are 
non-trivial bounds for both the under-determined $n<p$ and overdetermined cases. The bounds obtained from~\eqref{ubs-data-2} are upper and lower bounds
since we drop the cardinality constraint on $\B\beta$. The bounds are finite since for every $i \in \{1, \ldots, n\}$ the quantity 
$\langle \M{x}_{i}, \B\beta \rangle$ remains bounded in the feasible set for Problems~\eqref{ubs-data-2}.

The quantity ${ \v }_{i} = \max \{ |v^+_{i}| , |  v^{-}_{i} |  \}$  serves as an upper bound
to $| \langle \M{x}_{i}, \B\beta \rangle |$.  In particular, this leads to simple upper bounds on
%$\|\M{X}\widehat{\B\beta}\|_{1}$ and $\|\M{X}\widehat{\B\beta}\|_{\infty}$:
%\begin{equation}\label{bound-l1-1}
%\|\M{X}\widehat{\B\beta}\|_{\infty} \leq  \max_{i=1, \ldots, n} { \v }_{i} \quad \text{and} \quad 
%\|\M{X}\widehat{\B\beta}\|_{1}  \leq  \sum_{i=1}^{n} { \v }_{i},
% \end{equation}
$\|\M{X}\widehat{\B\beta}\|_{\infty} \leq  \max_{i} { \v }_{i}$ and $\|\M{X}\widehat{\B\beta}\|_{1}  \leq  \sum_{i} { \v }_{i}$
and can be thought of completely data-driven methods to estimate bounds appearing in~\eqref{l1-norm-bd-xbeta} and~\eqref{upper-bd-linxbet011}.

We note that Problems~\eqref{ubs-data-1} and~\eqref{ubs-data-2} have nice structure amenable to efficient computation as we discuss in Section~\ref{app:subsub-opt-cvx-1}.

%If they are solved sequentially, then it is possible to take advantage of the advanced warm-start capabilities of simplex type methods, with a total computation time that is orders of magnitude smaller than solving all the problems separately. 
%The above problems are well-structured and hence can also be solved by using specialized first order methods in convex optimization~\cite{nesterov2004introductory,nest-07}.
%In summary, the times to compute Problems~\eqref{ubs-data-1} and~\eqref{ubs-data-2} is substantially smaller when compared to solving a MIO, for the problem sizes we consider herein.
%
%\textcolor{red}{RM: Dimitris, please check if the above is correct}

\subsubsection{Parameter Specifications from Advanced Warm-Starts}\label{spec-warm-start-1}
The methods described above in Sections~\ref{subsub-nu-eta-1} and~\ref{subsub-opt-cvx-1} lead to \emph{provable} bounds on the parameters:
with these bounds Problem~\eqref{prob-bs-nLep-1} is provides an optimal solution to Problem~\eqref{eq-card-k}, and vice-versa.
We now describe some other alternatives that lead to excellent parameter specifications in practice. 

The discrete first order methods described in the following section~\ref{sec:DFOM} provide good upper bounds to Problem~\eqref{eq-card-k}.
These solutions when supplied as a warm-start to the MIO formulation~\eqref{eq-card-form2} are often improved by MIO, 
thereby leading to high quality solutions to Problem~\eqref{eq-card-k} within several minutes.  If $\hat{\B\beta}_{\text{hyb}}$  denotes an 
estimate obtained from this hybrid approach, then 
${\mathcal M}_{U} := \tau \| \hat{\B\beta}_{\text{hyb}}\|_{\infty}$ with $\tau$ a multiplier greater than one (e.g., $\tau \in \{ 1.1, 1.5, 2\}$) provides a good estimate for the parameter ${\mathcal M}_{U}$. 
A reasonable upper bound to $\|\widehat{\B\beta}\|_{1}$ is $k {\mathcal M}_{U}$.
Bounds on the other quantities: $\|\M{X}\widehat{\B\beta}\|_{1}, \|\M{X}\widehat{\B\beta}\|_{\infty}$ can be derived by using 
expressions appearing in~Theorem~\ref{prop-beta-bound1}, with aforementioned bounds on $\|\widehat{\B\beta}\|_{1}$ and 
$\|\widehat{\B\beta}\|_{\infty}$.

\subsubsection{Some Generalizations and Variants}\label{general-1-variants}
Some variations and improvements of the procedures described above are presented in Section~\ref{app:general-1-variants} (appendix).

\section{Discrete First Order Algorithms} \label{sec:DFOM}
In this section, we develop a discrete extension of  first order methods    in convex optimization~\cite{nesterov2004introductory,nest-07}
to obtain near optimal solutions for Problem~\eqref{eq-card-k} and its variant for the 
least absolute deviation (LAD) loss function. Our approach   applies to the problem of minimizing 
any smooth convex function subject to cardinality constraints. 

We will use these discrete first order methods to obtain solutions to warm start the MIO formulation. In Section~\ref{sec:computation-ls}, we will demonstrate how these methods greatly enhance the performance of the MIO.

%\textcolor{red}{Apart from warm starting, we also use the DFO methods to create local bounding boxes and select Big-M parameters (as done in some of the experiments)}

\subsection{Finding stationary solutions for minimizing smooth convex functions with cardinality constraints}\label{sec:generic-algo-smooth1}

\paragraph{Related work and contributions} In the signal processing literature~\cite{blumensath2008iterative,blumensath2009-acha} proposed iterative hard-thresholding algorithms,
 in the context of $\ell_{0}$-regularized least squares  problems, i.e., Problem~\eqref{L0-lag-1}. The authors establish convergence properties of the algorithm under 
the assumption that $\M{X}$ satisfies coherence~\cite{blumensath2008iterative} or Restricted Isometry Property~\cite{blumensath2009-acha}. 
The method we propose here applies to a larger class of cardinality constrained optimization problems of the form~\eqref{eq:subset-problem}, 
in particular, in the context  of Problem~\eqref{eq-card-k}
 our algorithm and its convergence analysis do not require any form of restricted isometry property on the model matrix $\M{X}$.
%to Problem~\eqref{eq-card-k}  with arbitrary $\M{X}$. 

Our proposed algorithm borrows ideas from projected gradient descent methods in first order convex optimization problems~\cite{nesterov2004introductory}
and generalizes it to the discrete optimization Problem~\eqref{eq:subset-problem}.
We also derive new global convergence results for our proposed algorithms as presented in~Theorem~\ref{FO - complexity-bound1}.
Our proposal, with some novel modifications also applies to 
the non-smooth least absolute deviation loss with cardinality constraints as discussed in Section~\ref{sec:appli-ls-bs-1}.

%In particular~\cite{nest-07} proposed a 
%generic gradient descent stylized method for problems of the form 
%$\mini_{s \in \M{S}} \;\left( f(s) + \Phi(s)\right)$, where $f$ is smooth (non-convex), $\Phi$ is closed and convex (non-differentiable), and $\M{S}$ is convex.
%However, none of the aforementioned techniques can be readily applied to the class of problems~\eqref{eq:subset-problem}.

Consider the   following optimization problem:
\begin{equation}\label{eq:subset-problem}
\min_{\B\beta} \;\;\;\;   g(\B\beta)  \;\;\; \sbt \;\;\; \| \B\beta \|_{0} \leq {\mmk},
\end{equation}
where $g(\B\beta) \geq 0 $  is convex and has Lipschitz continuous gradient:
\begin{equation}\label{def-lip-cont-1}
\| \nabla g(\B\beta) - \nabla g(\widetilde{\B\beta}) \| \leq \ell \|  \B\beta - \widetilde{\B\beta} \|.
\end{equation}
The first ingredient of our approach is the observation that when $ g(\B\beta) =\left \| \B\beta - \M{c}  \right \|_2^2$ for a given $ \M{c} $,
Problem \eqref{eq:subset-problem} admits a closed form solution. 
%\medskip
%\begin{prop}\label{hard-thresh-char-1}
%An optimal solution, denoted as $\M{H}_{{\mmk}} (\M{c})$,
%to the problem 
%\begin{equation}\label{defn:HT1}
% \min_{ \| \B\beta \|_{0} \leq \mmk } \left \| \B\beta - \M{c}  \right \|_2^2,
%\end{equation}
%can be computed as follows:  $\M{H}_{{\mmk}} (\M{c})$  retains the $\mmk$ largest (in absolute value)
% elements of $\M{c} \in  \mathbb{R}^{p}$ and sets the rest to zero, i.e., if
%$|c_{(1)}|  \geq |c_{(2)}|  \geq \ldots \geq |c_{(p)}|,$ denote the ordered values of the absolute values of the vector $\M{c}$, then:
%\begin{equation} \label{eq:Hkc}
% (\M{H}_{{\mmk}} (\M{c}))_{i} = \begin{cases} 
% c_{i},  & \text{if $i \in \{ (1), \ldots,  ({\mmk}) \}, $}\\
% 0, &\text{otherwise}.   
%\end{cases}
%\end{equation}
%\end{prop}
\begin{prop}\label{hard-thresh-char-1}
%An optimal solution, denoted as $\M{H}_{{\mmk}} (\M{c})$,
%to the problem 
If $\hat{\B\beta}$ is an optimal solution to the following problem:
\begin{equation}\label{defn:HT1}
\hat{\B\beta} \in  \argmin_{ \| \B\beta \|_{0} \leq \mmk } \;\; \left \| \B\beta - \M{c}  \right \|_2^2,
\end{equation}
then it can be computed as follows:  $\hat{\B\beta}$  retains the $\mmk$ largest (in absolute value)
 elements of $\M{c} \in  \mathbb{R}^{p}$ and sets the rest to zero, i.e., if
$|c_{(1)}|  \geq |c_{(2)}|  \geq \ldots \geq |c_{(p)}|,$ denote the ordered values of the absolute values of the vector $\M{c}$, then:
\begin{equation} \label{eq:Hkc}
\hat{\beta}_{i} = \begin{cases} 
 c_{i},  & \text{if $i \in \left\{ (1), \ldots,  ({\mmk}) \right\}, $}\\
 0, &\text{otherwise},   
\end{cases}
\end{equation}
where, $\hat{\beta}_{i}$ is the $i$th coordinate of $\hat{\B\beta}$.  We will denote the set of solutions to Problem~\eqref{defn:HT1} 
by the notation $\M{H}_{{\mmk}} (\M{c})$.
\begin{proof}
We provide a proof of this in Section~\ref{app:proof-hard-thresh-char-1}, for the sake of completeness.
%
%We provide a proof of this simple observation, for the sake of completeness.
%
%It suffices to consider $ |c_{i}| > 0$  for all $i$. Let $\B\beta$ be an optimal solution to Problem~\eqref{defn:HT1}
%and let $S :=\{ i: \beta_{i} \neq 0 \}$. The 
%objective function is given by 
% $\sum_{i\not\in S } |c_{i}|^2 + \sum_{i\in S} (\beta_{i} - c_{i})^2$. 
% Note that by selecting $\beta_i=c_i$ for $i\in S$, we can make the objective function 
%  $\sum_{i\not\in S } |c_{i}|^2 $.
%Thus, to minimize the objective function,  $S$ must correspond to the indices of the 
%largest $k$ values of $|c_{i}|, i \geq 1. \hfill$
%
\end{proof}
\end{prop}

Note that, we use the notation ``argmin'' (Problem~\eqref{defn:HT1} and in other places that follow) to denote the set of minimizers of the optimization Problem.

%\textcolor{red}{Clearly, for a given $\M{c}$, the operator $\M{H}_{{\mmk}} (\M{c})$ need not be unique. 
%From now on $\M{H}_{{\mmk}} (\M{c})$ will be used to denote the set of solutions to Problem~\eqref{defn:HT1}.}

The operator~\eqref{eq:Hkc} is also known as the hard-thresholding operator~\cite{Donoho93idealspatial}---a notion that
arises in the context of the following related optimization problem:
\begin{equation}\label{hard-thresh-lag-1}
\hat{\B\beta} \in \argmin_{\B\beta} \;\; \frac12 \| \B\beta  - \M{c} \|_{2}^2 + \lambda \| \B\beta \|_{0},
\end{equation}
where $\hat{\B\beta}$ admits a simple closed form expression given by $\hat{\beta}_{i} = c_{i}$ if $|c_{i}| > \sqrt{\lambda}$ and 
$\hat{\beta}_{i} = 0$ otherwise, for $i = 1, \ldots, p$.

%\begin{rem}\label{rem-LB-1} There is an important difference between the minimizers of Problems~\eqref{defn:HT1} and~\eqref{hard-thresh-lag-1}. In the former, 
%there is no lower bound to the minimum (in absolute value) non-zero element of a minimizer, i.e., $\M{H}_{{\mmk}} (\M{c})$. On the other hand, for Problem~\eqref{hard-thresh-lag-1}, 
%the smallest (in absolute value) non-zero element in $\hat{\B\beta}$ is greater than $\lambda$ in absolute value. 
%This needs to be taken care of using subtle techniques in our proof.
%\end{rem}

%\medskip

\begin{rem}\label{rem-LB-1} There is an important difference between the minimizers of Problems~\eqref{defn:HT1} and~\eqref{hard-thresh-lag-1}. For Problem~\eqref{hard-thresh-lag-1}, 
the smallest (in absolute value) non-zero element in $\hat{\B\beta}$ is greater than $\lambda$ in absolute value. 
On the other hand, in Problem~\eqref{defn:HT1} there is no lower bound to the minimum (in absolute value) non-zero element of a minimizer.
%, i.e., $\M{H}_{{\mmk}} (\M{c})$. 
This needs to be taken care of while analyzing the convergence properties of Algorithm~1 (Section~\ref{sec:conv-analysis1}).
\end{rem}

%%----------------
%Proposition~\ref{hard-thresh-char-1} is  inspired by iterative hard thresholding, a notion that can be traced back 
%to at least~\cite{Donoho93idealspatial}, who used it  in the context of $\ell_{0}$ regularized problems with orthogonal design, which is  
%equivalent to Problem~\eqref{L0-lag-1} with $\M{X} =\M{I}_{n \times n}$ (with $n=p$). The notion of hard 
%thresholding~\cite{Donoho93idealspatial} 
%arises in the context of the following optimization problem:
%\begin{equation}\label{hard-thresh-lag-1}
%\hat{\B\beta} \in \argmin_{\B\beta} \;\; \frac12 \| \B\beta  - \M{c} \|_{2}^2 + \lambda \| \B\beta \|_{0},
%\end{equation}
%where $\hat{\B\beta}$ admits a simple closed form expression given by $\hat{\beta}_{i} = c_{i}$ if $|c_{i}| > \lambda$ and 
%$\hat{\beta}_{i} = 0$ otherwise, for $i = 1, \ldots, p$.
%%----------------

%for Problem~\eqref{L0-lag-1}, which as we noted before is not the same as Problem~\eqref{eq-card-k} being investigated in this paper.
%\cite{blumensath2008iterative} also analyze their algorithm under the assumption that $\M{X}$ satisfies the restricted isometry property.
%The method we propose here applies to a large class of cardinality constrained optimization problems of the form~\eqref{eq:subset-problem},
%and in particular, to Problem~\eqref{eq-card-k}  with arbitrary $\M{X}$. 
% Our method, with some modifications as we discuss  in Section~\ref{sec:appli-ls-bs-1}, also applies to 
%the least absolute deviation problem with cardinality constraints.

Given a current solution $\B\beta$, 
the second ingredient of our approach is to upper bound the function $g(\B\eta)$ around $g(\B\beta)$.
To do so, 
we use ideas from projected gradient descent methods in first order convex optimization problems~\cite{nesterov2004introductory,nest-07}.

%\medskip

\begin{prop}\label{prop-lipsh-bound1}(\cite{nesterov2004introductory,nest-07})
For a convex function $g(\B\beta)$ satisfying condition~\eqref{def-lip-cont-1} and for any $L \geq \ell$ we have :
\begin{equation}\label{major-1}
g ( \B\eta )\leq Q_{L}(\B\eta, \B\beta) :=g(\B\beta) +  \frac{L}{2} \| \B\eta - \B\beta \|_2^2 + \langle \nabla g(\B\beta), \B\eta - \B\beta  \rangle  
\end{equation}
 for all $\B\beta, \B\eta$ with equality holding at $\B\beta = \B\eta$.
\end{prop}

Applying Proposition \ref{hard-thresh-char-1} to the upper bound $Q_{L}(\B\eta, \B\beta)$ in Proposition \ref{prop-lipsh-bound1} we obtain
\begin{align}
\argmin_{\| \B\eta \|_{0} \leq {\mmk} } \;\; Q_{L}(\B\eta, \B\beta) =& \argmin_{\| \B\eta \|_{0} \leq {\mmk} } \;\; \left (  \frac{L}{2} \left\| \B\eta - \left ( \B\beta - \frac{1}{L} \nabla g(\B\beta) \right) \right\|_2^2  - \frac{1}{2L}\left \|\nabla g(\B\beta) \right\|_2^2  + g(\B\beta) \right)   \nonumber\\
=&\argmin_{\| \B\eta \|_{0} \leq {\mmk} } \;\; \left \| \B\eta - \left (\B\beta -  \frac{1}{L}\nabla g(\B\beta) \right) \right \|_2^2  \nonumber \\
=&  \M{H}_{{\mmk}} \left(\B\beta - \frac1L \nabla g(\B\beta) \right) \label{eta-hat-HT1},
\end{align}
where $\M{H}_{{\mmk}} (\cdot)$ is   defined in \eqref{eq:Hkc}.  
In  light of \eqref{eta-hat-HT1} we are now ready to present Algorithm~1 to find a stationary point (see Definition~\ref{defn-FO-stat-1})
of Problem \eqref{eq:subset-problem}.

\subsection*{Algorithm~1}
{\bf Input:} $g(\B \beta)$, $L$, $\epsilon$.\\
{\bf Output:} A first order stationary solution  $\B\beta^*$.\\
{\bf Algorithm:}
\begin{enumerate}
\item[{\bf 1.}] Initialize with ${\B\beta}_{1} \in  \mathbb{R}^{p}$ such that $\|  {\B\beta}_{1} \|_{0} \leq \mmk$.
\item[{\bf 2.}] For $m\geq 1$, apply  \eqref{eta-hat-HT1} with 
$\B\beta = {\B\beta}_{m}$ to obtain ${\B\beta}_{m+1}$ as:
\begin{equation}
{\B\beta}_{m+1} \in  \M{H}_{{\mmk} } \left(\B\beta_{m} - \frac1L \nabla g(\B\beta_{m}) \right)   \label{update-form-1}
\end{equation}
\item[{\bf 3.}]  Repeat Step 2, until $\| {\B\beta}_{m+1} - {\B\beta}_{m}\|_{2}  \leq \epsilon$. 
\item[{\bf 4.}]  Let $\B\beta_{m}:=(\beta_{m1}, \ldots, \beta_{mp})$ denote the current estimate and let $I=\text{Supp}({\B\beta}_{m}):=\{i:~\beta_{mi}\neq 0\}$. 
Solve the continuous 
optimization problem: 
\begin{equation} \label{eq:cleaning1}
\min_{\B\beta, \beta_i=0,~i\notin I} g(\B \beta),\end{equation}
and let $\B\beta^*$ be a minimizer. 
\end{enumerate}
The convergence properties of Algorithm 1 are presented in Section~\ref{sec:conv-analysis1}. 
We also present Algorithm~2, a variant of Algorithm~1 with better empirical performance. 
Algorithm~2 modifies Step~2 of Algorithm~1 by using a line search. It obtains
$\B\eta_{m}  \in \M{H}_{{\mmk} } \left(\B\beta_{m} - \frac1L \nabla g(\B\beta_{m}) \right)$ and 
$\B\beta_{m+1} = \lambda_{m}   \B\eta_{m}  +(1-\lambda_m)  \B\beta_{m},$ 
where $\lambda_m \in \argmin_{\lambda} g \left( \lambda  \B\eta_{m}  +(1-\lambda)  \B\beta_{m} \right)$.

%A variant of Algorithm 1 that has better empirical performance 
%and   uses line searches is presented next. 
%\subsection*{Algorithm~2 (with Line Search)}
%\begin{enumerate}
%\item[{\bf 1.}] Initialize with ${\B\beta}_{1} \in  \mathbb{R}^{p}$ such that $\|  {\B\beta}_{1} \|_{0} \leq \mmk$.
%\item[{\bf 2.}]  For $m\geq 1$, 
%\begin{align}
%\B\eta_{m}  \in & \;\;\; \M{H}_{{\mmk} } \left(\B\beta_{m} - \frac1L \nabla g(\B\beta_{m}) \right),\nonumber \\
%\B\beta_{m+1} =& \;\;\; \lambda_{m}   \B\eta_{m}  +(1-\lambda_m)  \B\beta_{m}  \label{move-in-grad1}, 
%\end{align}
%where $\lambda_m$ is chosen to minimize the one-dimensional optimization problem:
%\begin{equation}\label{mini-line-search-alpha}
%\lambda_m \in \argmin_{\lambda} \;\; g \left( \lambda  \B\eta_{m}  +(1-\lambda)  \B\beta_{m} 
%\right).
%\end{equation}
%\item[{\bf 3.}]  Repeat Step 2, until $\| {\B\eta}_{m+1} - {\B\eta}_{m}\|_{2}  \leq \epsilon$. 
%\item[{\bf 4.}]  Let $\B\eta_{m}$ denote the current estimate and let $I=\text{Supp}({\B\eta}_{m})$. Solve  problem \eqref{eq:cleaning1}
%and let $\B\beta^*$ be a mininizer. 
%\end{enumerate}

Note that the iterate $\B\beta_{m}$ in Algorithm~2 need not be $\mmk$-sparse (i.e., need not satisfy: $\|\B\beta_{m}\|_{0} {\leq} \mmk$), however, 
$\B\eta_{m} $ is $\mmk$-sparse ($\|\B\eta_{m}\|_{0} \leq \mmk$). Moreover, 
the sequence may not lead to a decreasing set of objective values, but  it  satisfies:
$ g(\B\beta_{m+1})  \leq   g(\B\eta_{m}) \nleq g(\B\beta_{m}) .$

\subsection{Convergence Analysis of Algorithm~1}\label{sec:conv-analysis1}
In this section, we study convergence properties for Algorithm~1. 
Before we embark on the analysis, we need to define the  notion of 
first order optimality for Problem~\eqref{eq:subset-problem}.

%\medskip

\begin{mydef}\label{defn-FO-stat-1}
Given an $L \geq \ell$, the vector
$\B\eta \in  \mathbb{R}^{p}$ is said to be a  first order stationary point of  Problem~\eqref{eq:subset-problem} if 
$\| \B\eta\|_{0} \leq \mmk$ and it satisfies the following fixed point equation:
\begin{equation}\label{fixed-points-3}
\B\eta \in \M{H}_{{\mmk} } \left(\B\eta - \frac1L \nabla g(\B\eta) \right).
\end{equation}
\end{mydef}
%\textcolor{red}{Note that $\M{H}_{{\mmk} } \left(\B\eta - \frac1L \nabla g(\B\eta) \right)$ need not be unique. Thus the equality in~\eqref{fixed-points-3}
%should be taken to mean that if  a 
%value of  $\M{H}_{{\mmk} } \left(\B\eta - \frac1L \nabla g(\B\eta) \right)$ is equal to $\B\eta$ (where, $\B\eta$ is $k$-sparse), 
%then $\B\eta$ is a first order stationary point of Problem~\eqref{eq:subset-problem}.}

%\textcolor{red}{--------------EDITS----------------------------start}

Let us give some intuition associated with the above definition.

Consider $\B\eta$ as in Definition~\ref{defn-FO-stat-1}. Since $\|\B\eta \|_{0} \leq k,$ it follows that there is a set $I \subset \{ 1, \ldots, p \}$ such that 
$\eta_{i} = 0 $ for all $i \in I$ and the size of $I^c$ (complement of $I$) is $k$.
Since  $\B\eta \in \M{H}_{{k} } \left(\B\eta - \frac1L \nabla g(\B\eta) \right),$
it follows that for all $i \notin I$, we have:
$ \eta_{i} = \eta_{i} - \frac 1 L \nabla_{i} g(\B\eta),$
where, $\nabla_{i} g(\B\eta)$ is the $i$th coordinate of $\nabla g(\B\eta)$.
It thus follows that:
$ \nabla_{i} g(\B\eta) = 0$ for all $i \notin I$. Since $g(\B\eta)$ is convex in $\B\eta$, this means that $\B\eta$ solves the following convex optimization problem:
\begin{equation}\label{restr-1-}
 \min_{\B\eta} \;\;\; g(\B\eta) \;\;\; s.t. \;\;\; \eta_{i} = 0 , i \in I.
 \end{equation}
Note however, that the converse of the above statement is not true. That is, if $\tilde{{I}}\subset \{ 1, \ldots, p\}$ is an arbitrary subset with $|\tilde{{I}}^{c}| = \mmk$ then a 
solution $\hat{\B\eta}_{\tilde{I}}$ to the restricted convex problem~\eqref{restr-1-} with $I = \tilde{{I}}$ need \emph{not} correspond to a first order 
stationary point. 

Note that any global minimizer to Problem~\eqref{eq:subset-problem} is also a first order stationary point, as defined above (see Proposition~\ref{singleton-set-1}).

We present the following proposition (for its proof see Section~\ref{proof-prop-n-1-1}), which sheds light on a first order stationary point  $\B\eta$ for which $\|\B\eta\|_{0} < k$.
\begin{prop}\label{prop-n-1-1}
Suppose $\B\eta$ satisfies the first order stationary condition~\eqref{fixed-points-3} and $\|\B\eta\|_{0} < k$. Then
$ \B\eta \in \argmin\limits_{\B\beta} \;\; g( \B\beta)$.
%\begin{proof}
%See Section~\ref{proof-prop-n-1-1}.
%\end{proof}
\end{prop}

We next define the notion of an $\epsilon$-approximate first order stationary point of Problem~\eqref{eq:subset-problem}:

%\medskip

\begin{mydef}\label{defn-FO-stat-2}
Given an $\epsilon > 0 $ and  $L \geq \ell$ we say that $\B\eta$ satisfies an $\epsilon$-approximate first order optimality condition of Problem~\eqref{eq:subset-problem} if
$\|\B\eta\|_{0} \leq \mmk$ and for some $\hat{\B\eta} \in \M{H}_{{\mmk} } \left(\B\eta - \frac1L \nabla g(\B\eta) \right)$, we have 
$\| \B\eta - \hat{\B\eta} \|_{2} \leq \epsilon$.
%$$\left\| \B\eta - \M{H}_{{\mmk} } \left(\B\eta - \frac1L \nabla g(\B\eta) \right) \right\|_{2} \leq \epsilon.$$
%$$\left\| \B\eta - \M{H}_{{\mmk} } \left(\B\eta - \frac1L \nabla g(\B\eta) \right) \right\|_{2} \leq \epsilon.$$
\end{mydef}

%\textcolor{red}{--------------EDITS----------------------------end}

%Clearly, $\B\eta$ is a fixed point of the algorithm iff $G_{\mmk, L}(\B\eta) =0$.

%\subsubsection*{Convergence Analysis of Algorithm~1}

%\subsubsection*{Global Convergence Analysis}

%\medskip
Before we dive into the convergence properties of Algorithm~1, we need to introduce some notation. 
Let $\B\beta_m=(\beta_{m1},\ldots, \beta_{mp})$ and  $\M{1}_{m} =(e_1,\ldots,e_p)$ with $e_j=1$, if $\beta_{mj}\neq 0$, and $e_j=0$, if $\beta_{mj}= 0$, $j=1,\ldots, p$, i.e., 
$\M{1}_{m}$ represents the sparsity pattern  of   the support of $\B\beta_{m}$.

%\textcolor{red}{--------------EDITS----------------------------start}

Suppose, we order the coordinates of $\B\beta_{m}$ by their absolute values: $ |\beta_{(1), m}| \geq |\beta_{(2), m}| \geq \ldots \geq |\beta_{(p), m}|$.
Note that by definition~\eqref{update-form-1}, $\beta_{(i),m} = 0$ for all $i > k$ and $m \geq 2$.
We denote $\alpha_{k,m}= |\beta_{(k),m}|$ to be the $k$th largest (in absolute value) entry in $\B\beta_{m}$ for all $m\geq 2$. 
Clearly if $\alpha_{k,m}>0$ then $\|\B\beta_{m}\|_{0} = k$ and if 
$\alpha_{k,m}=0$ then $\|\B\beta_{m}\|_{0} < k$. Let 
$\overline{\alpha}_{k} := \limsup\limits_{m \rightarrow \infty} \; \alpha_{k,m}$ and 
$\underline{\alpha}_{k} :=  \liminf\limits_{m \rightarrow \infty} \; \alpha_{k,m}$.

%We will also introduce a condition on the boundedness of the level sets of $k$-sparse vectors.

%\medskip
\begin{prop}\label{prop-of-suff-dec1}
Consider $g(\B\beta) $ and $\ell$ as defined in~\eqref{eq:subset-problem} and~\eqref{def-lip-cont-1}.
Let $\B\beta_{m}, m \geq 1$ be  the sequence generated by Algorithm~1. Then we have :
\begin{enumerate}
\item[{\bf (a)}]
For any $L \geq \ell$, the sequence $g(\B\beta_{m})$ satisfies 
\begin{equation}\label{eqn-suff-decrease-1}
g(\B\beta_{m}) -  g(\B\beta_{m+1})   \geq \frac{L - \ell}{2} \left \|\B\beta_{m+1} - \B\beta_{m} \right\|_2^2,
\end{equation}
is decreasing and   converges. 
\item[{\bf (b)}] If  $L > \ell$, then $\B\beta_{m+1} - \B\beta_{m} \rightarrow \M{0}$ as $m \rightarrow \infty$. 
\item[{\bf (c)}]
If $L > \ell$ and $\underline{\alpha}_{k}>0$ then the sequence $\M{1}_{m}$ converges after finitely many iterations, i.e., 
there exists an iteration index $M^*$ such that $\M{1}_{m}= \M{1}_{m+1}$  for all $m \geq M^*$.
Furthermore, the sequence $\B\beta_{m}$ is bounded and converges to a first order stationary point.
\item[{\bf (d)}]
 If $L > \ell$ and $\underline{\alpha}_{k}=0$ then $\liminf\limits_{m \rightarrow \infty} \| \nabla g(\B\beta_{m})  \|_{\infty} =0$.
\item[{\bf (e)}]
Let $L > \ell$, $\overline{\alpha}_{k}=0$ and suppose that  the sequence $\B\beta_{m}$ has a limit point. 
Then $g(\B\beta_{m}) \rightarrow \min\limits_{\B\beta} \;\; g(\B\beta)$.
%then every limit point $\B\beta_{\infty}$ of the sequence $\B\beta_{m}$ is a solution to the 
%unconstrained convex optimization problem:  $\B\beta_{\infty} \in \argmin_{\B\beta} \;\; g(\B\beta)$. 
 %$\| \liminf_{m \rightarrow \infty} \B\beta_{m} \|_{0} < \mmk$,
%then $\liminf_{m \rightarrow \infty} \;\; \nabla g( \B\beta_{m}) \rightarrow \M{0}$. In other words
%then $g( \B\beta_{m})  \rightarrow g(\B\beta^*)$ where  $\B\beta^* \in \argmin g(\B\beta)$ is an unconstrained minimizer.
\end{enumerate}
\begin{proof}
See Section~\ref{proof-prop-of-suff-dec1}.
\end{proof}
\end{prop}

\begin{rem}
Note that the existence of a limit point in Proposition~\ref{prop-of-suff-dec1}, Part (e) is guaranteed under fairly weak conditions. 
One such condition is that 
$\sup \left (\left \{  \B\beta :   \| \B\beta\|_{0} \leq k, f(\B\beta) \leq f_{0}   \right \} \right) < \infty,$ for 
any finite value $f_{0}$.
In words this means that the $k$-sparse level sets of the function $g(\B\beta)$ is bounded. 
%This implies the existence of limit points of the sequence 
%$\B\beta_{m}$ for Part (e).

In the special case where $g(\B\beta)$ is the least squares loss function, the above condition is equivalent to
every $k$-submatrix ($\M{X}_{J}$) of $\M{X}$ comprising of $k$ columns being full rank. In particular, this holds with probability one when
the entries of $\M{X}$ are drawn from a continuous distribution and $k<n$.
%$\cup_{J} \left \{ \B\beta :   \B\beta \in {\mathcal A}_{J}, f(\B\beta) \leq f_0  \right \}$
%is bounded, where, ${\mathcal A}_{J} =\left \{ \B\beta \in \Re^p :  \beta_{i} = 0 , i \in I_{J} \right\}$ and 
%$I_{J} \subset \{ 1, 2, \ldots, p \}$ with $| I^c_{J} | = k$.
\end{rem}

\begin{rem}
Parts (d) and (e) of Proposition~\ref{prop-of-suff-dec1} are probably not statistically interesting cases, since they correspond to un-regularized solutions of 
the problem $\min g(\B\beta)$. However, we include them since they shed light on the properties of Algorithm~1.

The conditions assumed in Part (c) imply that the support of $\B\beta_{m}$ stabilizes and Algorithm~1 behaves like vanilla gradient descent thereafter. 
The support of $\B\beta_{m}$ need not stabilize for Parts (d), (e) and thus Algorithm~1 may not behave like vanilla gradient descent after finitely many iterations.  
However, the objective values (under minor regularity assumptions) converge to $\min \; g(\B\beta)$.
\end{rem}

We present the following Proposition (for proof see Section~\ref{proof-singleton-set-1}) about a uniqueness property of the fixed point equation~\eqref{defn-FO-stat-1}.
\begin{prop}\label{singleton-set-1}
Suppose $L> \ell$ and let $\B\eta$ satisfy a first order stationary point as in Definition~\ref{defn-FO-stat-1}.
Then the set $\M{H}_{{k}} \left(\B\eta - \frac1L \nabla g(\B\eta) \right)$ has exactly one element: $\B\eta$. 
%\begin{proof}
%See Section~\ref{proof-singleton-set-1}
%\end{proof}
\end{prop}

The following proposition (for a proof see Section~\ref{proof-singleton-set-2}) shows that a global minimizer of the Problem~\eqref{eq:subset-problem} is also a first order stationary point.
\begin{prop}\label{singleton-set-2}
Suppose $L> \ell$ and let $\widehat{\B\beta}$ be a global minimizer of Problem~\eqref{eq:subset-problem}.
Then $\widehat{\B\beta}$ is a first order stationary point.
%\begin{proof}
%See Section~\ref{proof-singleton-set-2}
%\end{proof}
\end{prop}
%Proofs of the above propositions are presented in Section~\ref{proof-singleton-set-1} and~\ref{proof-singleton-set-2} respectively.

Proposition~\ref{prop-of-suff-dec1} establishes that Algorithm~1 either converges to a first order stationarity point (part (c)) or it 
converges\footnote{under minor technical assumptions} to a 
global optimal solution (Parts (d), (e)), 
 but does not quantify the rate of convergence.
We next characterize   the  rate of convergence of the algorithm to an $\epsilon$-approximate 
first order stationary point.

%\medskip

\begin{thm}\label{FO - complexity-bound1}
Let $L > \ell$ and $\B\beta^*$ denote a first order stationary point of Algorithm~1.
 After $M$ iterations  Algorithm~1 satisfies
\begin{equation}\label{frts-order-rate}
\min_{m = 1, \ldots, M}   \|\B\beta_{m+1} - \B\beta_{m}\|_2^2 \leq  \frac{2 ( g ( \B\beta_{1} ) -  g( \B\beta^* ))}{M (L - \ell)},  
\end{equation}
where $g(\B\beta_{m}) \downarrow g( \B\beta^*)$ as $m \rightarrow \infty$.
\begin{proof}
See Section~\ref{proof-FO-complexity-bound1}.
\end{proof}
\end{thm}

Theorem~\ref{FO - complexity-bound1} implies that for any $\epsilon >0$  there exists   $M=O(\frac1\epsilon)$ such that for some 
$ 1 \leq m^* \leq M$, we have: $   \|\B\beta_{m^*+1} - \B\beta_{m^*} \|_2^2   \leq \epsilon.$
Note that the convergence rates derived above apply for a large class of problems~\eqref{eq:subset-problem}, where, 
the function $g(\B\beta) \geq 0$ is convex with Lipschitz continuous gradient~\eqref{def-lip-cont-1}. Tighter rates may be obtained 
under additional structural assumptions on $g(\cdot)$. 
For example, the adaptation of Algorithm~1 for Problem~\eqref{L0-lag-1} was 
analyzed in~\cite{blumensath2008iterative,blumensath2009-acha} with 
$\M{X}$ satisfying coherence~\cite{blumensath2008iterative} or Restricted Isometry Property (RIP)~\cite{blumensath2009-acha}. 
In these cases, the algorithm can be shown to have  
 a linear convergence rate~\cite{blumensath2008iterative,blumensath2009-acha}, where the rate depends upon the RIP constants.

Note that by Proposition~\ref{prop-of-suff-dec1} the support of $\B\beta_{m}$ stabilizes after finitely many iterations, after which Algorithm~1 
behaves like gradient descent on the stabilized support. If $g(\B\beta)$ restricted to this support is strongly convex, then Algorithm~1 will enjoy a 
linear rate of convergence~\cite{nesterov2004introductory}, as soon as the support stabilizes. This behavior is adaptive, i.e., Algorithm~1 does not need to be modified 
after the support stabilizes. 

The next section describes practical post-processing schemes via which first order stationary points of
Algorithm~1 can be obtained by solving a low dimensional convex optimization problem, as soon as the support is found to stabilize, numerically.
In our numerical experiments, we this version of Algorithm~1 (with multiple starting points) took at most  
a few minutes for $p=2000$ and a few seconds for smaller values of $p$.

\subsubsection*{Polishing coefficients on the active set}
%Proposition \ref{prop-of-suff-dec1}(b)  states that the active set of Algorithm 1  converges after finitely many iterations. 
 Algorithm~1 \emph{detects} the active set after a few iterations.
% though  it is not clear how to estimate 
%when the algorithm has \emph{stabilized} onto the active set.
Once the active set 
stabilizes, the   algorithm may take a number of  iterations to estimate the values of the regression coefficients on the active set to a 
high accuracy level.

In this context, we found the following simple polishing of coefficients to be useful. 
When the  algorithm has converged to a tolerance of $\epsilon$ ($\approx 10^{-4}$), we fix the current 
active set, ${\mathcal I}$,  and solve the following lower-dimensional convex optimization problem:
\begin{equation}\label{polish-1}
 \min_{\B\beta, \beta_{i} = 0 , i \notin {\mathcal I}} \;\;\;\; g(\B\beta).
\end{equation}
In the context of the least squares  and the least absolute deviation problems, 
Problem~\eqref{polish-1} reduces to  to a smaller dimensional least squares and a linear optimization problem respectively, which can be 
solved very efficiently up to a very high level of accuracy.

\subsection{Application to Least Squares}\label{sec:appli-ls-bs-1}
For the support constrained problem with squared error loss, we have $g(\B\beta) = \half \| \M{y} - \M{X}\B\beta \|_2^2$
and $\nabla g(\B\beta)=  -\M{X}'(\M{y} - \M{X}\B\beta )$.
%$$ g(\B\beta) = \half \| \M{y} - \M{X}\B\beta \|_2^2, \;\; \nabla g(\B\beta)=  -\M{X}'(\M{y} - \M{X}\B\beta )$$
The general algorithmic framework developed above applies in a straightforward fashion for this special case.
Note that for this case $\ell = \lambda_{\max}(\M{X}'\M{X})$.
%
%$ L_{\mmk} = \lambda_{\max}(\M{X}_{I}'\M{X}_{I})$ and 
%$\mu_{\mmk} =  \lambda_{\min}(\M{X}_{I}'\M{X}_{I})$.

The polishing of the regression coefficients in the active set can be performed via a least squares problem on $\M{y}, \M{X}_{I}$, where 
 $I$ denotes the support of the regression coefficients.

\subsection{Application to Least Absolute Deviation}\label{sec:appli-ls-bs-2}
We will now show how the method proposed in the previous section applies to the least absolute deviation problem with 
support constraints in $\B\beta$:
%\begin{equation}\label{lad-l0}
%\begin{array}{rr}
%\min_{\B\beta} &  g_{1} (\B\beta) := \| \M{Y} - \M{X}\B\beta\|_1 \\
% s.t. & \| \B\beta \|_0 \leq \mmk.
%\end{array}
%\end{equation}
\begin{equation}\label{lad-l0}
%\begin{array}{rr}
\min_{\B\beta} \;\;  g_{1} (\B\beta) := \| \M{Y} - \M{X}\B\beta\|_1 \;\; s.t. \;\;  \| \B\beta \|_0 \leq \mmk.
% s.t. & \| \B\beta \|_0 \leq \mmk.
%\end{array}
\end{equation}

Since $g_{1}(\B\beta)$ is non-smooth,  our framework  does not apply directly.
 We smooth the non-differentiable $g_{1} (\B\beta)$  
 so that   we can apply  Algorithms 1 and 2. 
 %This idea has been put to use in the literature of fast first order methods for convex optimization, pioneered by Nesterov~\cite{nest_05}.
% For smoothing the non-smooth objective function appearing in~\eqref{lad-l0}, we will use ideas pioneered by Nesterov  for developing first order methods in the context of  minimizing \emph{non-smooth convex} functions 
 % having favorable geometrical properties. 
 % We  use this idea in the context of the non-smooth non-convex optimization Problem~\eqref{lad-l0}.
Observing that $g_{1}(\B\beta)  = \sup_{ \| \M{w} \|_\infty \leq 1} \langle \M{Y} - X\B\beta , \M{w}     \rangle$ 
we make use of the smoothing technique of~\cite{nest_05} to obtain 
$g_{1}(\B\beta ; \tau ) =  \sup_{\| \M{w} \|_\infty \leq 1} ( \langle \M{Y} - X\B\beta , \M{w}     \rangle    - \frac{\tau}{2} \| \M{w} \|_2^2)$;
which is a smooth approximation of $g_{1}(\beta)$, with $\ell =  \frac{\lambda_{\max}(\M{X}'\M{X})}{\tau}$ for which Algorithms~1 and 2 apply.

In order to obtain a good approximation to Problem~\eqref{lad-l0}, we found the following   strategy to be useful in practice:
\begin{enumerate}
\item[1.] Fix $\tau>0$, initialize with $\B\beta_0 \in  \mathbb{R}^{p}$ and repeat the following steps [2]---[3] till convergence:
\item[2.] Apply  Algorithm~1 (or Algorithm~2) to the smooth function $g_{1}(\B\beta; \tau)$. 
Let $\B\beta_{\tau}^{*}$ be the limiting solution. 
\item[3.] Decrease $\tau \leftarrow \tau\gamma$ for some pre-defined constant $\gamma = 0.8$ (say), and 
go back to step [1] with $\B\beta_{0} = \B\beta_{\tau}^{*}$. Exit if $\tau < \text{TOL},$ for some pre-defined tolerance.
%or the objective value of the Problem~\eqref{lad-l0} does not change across successive values of $\tau$.
\end{enumerate}

\section{A Brief Tour of the Statistical Properties of Problem~\eqref{eq-card-k}} \label{sec:stat-prop1}
As already alluded to in the introduction, there is a substantial body of impressive work characterizing the theoretical 
properties of best subset solutions in terms of various metrics: predictive performance, estimation of regression coefficients, and variable selection properties. 
For the sake of completeness, we present a brief review of some of the properties of solutions to Problem~\eqref{eq-card-k} in Section~\ref{app:sec:stat-prop1}.

\section{Computational Experiments for Subset Selection with Least Squares Loss} \label{sec:computation-ls}

In this section, we    present a variety of computational experiments to assess the algorithmic and statistical performances of  our  approach. We consider
both the classical overdetermined  case with $n > p$ (Section~\ref{sec:n>p}) and the high dimensional $ p\gg  n$ case (Section~\ref{sec:n<p})
for  the least squares loss function with support constraints. 

\subsection{Description of Experimental Data}\label{sec:expt-data-desc}
%We \textcolor{red}{consider} a series of both  synthetic and real datasets for our experiments.
We demonstrate the performance of our proposal via a series of experiments on both  synthetic and real data.
\paragraph{Synthetic Datasets.}
We consider a collection of problems where
%\textcolor{red}{We consider a collection of problems} where 
$\M{x}_{i} \sim \text{N} (\M{0} , \M{\Sigma}), i = 1, \ldots, n$ are independent realizations from a 
$p$-dimensional multivariate normal distribution with mean zero and covariance matrix $\M{\Sigma}:= (\sigma_{ij})$.  
The columns of the $\M{X}$ matrix were subsequently standardized to have unit $\ell_{2}$ norm.
For a fixed $\M{X}_{n\times p},$ we generated the response $\M{y}$ as follows:
 $\M{y} = \M{X} \B\beta^{0} + \B\epsilon$, where 
 $ \epsilon_{i} \stackrel{\text{iid}}{\sim}N(0 , \sigma^2)$. We denote the number of nonzeros in $\B\beta^0$ by $k_0$.
The choice of $\M{X}, \B\beta^{0},\sigma$ determines the Signal-to-Noise Ratio (SNR) of the problem, which is defined as:
$\text{SNR} = \frac{\text{var}(\M{x}'\B\beta^{0})}{ \sigma^2}.$
%where $\M{x}\sim \text{N} (\M{0} , \M{\Sigma})$.

We considered the following four different examples: %types of datasets:
%\paragraph{Example 1}

\noindent {\bf{Example 1:  }}
We took $\sigma_{ij} = \rho^{| i - j |}$ for $i, j \in \{1, \ldots, p \} \times \{ 1, \ldots, p\}$. 
%Here, $k_0 = 5$ and $\beta^0_{i} = 1$ for $i \in \{ \kappa_{1}, \ldots, \kappa_{5}\}$ for five equi-spaced
We consider different values of $k_0 \in \{ 5, 10 \} $ and $\beta^0_{i} = 1$ for $k_{0}$ equi-spaced values. In the case where exactly equi-spaced values are not possible we rounded the indices to the nearest large integer value.
of $i$ in the range $\{1,2, \ldots, p \}$.

%Different values of $n,p,\rho$ and $\text{SNR}$ were considered.

%\paragraph{Example 2}
\noindent {\bf{Example 2:}}
We took $\B\Sigma = \M{I}_{p \times p}$, 
%i.e., the entries $\M{X}_{n \times p}$ formed an iid standard Gaussian ensemble.
 $k_0 = 5$ and $\B\beta^{0} = (\M{1}'_{ 5\times 1}, \M{0}'_{ p - 5 \times 1})' \in  \mathbb{R}^{p}$.

%[ {\texttt iidGN case 2} ]

%\paragraph{Example 3}
\noindent {\bf{Example 3:}} We took $\B\Sigma = \M{I}_{p \times p}$,
 $k_0 = 10$ and  $\beta_{i}^{0} = \frac12 + (10 - \frac12)\frac{(i-1)}{k_0}, i = 1, \ldots, 10$ and $\beta^0_i= 0, \forall i > 10$ --- i.e., a vector with ten nonzero entries, with the nonzero values being 
equally spaced in the interval $[\frac12, 10]$. 

%[ {\texttt iidGN case 3} ]

%beta0=0*randn(pp,1); beta0(1:kpop) = linspace(10,.5,kpop); 
%kpop = 10;

%\paragraph{Example 4}
\noindent {\bf{Example 4:  }}
We took $\B\Sigma = \M{I}_{p \times p}$,
$k_0 = 6$ and $\B\beta^{0} = (-10,    -6 ,   -2,     2,     6,    10, \M{0}_{p - 6})$, i.e., a vector with six nonzero entries, equally spaced in the interval $[-10,10]$.

\paragraph{Real Datasets}
We considered the Diabetes dataset analyzed in~\cite{LARS}. We used the dataset with all the second order interactions included in the model, which resulted in 64 predictors. 
We reduced the sample size to $n = 350$ by taking a random sample and standardized the response and the columns of the model matrix to have zero means and unit $\ell_{2}$-norm.  

In addition to the above, we also considered a real microarray dataset: the Leukemia data~\cite{dettling2004bagboosting}. We downloaded 
the processed dataset from~\url{http://stat.ethz.ch/~dettling/bagboost.html}, which had $n = 72$ binary responses and more than 3000 predictors. 
We standardized the response and columns of features to have zero means and unit $\ell_{2}$-norm. 
We reduced the set of features to 1000 by retaining the features maximally correlated (in absolute value) to the 
response. We call the resulting  feature matrix $\M{X}_{n \times p}$ with $n = 72, p = 1000$.  We then generated a semi-synthetic dataset with 
continuous response as $\M{y} = \M{X}\B\beta^{0} + \epsilon$, where the first five coefficients of $\B\beta^{0}$ were taken as one and the rest as zero. The noise was distributed as
$\epsilon_{i} \stackrel{\text{iid}}{\sim} N(0 , \sigma^2)$, with $\sigma^2$ chosen to get a SNR=7.

%Leukemia results:
%The properties of the estimator (after doing CV etc)
%            pred-accuracy | stand-err | nnz 
%{Lasso}_soln =[ 0.0458    0.0067   27.0000 ]
%FO_soln : [0.0084    0.0016    5.0000]
%BS_soln = [0.0663    0.0141    3.0000 ]
%what units are prediction accuracy in?

\paragraph{Computer Specifications and Software} Computations were carried out in a linux 64 bit server---Intel(R) Xeon(R) eight-core processor @ 1.80GHz,  16 GB of RAM 
% cache size 20480 KB
for the overdetermined $n>p$  case and in a Dell Precision T7600 computer with an Intel Xeon E52687 sixteen-core processor @ 3.1GHz, 128GB of Ram for the high-dimensional $p\gg n$ case. 
The discrete first order methods were implemented in \textsc{Matlab} 2012b. We used {\textsc{Gurobi}}~\cite{gurobi} version 5.5 and 
the \textsc{Matlab} interface to {\textsc{Gurobi}} for all of our experiments, apart from the 
computations for synthetic data for $n > p$, which were done in {\textsc{Gurobi}} via its Python 2.7 interface. 
%Some of the advanced ``Callback'' features were implemented via the  interface to {\textsc{Gurobi}}.
%To implement some of the advanced Callback features in {\textsc{Gurobi}}, we used 
%In the $p>n$ examples, we used the Python 2.7 interface to {\textsc{Gurobi}}  and 
%We used R 3.0.2 to compute 
%{\texttt {Lasso}},    {\texttt {Sparsenet}}  and stepwise regression  
%using the glmnet 1.7.3, {\texttt {Sparsenet}}  and Stats 3.0.2 packages respectively. 

%\subsection{The $n > p$ Regime}\label{sec:n>p}
\subsection{The Overdetermined Regime: $n > p$}\label{sec:n>p}

%\section{Algorithmic Performance of MIO} \label{sec:comp-expt-algo1}

%In this section, we illustrate the performance of the MIO approach for the best subset problem. We perform a variety of experiments to: (a) assess the quality of upper bounds obtained by Algorithm~2, the MIO formulation~\eqref{}, and the MIO formulation warm started with advanced solutions obtained by Algorithm~2 and (b) describe the ability of the MIO to deliver certificates of global optimality.

Using the Diabetes dataset and synthetic datasets, we demonstrate the combined  effect of using the discrete first order methods with the MIO approach. Together, these methods show improvements in obtaining good upper bounds and in closing the MIO gap to certify global optimality. Using synthetic datasets where 
we know the true  linear regression model, we perform side-by-side comparisons of this method with several other state-of-the-art algorithms designed to  estimate  sparse linear models. 
%Finally, we demonstrate the performance of these methods in the LAD case.

%%%%%%%%%%%%%%%%%%%%%
%% Obtaining Good Upper Bounds
%%%%%%%%%%%%%%%%%%%%%

\subsubsection{Obtaining Good Upper Bounds}\label{sec:good-ub1}

We conducted experiments to evaluate the performance of our
methods
% First Order algorithm, the MIO formulation and the MIO formulation with warm starts 
 in terms of obtaining high quality solutions for Problem~\eqref{eq-card-k}.

We considered the following three algorithms:

\begin{enumerate}
\item[{\bf (a)}]  Algorithm~2 with fifty random initializations\footnote{we took fifty random starting values around $\M{0}$ of the form 
$\min(i-1,1) \epsilon, i = 1, \ldots, 50$, where 
$\B\epsilon \sim N(\M{0}_{p \times 1}, 4\M{I})$. We found empirically that  Algorithm~2 provided better upper bounds than Algorithm~1.}. 
We took the solution corresponding to the best objective value.
\item[{\bf (b)}]  MIO with cold start, i.e., formulation~\eqref{eq-card-form2-0} with a time limit of $500$ seconds.
\item[{\bf (c)}]  MIO with warm start. This was the MIO formulation initialized with the discrete first order optimization solution obtained from {\bf (a)}. This was 
run for a total of 500 seconds.
\end{enumerate}

%summarized in Table~\ref{table-gran-1-small}. 
To compare the different algorithms in terms of the quality of upper bounds,  
we run for every instance  all the algorithms and obtain the best solution among them, say, $f_*$. If 
$f_{\text{alg}}$ denotes the value of the best subset objective function for  method ``alg'', then we define the relative accuracy of the solution obtained by ``alg'' as:
\begin{equation}\label{rel-accu}
\text{Relative Accuracy} = (f_{\text{alg}} - f_{*})/f_{*},
\end{equation}
where $\text{alg} \in \{ \text {(a)}, \text{(b)}, \text{(c)}\}$ as described above.

We did experiments for the Diabetes dataset for different values of $\mmk$ (see Table~\ref{tab:ub-diab-1}). For each of the algorithms we report 
 the amount of time taken by  the  algorithm to reach the best objective value during the time of $500$ seconds.

%Results comparing these methods are

\begin{table}[ht]
\centering
\scalebox{1.0}[.9]{\begin{tabular}{|c | cc | cc | cc |}
  \hline
\multirow{2}{*}{$\mmk$} &  \multicolumn{2}{ c| }{Discrete First Order} &  \multicolumn{2}{ c| }{MIO Cold Start} &  \multicolumn{2}{ c| }{MIO Warm Start} \\  
  &Accuracy& Time  &  Accuracy& Time &  Accuracy& Time \\  \hline
  9 & 0.1306 & 1 & 0.0036 & 500 & 0 & 346 \\ 
  20& 0.1541 & 1 & 0.0042 & 500 & 0 & 77 \\ 
%  27 & 0.1709 & 1 & 0 & 498 & 0.0021 & 463 \\ 
%  42 & 0.1875 & 1 & 0.0005 & 500 & 0 & 458 \\ 
  49 & 0.1915 & 1 & 0.0015 & 500 & 0 & 87 \\ 
  57 & 0.1933 & 1 & 0 & 500 & 0 & 2 \\   \hline
\end{tabular}}
\caption{Quality of upper bounds for  Problem~\eqref{eq-card-k} for the Diabetes dataset, for different values of $\mmk$. 
We see that the MIO equipped with warm starts deliver the best upper bounds in the shortest overall times.
The run time for the MIO with warm start includes the time taken by the discrete first order method (which were all less than a second).}\label{tab:ub-diab-1}
\end{table}

Using the discrete first order methods in combination with the MIO algorithm resulted in finding the best possible relative accuracy in a matter of a few minutes.

\subsubsection{Improving MIO Performance via Warm Starts}

%\paragraph{Performance on the Diabetes dataset}

\begin{figure}[h]
\centering
\scalebox{1.05}[.95]{\begin{tabular}{ c c c c}
\scriptsize { \sf { k=9} } &\scriptsize { \sf { k=20} } & \scriptsize{ \sf { k=31} } &\scriptsize{ \sf { k=42} }\\
\includegraphics[height = .3\textheight, width = 0.22\textwidth, trim = 0mm 12mm 5mm 15mm, clip]{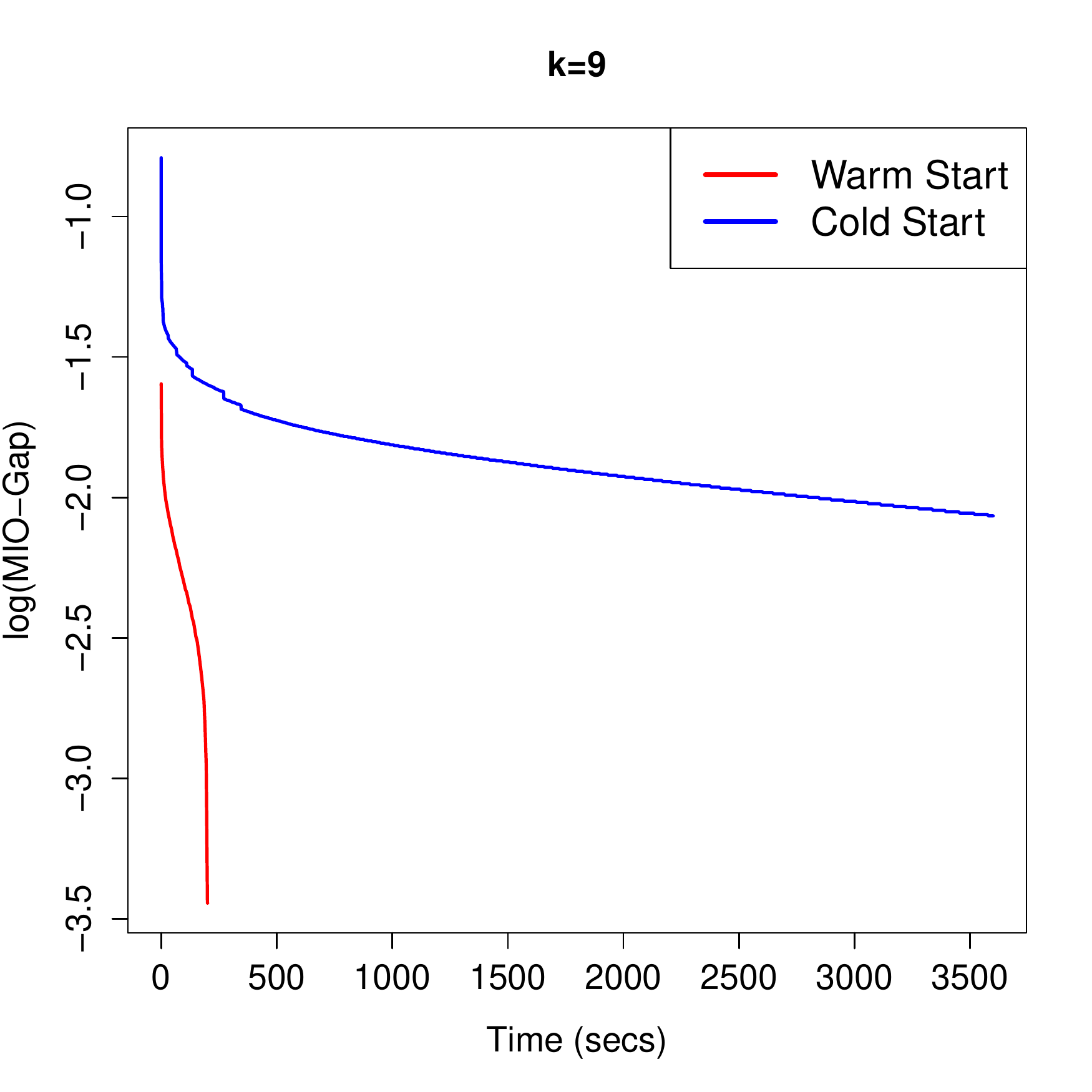}&
\includegraphics[height = .3\textheight, width = 0.22\textwidth, trim = 10mm 12mm 5mm 15mm, clip]{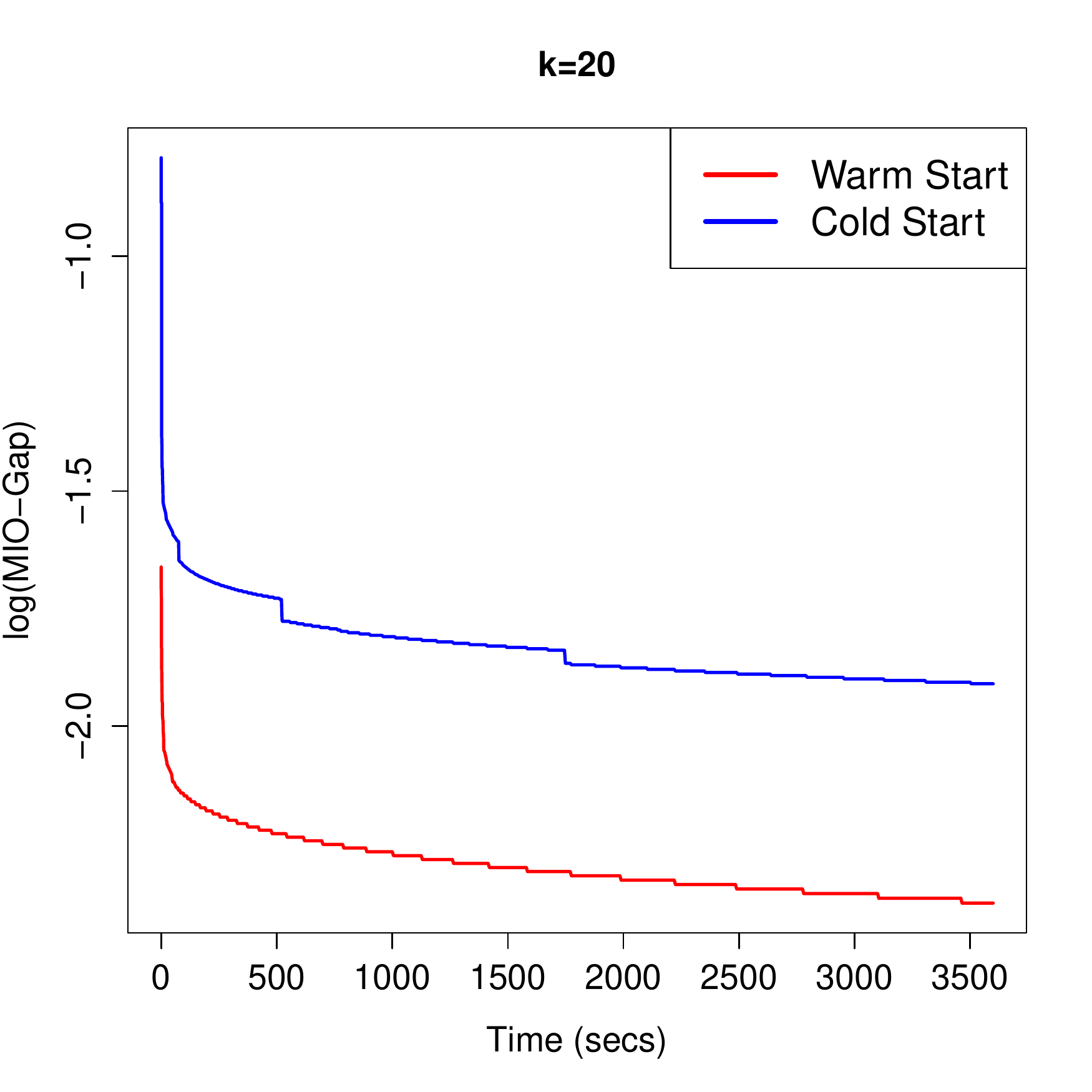}&
\includegraphics[height = .3\textheight, width = 0.22\textwidth, trim = 10mm 12mm 5mm 15mm, clip]{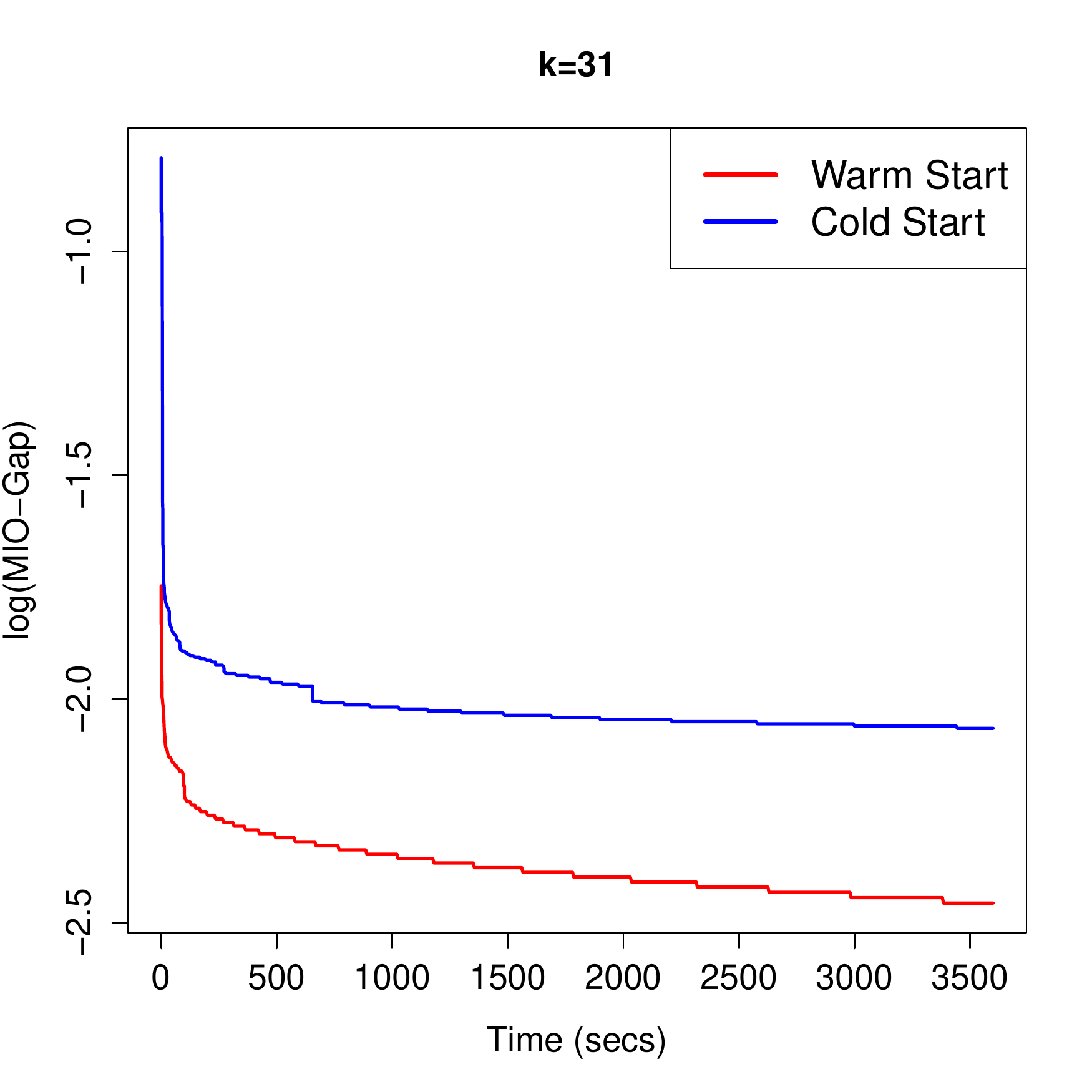}&
\includegraphics[height = .3\textheight, width = 0.22\textwidth, trim = 10mm 12mm 5mm 15mm, clip]{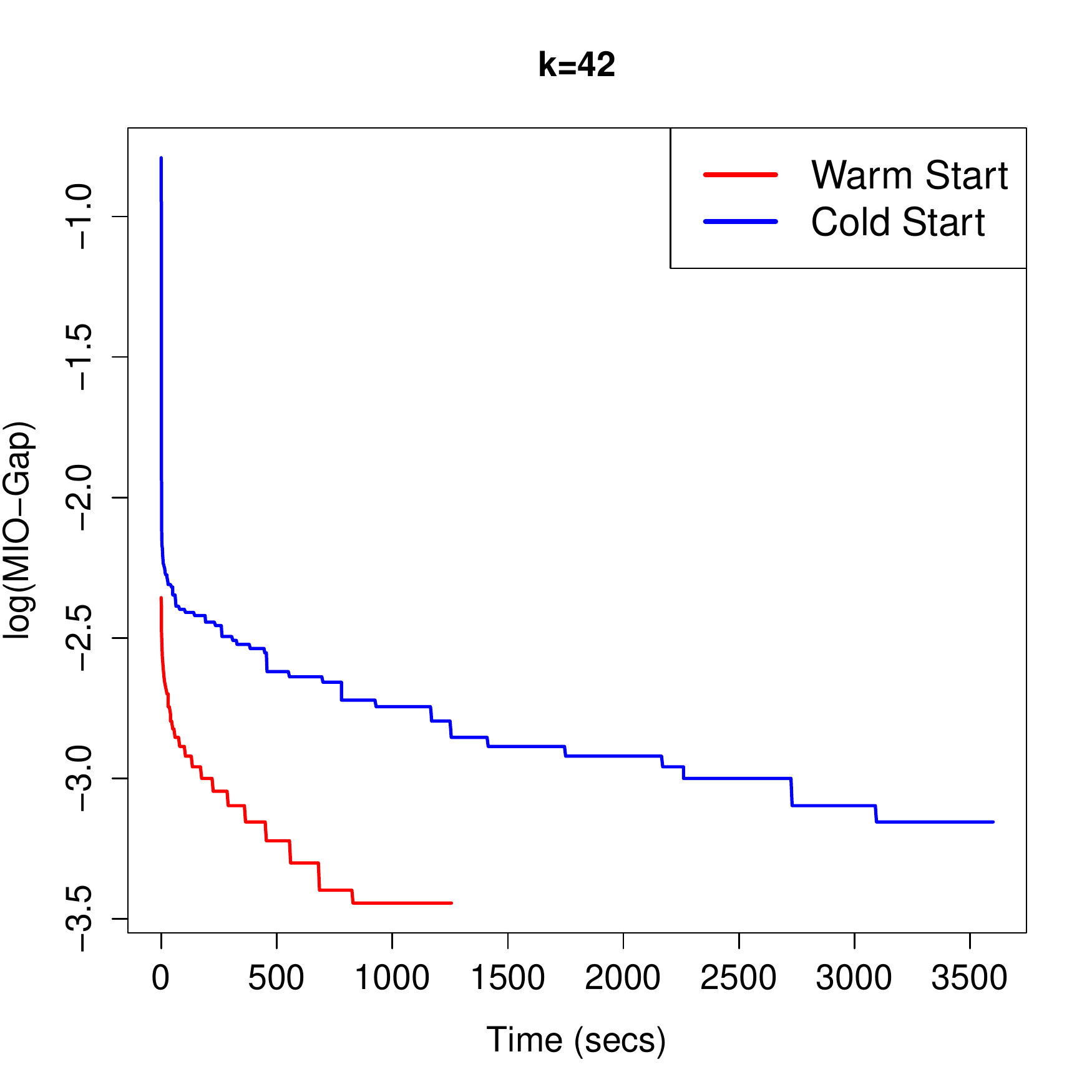} \\
\scriptsize {\sf {Time (secs)} } &\scriptsize { \sf { Time (secs)} } & \scriptsize{ \sf { Time (secs)} } &\scriptsize{ \sf {Time (secs)} }\\
\end{tabular}}
\caption{{ \small {The evolution of the MIO optimality gap (in $\log_{10}(\cdot)$ scale) for Problem~\eqref{eq-card-k}, for the Diabetes 
dataset with $n = 350 , p = 64$ with and without warm starts (and parameter specifications as in Section~\ref{sec:enhanced_form-1}) for different values of $\mmk$.
The MIO  significantly benefits by 
advanced warm starts delivered by Algorithm~2. In all of these examples, the global optimum was found within a very small 
fraction of the total time, but the proof of global optimality came later. 
%As the number of possible solutions
%grows as ${p \choose k}$, it takes longer to prove optimality for $k=31, 35$ compared to $k=42$.
} }  }\label{fig-diab-data2}
\end{figure}

We performed a series of experiments on the Diabetes dataset to obtain  a globally  optimal  solution  to Problem~\eqref{eq-card-k} via our approach  and  to understand the implications 
of using advanced warm starts to the MIO formulation in terms of certifying  optimality.
For each  choice of $\mmk$ we ran Algorithm~2  with fifty random initializations. They took less than a few seconds to run. 
We used the best solution  as an advanced warm start to the MIO formulation~\eqref{eq-card-form2-0}. 
For each of these examples, we also ran the MIO formulation 
without any warm start information and also without the parameter specifications in Section~\ref{sec:enhanced_form-1} (we refer to this as ``cold start'').
Figure~\ref{fig-diab-data2} summarizes the results. The figure shows that 
in the presence of warm starts and problem specific side information, the MIO closes the  optimality gap  significantly  faster.
%Some additional  are shown in Section~\ref{more-expts-1}.

\subsubsection{Statistical Performance}\label{stat-perf-n>p}

%In the overdetermined case, we considered synthetic datasets for values of $n$ ranging from $100-100,000$ and values of $p$ ranging from $30-100$. We took $\sigma_{ij} = \rho^{| i - j |}$ for $i, j \in \{1, \ldots, p \} \times \{ 1, \ldots, p\}$ and varied $\rho$ in the interval $[0,1]$ and $\sigma$ in the interval $[0.5, 15]$ in order to test different signal-to-noise ratios. In each case, we generated a dataset with ten underlying true variables by setting $\beta_i = 1$ for all $i$ such that $p$ mod $i = 0$, $0$ otherwise.  We denote this vector by $\B\beta^0$. 

We considered datasets as described in Example~1, Section~\ref{sec:expt-data-desc}---we took
different values of  $n,p$ with $n > p$, $\rho$ with $k_{0} = 10$.
\paragraph{Competing Methods and Performance Measures}

For every example, we considered the following learning procedures for comparison purposes: 
(a) the MIO approach equipped warm starts from Algorithm~2 (annotated as ``MIO'' in the figure), (b) the {\texttt {Lasso}},  (c) {\texttt {Sparsenet}} and (d) stepwise regression (annotated as ``Step'' 
in the figure).

We used {\texttt R} to compute 
{\texttt {Lasso}},    {\texttt {Sparsenet}}  and stepwise regression  
using the glmnet 1.7.3, {\texttt {Sparsenet}}  and Stats 3.0.2 packages respectively, which were all downloaded from {\texttt {CRAN}} at~\url{http://cran.us.r-project.org/}.

In addition to the above, we have also performed comparisons with a \emph{debiased} version of the {\texttt{Lasso}}: i.e., performing unrestricted least squares on the
{\texttt{Lasso}} support to mitigate the bias imparted by  {\texttt{Lasso}} shrinkage.

%For this example, we used the Python 2.7 interface to {\textsc{Gurobi}}.
%\textcolor{red}{RM modifications begin----------------------}

We note that {\texttt {{Sparsenet}}}~\cite{mhf-09-jasa} considers a penalized likelihood formulation of the form~\eqref{non-convex-1}, where the penalty 
is given by the generalized MCP penalty family (indexed by $\lambda, \gamma$) for a family of 
values of $\gamma \geq 1$ and $\lambda \geq 0$. The family of penalties used by {\texttt {{Sparsenet}}} is thus given by:
%\begin{equation}\label{mcp-pen-0}
%p(t; \gamma; \lambda) = \lambda( | t | - \frac{t^2}{2\lambda\gamma}) \M{I} ( | t | < \lambda \gamma) + \frac{\lambda^2\gamma}{2} \M{I} ( |t| \geq \lambda \gamma),
%\end{equation}
$p(t; \gamma; \lambda) = \lambda( | t | - \frac{t^2}{2\lambda\gamma}) \M{I} ( | t | < \lambda \gamma) + \frac{\lambda^2\gamma}{2} \M{I} ( |t| \geq \lambda \gamma)$
for $\gamma, \lambda$ described as above. As $\gamma = \infty$ with $\lambda$ fixed, we get the penalty $p(t; \gamma; \lambda)= \lambda |t|$.
The family above includes as a special case ($\gamma = 1$), the hard thresholding penalty, a penalty recommended in the paper~\cite{Lv-2014} for its useful statistical properties.

%\textcolor{red}{RM modifications END----------------------}

For each procedure, we obtained the ``optimal'' tuning parameter by selecting the model that  achieved
the best predictive performance on  
a held out validation set. Once the model $\widehat{\B\beta}$ was selected, we obtained  the  
 prediction error as:
\begin{equation}\label{def-pred-error-1}
\text{Prediction Error} = \| \M{X} \widehat{\B\beta} - \M{X} \B{\beta}^0\|_2^2 /\|\M{X} \B{\beta}^0\|_2^2.
\end{equation}
We   report ``prediction error'' and number of non-zeros in the optimal model in our results.
 The results were averaged over ten random instances, for different realizations of $\M{X},\epsilon$.
For every run: the training and validation data had a fixed $\M{X}$ but random noise $\epsilon$.

%Figure~\ref{fig-n1000p50} presents results for $n$ = 1000 and $p$ = 50 and 
%The results show the superior statistical performance of the models delivered by MIO.
%This is apparent especially in cases of low noise, but continues even as the correlation between columns of $\M{X}$ increases. In addition to predictive performance, the MIO best subset algorithm performs competitively as a variable selection method. Each of the algorithms tested correctly identified the \textcolor{red}{set of 10 true nonzero elements of $\B\beta^0$}, but the MIO best subset algorithm typically includes a very few number of false positives.

\begin{figure}[!h]
\centering
\scalebox{.98}[.63]{\begin{tabular}{ c c c }
\includegraphics[height = .35\textheight, width = 0.36\textwidth, trim = 0mm 8mm 3mm 3mm, clip]{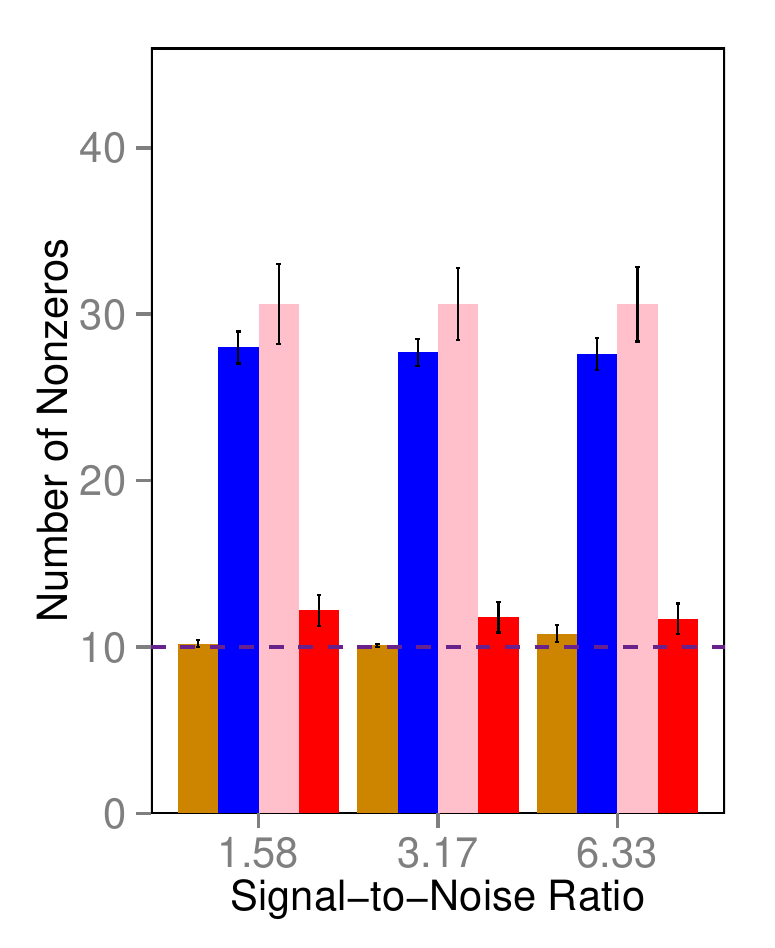}&
\includegraphics[height = .35\textheight,width = 0.29\textwidth, ,trim = 15mm 8mm 3mm 3mm, clip]{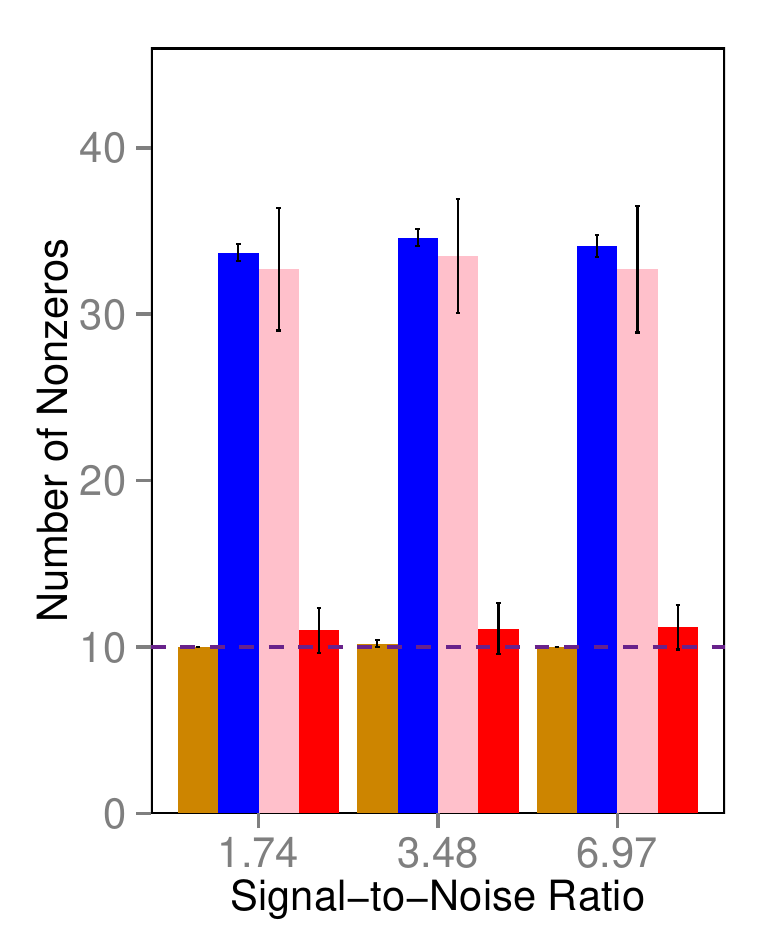}&
\includegraphics[height = .35\textheight,width = 0.29\textwidth, , trim = 15mm 8mm 3mm 3mm, clip]{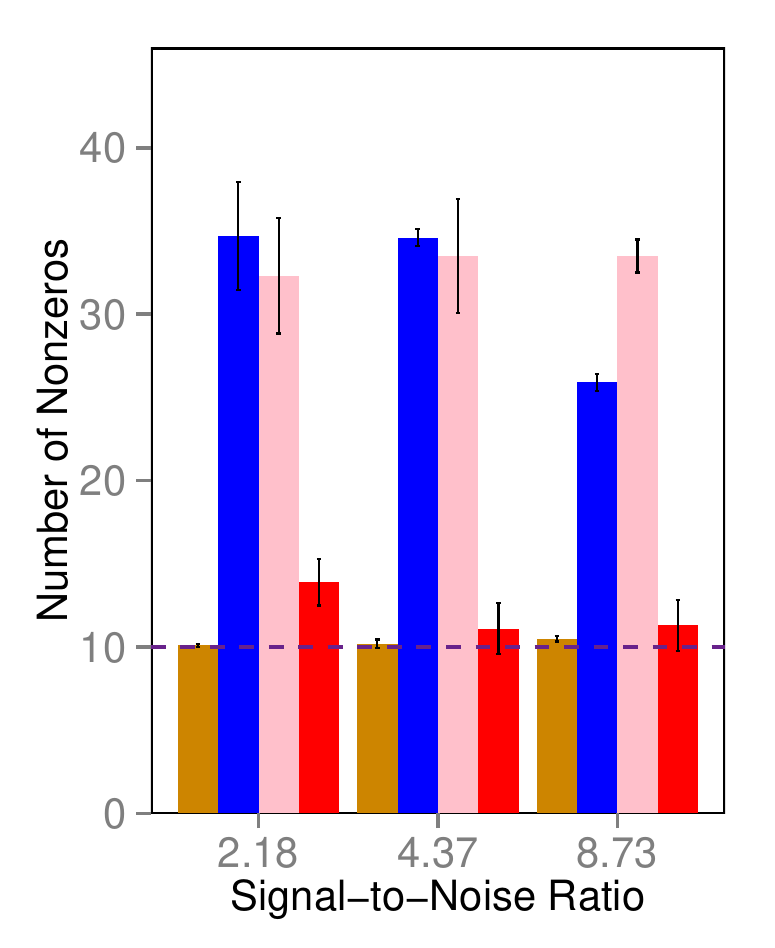}\\
\includegraphics[height = .35\textheight,width = 0.36\textwidth, , trim = 3.5mm 2mm 3mm 3mm, clip]{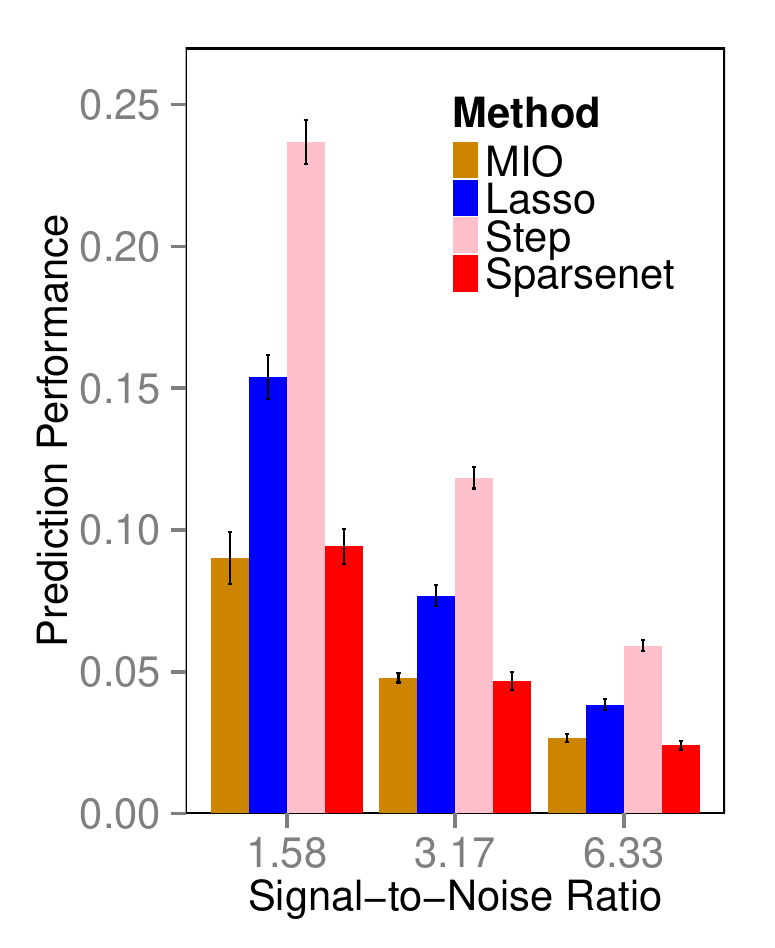}&
\includegraphics[height = .35\textheight, width = 0.29\textwidth, trim = 18.5mm 2mm 3mm 3mm, clip]{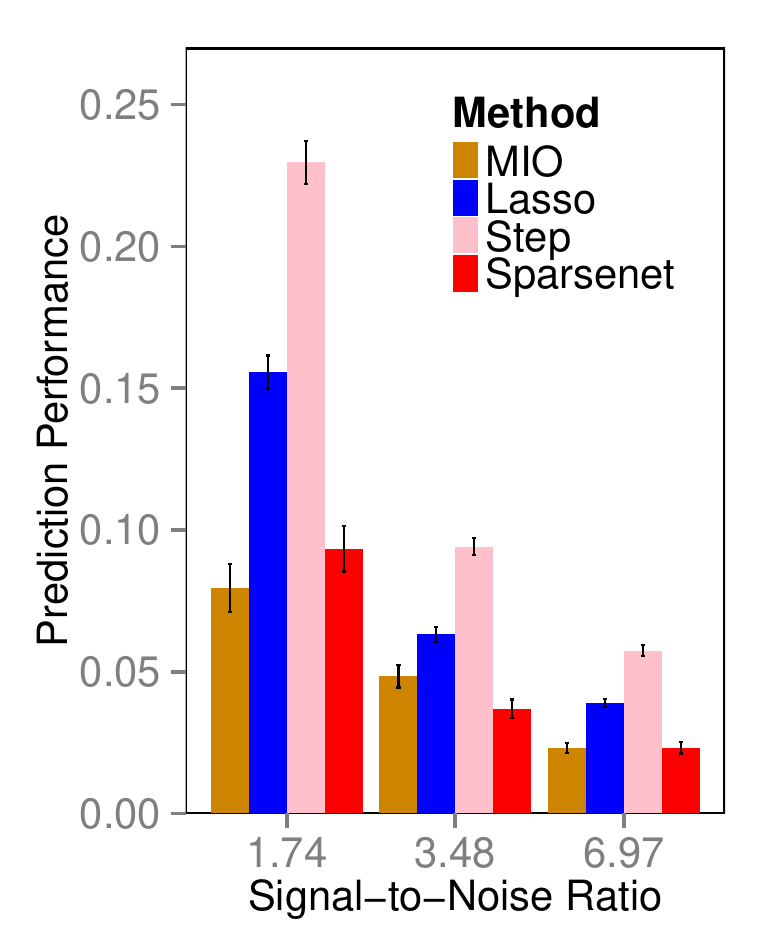}&
\includegraphics[height = .35\textheight,width = 0.29\textwidth, ,trim = 18.5mm 2mm 3mm 3mm, clip]{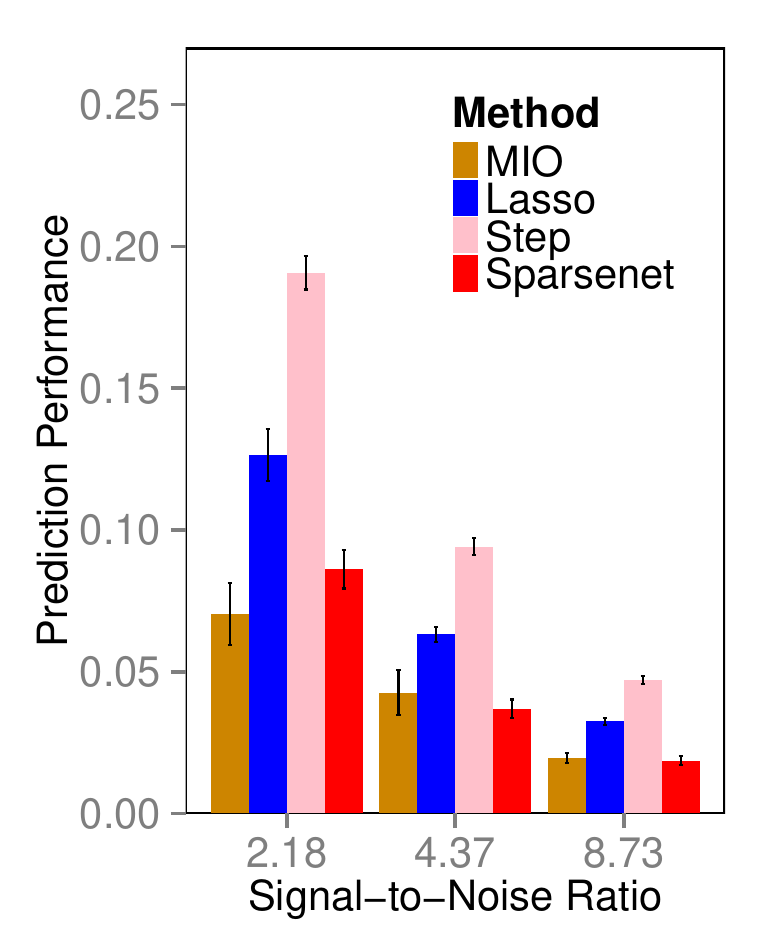}
\end{tabular}}
\caption{{ \small {Figure showing the sparsity (upper panel) and predictive performances (bottom panel) for different subset selection procedures for the least squares loss.  
Here,  we consider data generated as per Example~1, with $n = 500, p = 100$, $k_{0} = 10$, for three different SNR values with [Left Panel] $\rho = 0.5$, [Middle Panel] $\rho = 0.8$, and [Right Panel] $\rho = 0.9$.  The dashed line in the top panel represents the true number of nonzero values. For each of the procedures, the optimal model was selected as the one which produced the best prediction accuracy on a separate validation set, as described in~Section~\ref{stat-perf-n>p}.   } } }
\label{fig-n500p100}
\end{figure}

Figure~\ref{fig-n500p100} presents results for data generated as per  Example~1 with $n$ = 500 and $p$ = 100. 
We see that the MIO procedure performs very well across all the examples. Among the methods, MIO performs the best, followed by {\texttt {Sparsenet}},
{\texttt {Lasso}} with {\texttt Step(wise)} exhibiting the worst performance. 
In terms of prediction error, the MIO performs the best, only to be marginally 
outperformed by {\texttt {Sparsenet}} in a few instances. This further illustrates the importance of using non-convex methods in sparse learning.
Note that the MIO approach, unlike {\texttt {Sparsenet}} certifies global optimality in terms of solving Problem~\ref{eq-card-k}.
However, based on the plots in the upper panel, {\texttt {Sparsenet}} selects a few redundant variables unlike MIO. 
{\texttt {Lasso}} delivers quite dense models and pays the price in predictive performance too, by selecting wrong variables. 
As the value of SNR increases, the predictive power of the methods improve, as expected. The differences in predictive errors between the methods diminish with increasing SNR values.
With increasing values of $\rho$ (from left panel to right panel in the figure), the number of non-zeros selected by the {\texttt {Lasso}} in the optimal model increases.

We also performed experiments with the debiased version of the {\texttt{Lasso}}. The unrestricted least squares solution on the optimal 
model selected by {\texttt{Lasso}} (as shown in Figure~\ref{fig-n500p100}) had worse predictive performance than the {\texttt{Lasso}}, 
with the same sparsity pattern.  This is probably due to overfitting since the model selected by the {\texttt{Lasso}} is quite dense compared to $n,p$.
We also tried some variants of debiased {\texttt{Lasso}} which led to models with better performances than the {\texttt{Lasso}} but  the results were inferior compared to MIO --- we provide a 
detailed description in Section~\ref{sec:debiased-lasso}.

%We thus experimented with another variant of the debiased Lasso, where for every $\lambda$ we computed the lasso solution~\eqref{lass-lag}
%and obtained $\hat{\B\beta}_{\text{Deb}, \lambda}$ by performing an unrestricted least squares fit on the support selected by the lasso solution at $\lambda$. 
%This method can be thought of delivering feasible solutions for Problem~\eqref{eq-card-k}, for a value of $k:=k(\lambda)$ determined by the lasso solution at $\lambda$.
%The success of this method makes a case in support of using criterion~\eqref{eq-card-k}. 
%The tuning parameter was then selected by minimizing predictive performance on a held out test validation set. The performance of this model 
%was comparable with {\texttt{Sparsenet}}---it was better than lasso in terms of obtaining a sparser model with better predictive accuracy. 
%However, the performance of MIO was significantly better than the debiased version of the lasso, especially for larger values of $\rho$ and smaller SNR values.
%The details of the results are provided in the appendix. 

We also performed experiments with $n = 1000, p = 50$ for data generated as per Example~1. We  solved  the problems to provable optimality and found that the MIO performed very well when compared to other competing methods. We do not report the experiments for brevity.

\subsubsection{MIO model training} 
We trained a sequence of best subset models (indexed by $\mmk$) by applying the MIO approach with warm starts. 
Instead of running the MIO solvers from scratch for different values of $\mmk$, we used \emph{callbacks}, a feature of integer optimization solvers.
%\footnote{To implement the callback feature we used the Python 2.7 interface to {\textsc{Gurobi}}.}.
Callbacks allow the user to solve an initial model, and then add additional constraints to the model one at a time. These ``cuts" reduce the size of the feasible region without having to rebuild the entire optimization model. Thus, in our case, we can save time by building the initial optimization model for $k = p$. Once the solution for $k = p$ is obtained, a cut can be added to the model: $\sum_{i=1}^{p} z_{i} \leq k$ for $k = p-1$ and the model can be re-solved from this point. We apply  this procedure until we arrive at a model with $k = 1$.

For each value of $\mmk$ tested, the MIO best subset algorithm was set to stop the first time either an optimality gap of 1\%
 was reached or a time limit of 15 minutes was reached. Additionally, we only tested values of $\mmk$ from 5 through 25, and used Algorithm~2 to warm start the MIO algorithm.
We observed  that it was possible to obtain speedups of a factor of 2-4 by carefully tuning the optimization solver for a particular problem, but chose to maintain generality by solving with default parameters. Thus, we do not report times with the intention of accurately benchmarking the best possible time but rather to show that it is computationally   tractable to solve problems to   optimality using modern    MIO solvers.

\subsection{The High-Dimensional Regime: $p\gg  n$}\label{sec:n<p}

In this section, we investigate  
{\bf (a)} the evolution of upper bounds in the high-dimensional regime,
{\bf (b)} the effect of a bounding box formulation on the speed of closing the optimality  gap and 
{\bf (c)}  the statistical performance of the MIO approach in comparison to other state-of-the art methods.

%\begin{enumerate}
% \item[{\bf (a)}]  the evolution of upper bounds in the high-dimensional regime,
% \item[{\bf (b)}]  the effect of a bounding box formulation on the speed of closing the optimality  gap,
% \item[{\bf (c)}]  the statistical performance of the MIO approach in comparison to other state-of-the art methods
% %\item[{\bf (d)}]  the performance of the MIO approach  in the high-dimensional LAD case.  
%\end{enumerate}
\subsubsection{Obtaining Good Upper Bounds}
%%\textcolor{red}{changed a minor thing here, in red}
We performed tests similar to those in Section~\ref{sec:good-ub1} for the $p\gg  n$ regime. We tested a synthetic dataset corresponding to Example~2 with
$n = 30 , p = 2000$ for varying  SNR  values (see Table~\ref{tab:ub-synth-1}) over a time of 500s.  
As before, using the discrete first order methods in combination with the MIO algorithm resulted in finding the best possible upper bounds in the shortest possible times.

% latex table generated in R 2.15.1 by xtable 1.7-3 package
% Mon Apr  7 11:47:07 2014

\begin{table}[ht]
\centering
\scalebox{1.2}[.7]{\begin{tabular}{| c | c| cc | cc | cc |}
  \hline
 &\multirow{2}{*}{$\mmk$} &  \multicolumn{2}{  c | }{Discrete First Order} &  \multicolumn{2}{ c |}{MIO Cold Start} &  \multicolumn{2}{ c | }{MIO Warm Start} \\  
 & &Accuracy& Time  &  Accuracy& Time &  Accuracy& Time \\  \hline
% & 4 & 0.1091 & 42.9 & 0.2910 & 500 & 0 & 65.9 \\ 
  &5 & 0.1647 & 37.2 & 1.0510 & 500 & 0 & 72.2 \\ 
 & 6 & 0.6152 & 41.1 & 0.2769 & 500 & 0 & 77.1 \\ 
 & 7 & 0.7843 & 40.7 & 0.8715 & 500 & 0 & 160.7 \\ 
 \rot{\rlap{SNR = 3}}
 & 8 & 0.5515 & 38.8 & 2.1797 & 500 & 0 & 295.8 \\ 
 & 9 & 0.7131 & 45.0 & 0.4204 & 500 & 0 & 96.0 \\ 
   \hline \hline
    %%%%% -----------------------------------------------------------------------
% & 4 & 0.2708 & 47.8 & 0 & 31 & 0 & 107.8 \\ 
 & 5 & 0.5072 & 45.6 & 0.7737 & 500 & 0 & 65.6 \\ 
 & 6 & 1.3221 & 40.3 & 0.5121 & 500 & 0 & 82.3 \\ 
 & 7 & 0.9745 & 40.9 & 0.7578 & 500 & 0 & 210.9 \\ 
  \rot{\rlap{SNR = 7}}
   & 8 & 0.8293 & 40.5 & 1.8972 & 500 & 0 & 262.5 \\ 
 & 9 & 1.1879 & 44.2 & 0.4515 & 500 & 0 & 254.2 \\\hline
\end{tabular}}
\caption{{\small{The quality of upper bounds for Problem~\eqref{eq-card-k} obtained by Algorithm~2, MIO with cold start and MIO warm-started with Algorithm~2.
We consider the synthetic dataset of Example~2 with $n = 30 , p = 2000$  and different values of SNR. The MIO method, when warm-started with the first order solution 
performs the best in terms of getting a good upper bound in the shortest time. The metric ``Accuracy'' is defined in~\eqref{rel-accu}. The first order methods are fast but need not lead to 
highest quality solutions on their own. MIO improves the quality of upper bounds delivered by the first order methods and their combined effect leads to the best performance.}}}\label{tab:ub-synth-1}
\end{table}

%% figure taken from: synthetic_iidGncase2_MIO_progress_expt3.RData, synthetic_iidGncase2_MIO_progress_expt4.RData, %%%synthetic_iidGncase2_MIO_progress_expt5.RData:
%% here n = 30 , p = 2K, kpop = 5m bet =rep(1,5) and zero o/w

\begin{figure}[!h]
\begin{center}
\scalebox{.9}[.7]{\includegraphics[height = .4\textheight, width = 0.7\textwidth, trim = 0mm 5mm 10mm 14mm, clip]{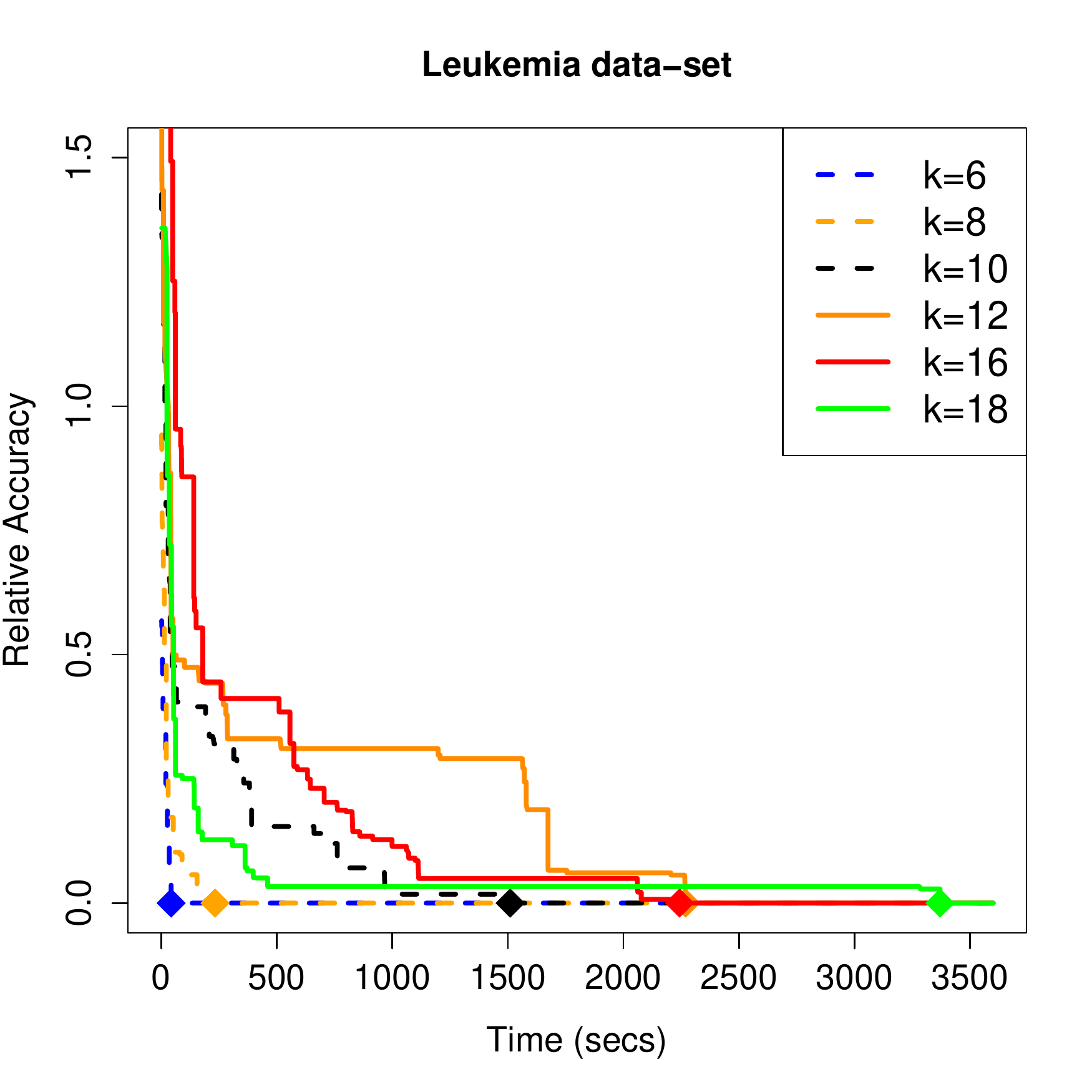}}
\caption{{ \small {Behavior of MIO aided with warm start in obtaining good upper bounds  over  time for the Leukemia dataset $(n=72, p = 1000$).
The vertical axis shows relative accuracy, i.e.,  $(f_{t} - f_*)/f_*$, where $f_t$ is the objective value obtained after $t$ seconds and 
$f_*$ denotes the best objective value obtained by the method after 4000 seconds. The colored diamonds correspond to the locations 
where the MIO (with warm start) attains the best solution. The figure shows that MIO improves the solution obtained by the first order method in all the instances.  
The time at which the best possible upper bound is obtained depends upon the choice of $\mmk$. Typically larger $\mmk$ values make the problem harder---hence the best solutions are obtained after a longer wait.} }  }\label{fig-leuk-data2}
\end{center}
\end{figure}

We also did experiments on the Leukemia dataset. In Figure~\ref{fig-leuk-data2} we demonstrate the evolution of the objective value of the best subset problem for 
different values of $\mmk$. For each value of $\mmk$, we warm-started the MIO with the solution obtained by Algorithm~2 and allowed the 
MIO solver to run for 4000 seconds. The best objective value obtained at the end of 4000 seconds is denoted by $f_*$. We plot 
the Relative Accuracy, i.e.,  $(f_{t} - f_*)/f_*$, where $f_t$ is the objective value obtained after $t$ seconds. The figure shows that 
the solution obtained by Algorithm~2 is improved by the MIO on various instances and the time taken to improve the upper bounds 
depends upon $\mmk$. In general, for smaller values of $\mmk$ the upper bounds obtained by the MIO algorithm stabilize earlier, i.e., the MIO 
finds improved solutions faster than larger values of $\mmk$. 

%%%%%%%%%%%%%%%%%%%%%
%% Effect of a Bounding Box
%%%%%%%%%%%%%%%%%%%%%
\subsubsection{Bounding Box Formulation}

%\paragraph{Performance in regimes with $p \gg n$}
With the aid of advanced warm starts as 
provided by Algorithm~2, the MIO obtains a very high quality solution very quickly---in most of the examples the solution thus obtained turns out to be the global minimum. However, in the typical ``high-dimensional''  regime, with $p\gg  n$, we observe that the certificate of global optimality comes later as the lower bounds of the problem ``evolve'' slowly.
This is observed even in the presence of warm starts and using the implied bounds as developed in Section~\ref{sec:formulations} and is aggravated  
for the cold-started MIO formulation~\eqref{eq-card-form2-1}.

To address this, we consider   the MIO formulation~\eqref{eq-card-form2-1-xbet} obtained by
adding bounding boxes around a local solution.
These restrictions \emph{guide} the MIO in restricting its \emph{search} space and enable the MIO to certify global optimality
inside that bounding box.
We consider the following additional bounding box constraints to the MIO formulation~\eqref{eq-card-form2-1}:
$$ \left \{ \B\beta : \| \M{X}\B\beta - \M{X}\B\beta_{0} \|_{1} \leq {\mathcal L}^{\zeta}_{\ell, \text{loc}} \right \} \; \cap \;  \left \{ \B\beta :  \| \B\beta - \B\beta_{0} \|_{1} \leq {\mathcal L}^{\beta}_{\ell, \text{loc}}\right\},$$
%\textcolor{red}{where $\B\beta_{0}$ is a candidate $k$-sparse solution ($\|\B\beta\|_{0} \leq \mmk$) which has been obtained, say,  after running the MIO} formulation~\eqref{eq-card-form2-1} %equipped with a warm start delivered by Algorithm~2 for a sufficiently long time (for example, $500$---$1000$ seconds).  
%The radii of the $\ell_{1}$-balls, namely ${\mathcal L}^{\zeta}_{\ell, \text{loc}}$ and ${\mathcal L}^{\beta}_{\ell, \text{loc}}$ 
%are user-defined parameters and control the size of the feasible set.
where, $\B\beta_{0}$ is a candidate sparse solution. The radii of the two $\ell_{1}$-balls above, namely, ${\mathcal L}^{\zeta}_{\ell, \text{loc}}$ and ${\mathcal L}^{\beta}_{\ell, \text{loc}}$ 
are user-defined parameters and control the size of the feasible set.

\begin{figure}[!h]
\centering
\scalebox{.99}[.7]{\begin{tabular}{ c  c }
 \multicolumn{2}{  c  }{ \sf Leukemia dataset: Effect of a Bounding Box for MIO formulation~\eqref{eq-card-form2-1-xbet}} \\
{\sf $\mmk = 5$} & {\sf $\mmk = 10$} \\
\includegraphics[height = .3\textheight, width = 0.45\textwidth, trim = 0mm 6mm 3mm 15mm, clip]{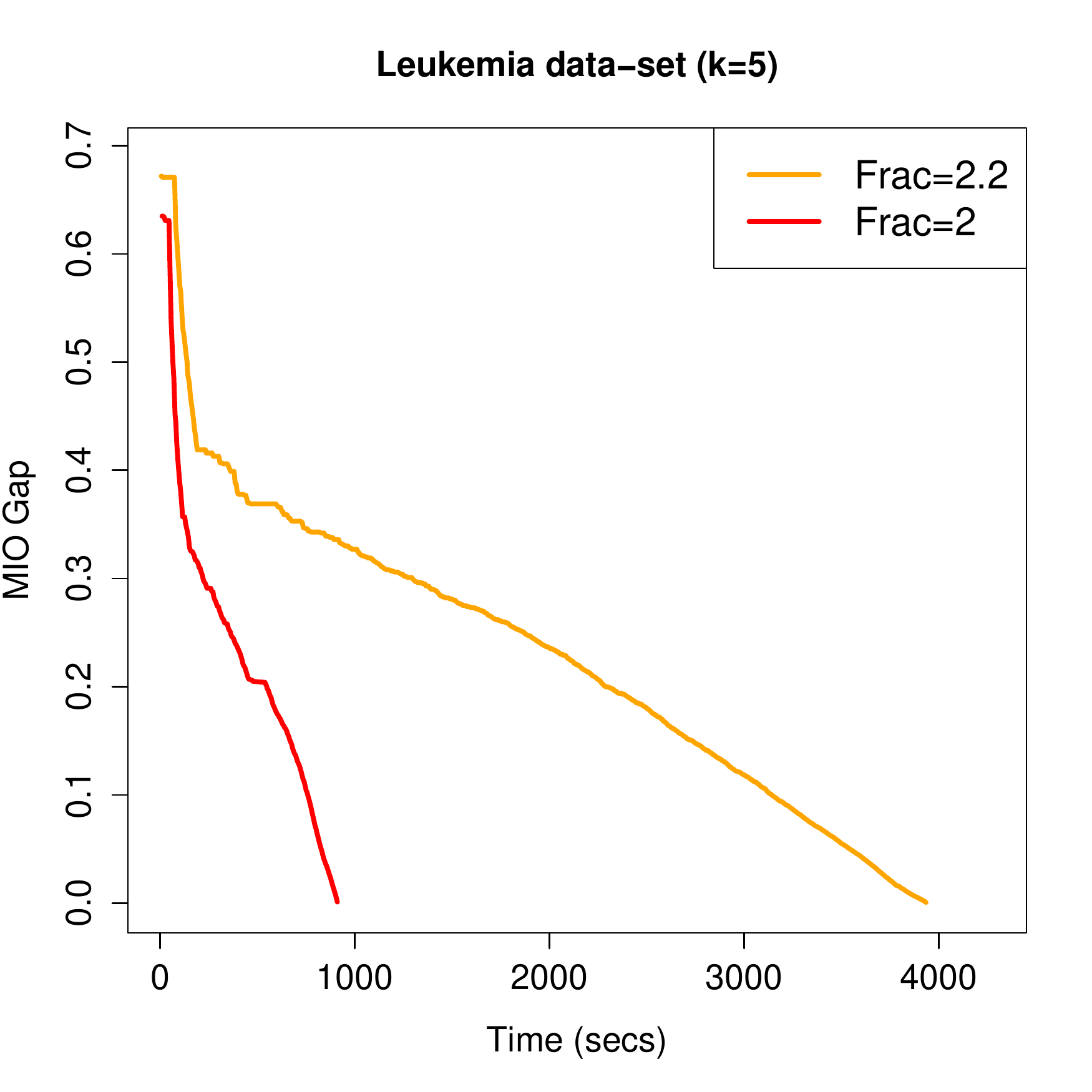} &
\includegraphics[height = .3\textheight,width = 0.45\textwidth, ,trim = 6mm 6mm 3mm 15mm, clip]{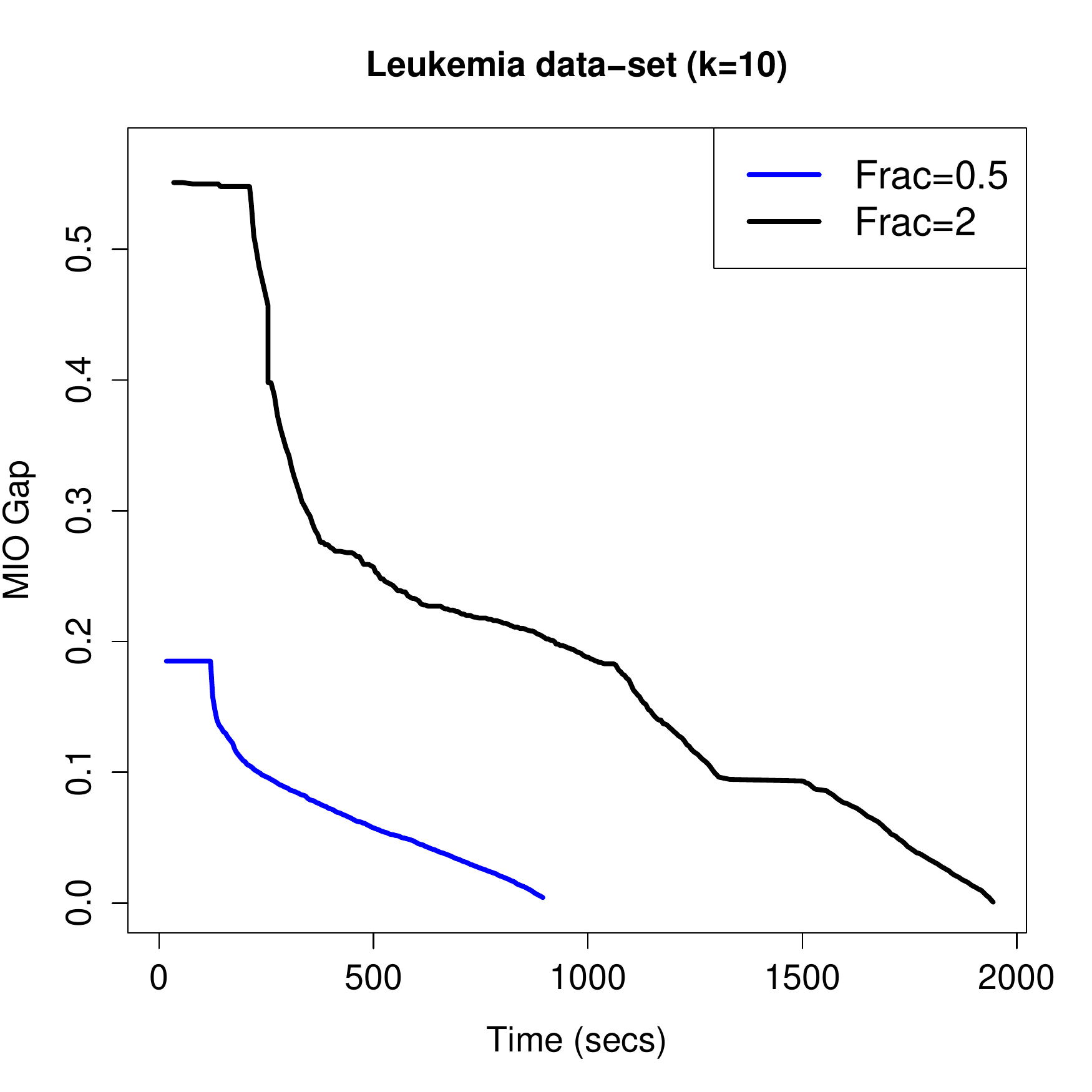}
\end{tabular}}
\caption{{ \small {The effect of the MIO formulation~\eqref{eq-card-form2-1-xbet} for the Leukemia dataset, for different values of $\mmk$.
 Here ${\mathcal L}^{\zeta}_{\ell, \text{loc}} = \infty$  and ${\mathcal L}^{\beta}_{\ell, \text{loc}} = \text{Frac}$. For each value of $\mmk$, 
 the global minimum obtained was the same for the different choices of ${\mathcal L}^{\beta}_{\ell, \text{loc}}$. } }  }\label{fig-leuk-data-bounding1}
\end{figure}

\begin{figure}[!h]
\centering
\scalebox{.98}[.62]{\begin{tabular}{ccc}\\
 \multicolumn{3}{  c  }{ { \sf Evolution of  the MIO gap for~\eqref{eq-card-form2-1-xbet}, effect of type of bounding box ($n=50, p = 500$).}} \\
\includegraphics[height = .4\textheight, width = 0.31\textwidth, trim = 0mm 6mm 3mm 15mm, clip]{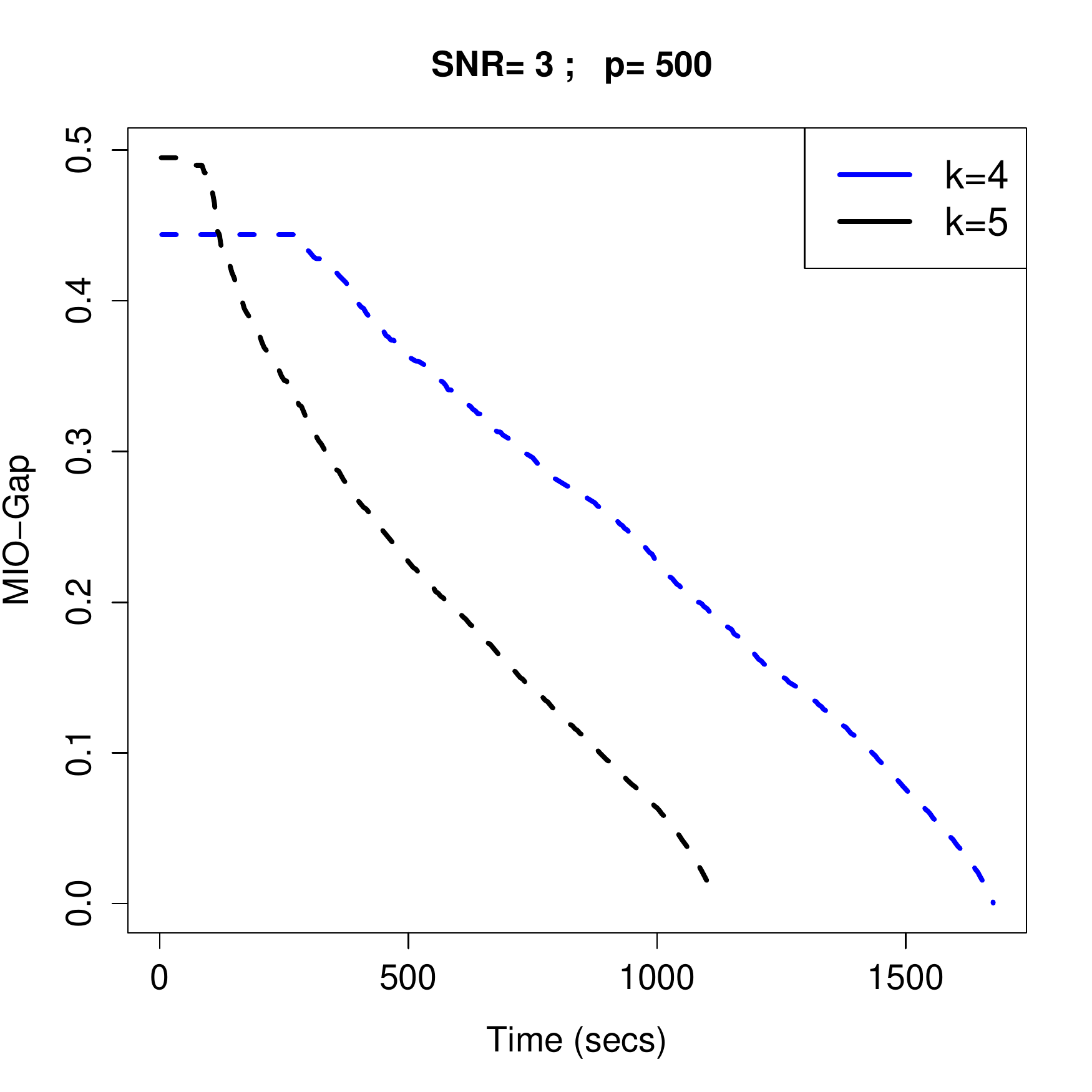}&
 \includegraphics[height = .4\textheight, width = 0.31\textwidth, trim = 6mm 6mm 3mm 15mm, clip]{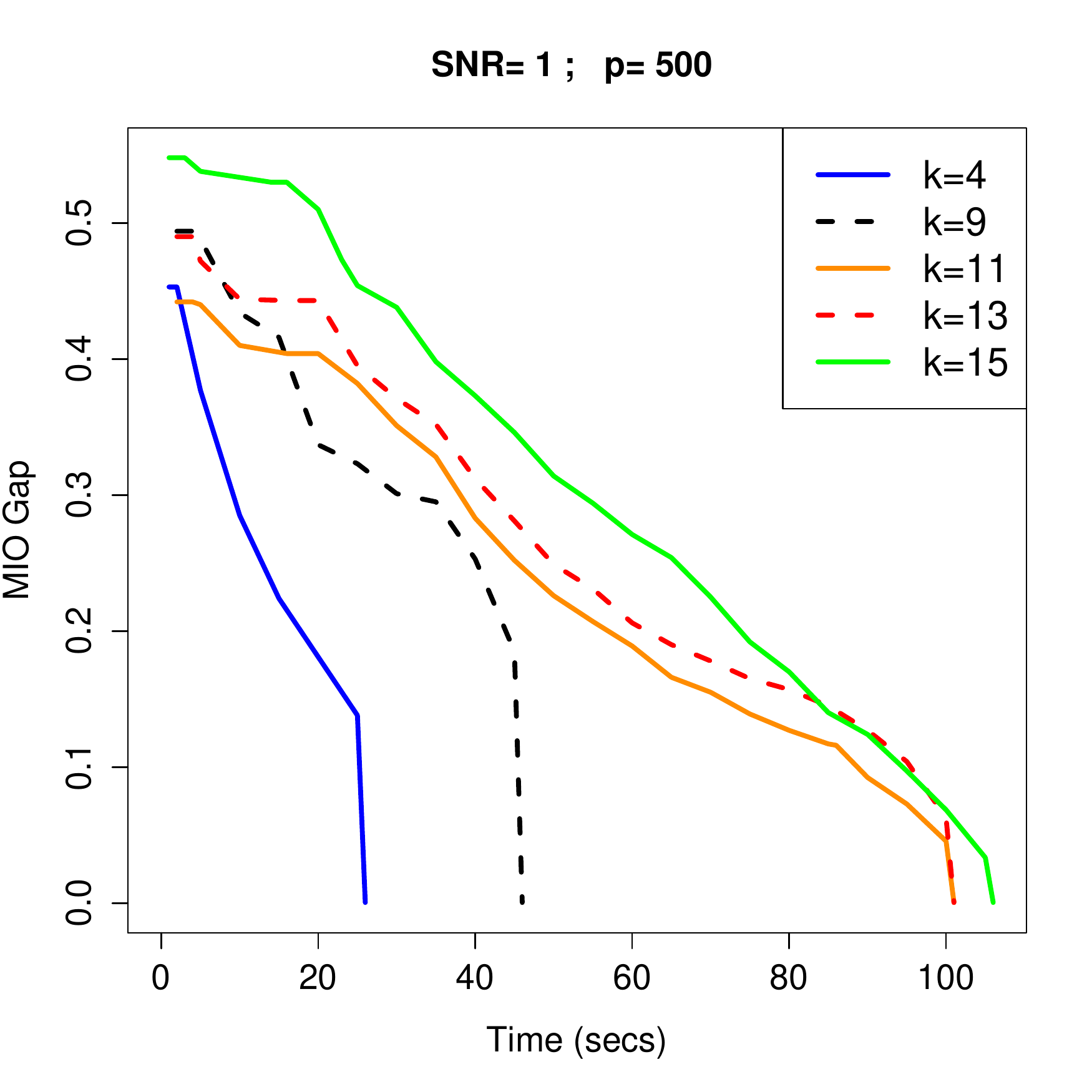}&
 \includegraphics[height = .4\textheight, width = 0.31\textwidth, trim = 6mm 6mm 3mm 15mm, clip]{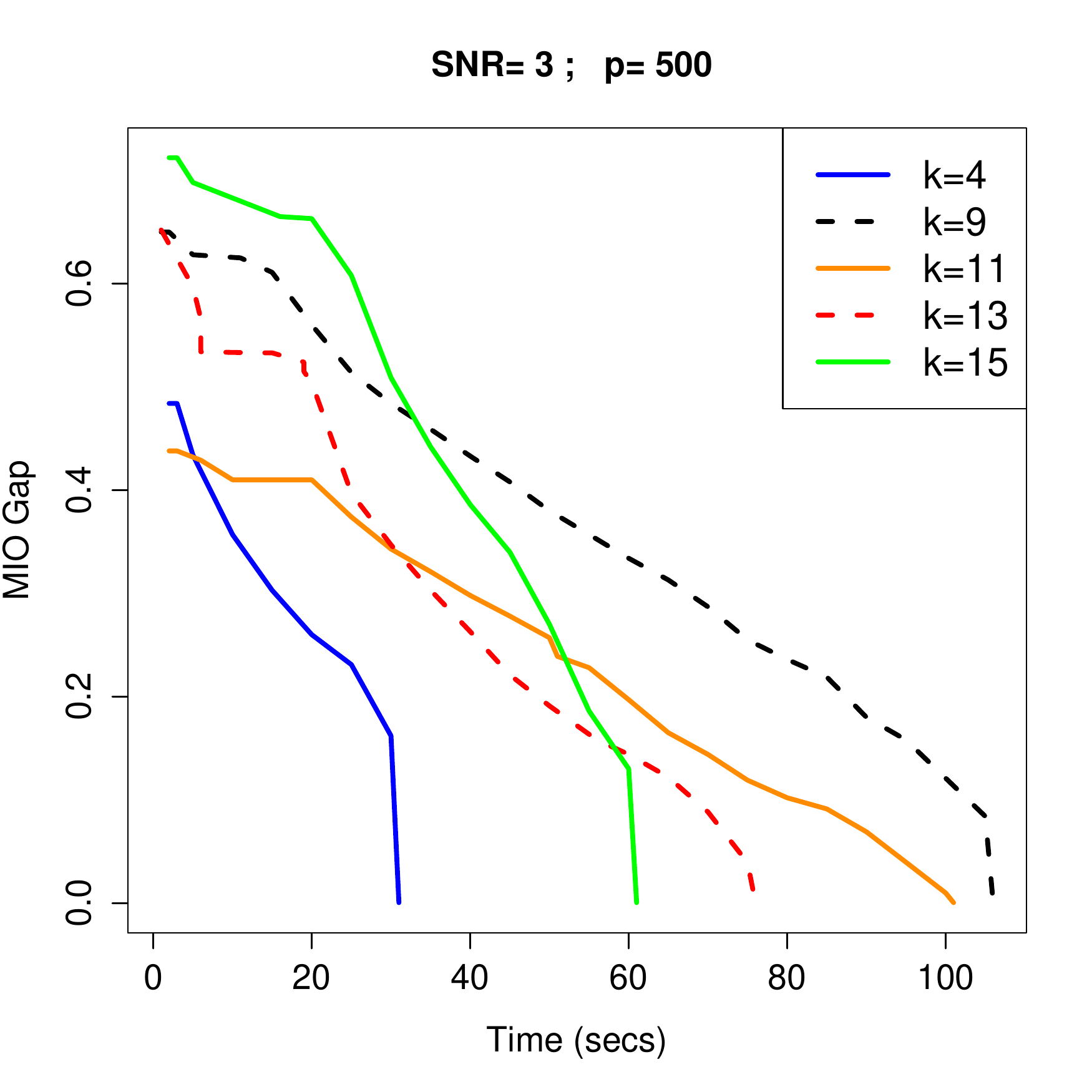} 
 \end{tabular}}
\caption{{ \small {The effect of the MIO formulation~\eqref{eq-card-form2-1-xbet} for a synthetic dataset as in Example~1 with $\rho= 0.9$, $k_{0} = 5$, $n = 50, p = 500$, 
for different values of $\mmk$. 
 [Left Panel] ${\mathcal L}^{\zeta}_{\ell, \text{loc}} = 0.5\| \M{X} \B\beta_{0} \|_{1}$  and ${\mathcal L}^{\beta}_{\ell, \text{loc}} = \infty$ for a data-set with SNR = 3.
 [Middle Panel] ${\mathcal L}^{\zeta}_{\ell, \text{loc}} = \infty $, ${\mathcal L}^{\beta}_{\ell, \text{loc}} =\| \B\beta_{0}\|_{1}/\mmk$ and SNR = 1.  
 [Right Panel] ${\mathcal L}^{\zeta}_{\ell, \text{loc}} = \infty $, ${\mathcal L}^{\beta}_{\ell, \text{loc}} =\| \B\beta_{0}\|_{1}/\mmk$ and SNR = 3. 
The figure shows that the bounding boxes in terms of $\M{X}\B\beta$ (left-panel) make the problem harder to solve, when compared to bounding boxes around $\B\beta$ (middle and right panels). 
A possible reason is due to the strong correlations among the columns of $\M{X}$. The SNR values do not seem to have a big impact on the run-times of the algorithms (middle and right panels). } }  }
 \label{fig-synth-data-bounding1}
\end{figure}

Using the notation $\B\zeta = \M{X}\B\beta$ we have the following MIO formulation (equipped with the additional bounding boxes):
\begin{equation}\label{eq-card-form2-1-xbet}
\begin{myarray}[1.1]{l  l }
\min\limits_{\B\beta, \M{z} , \B\zeta }&  \;\;  \frac{1}{2}\; \B\zeta^{T}\B\zeta - \langle  \M{X}'\M{y}, \B\beta   \rangle  + \frac{1}{2}\; \| \M{y}\|_2^2  \\
s.t. \; & \B\zeta = \M{X} \B\beta \\
& (\beta_{i}, 1 - z_{i}) : \text{SOS type-1,}\;\; i = 1, \ldots, p\\
& \;\; z_{i} \in \{ 0 , 1 \}, i = 1, \ldots, p\\
& \;\; \sum\limits_{i=1}^{p} z_{i} \leq \mmk \\
&  - {\mathcal M}_{U} \leq \beta_{i} \leq {\mathcal M}_{U} , i = 1, \ldots, p\\
&  \| \B\beta\|_{1} \leq {\mathcal M}_{\ell}\\
&  - {\mathcal M}^{\zeta}_{U} \leq \zeta_{i} \leq {\mathcal M}^{\zeta}_{U} , i = 1, \ldots, n \\
&  \| \B\zeta \|_{1} \leq {\mathcal M}^{\zeta}_{\ell} \\
&  \| \B\zeta - \B\zeta_{0} \|_{1} \leq {\mathcal L}^{\zeta}_{\ell, \text{loc}}\\
&  \| \B\beta - \B\beta_{0} \|_{1} \leq {\mathcal L}^{\beta}_{\ell, \text{loc}}.
\end{myarray} 
\end{equation}
For large values of  ${\mathcal L}^{\zeta}_{\ell, \text{loc}}$ (respectively, ${\mathcal L}^{\beta}_{\ell, \text{loc}}$) the constraints on $\M{X}\B\beta$ (respectively, $\B\beta$)
become ineffective and one gets back formulation~\eqref{eq-card-form2-1}. To see the impact of these additional cutting planes in the MIO formulation, we 
consider a few examples as illustrated in Figures~\ref{fig-leuk-data-bounding1},\ref{fig-synth-data-bounding1},\ref{fig-synth-data-bounding-p1k}.

\paragraph{Interpretation of the bounding boxes}
A local bounding box in the variable $\B\zeta=\M{X}\B\beta$ directs the MIO solver to seek for candidate solutions that deliver models with 
predictive accuracy ``similar'' (controlled by the radius of the ball)  to a reference predictive model, given by $\B\zeta_{0}$. 
In our experiments, we typically chose $\B\zeta_{0}$ as the solution delivered by running MIO (warm-started with a first order solution) for a few hundred to a few thousand seconds. 
More generally, $\B\zeta_{0}$ may be selected by any other sparse learning method. In our experiments, we found that 
the run-time behavior of the MIO depends upon how correlated the columns of $\M{X}$ are --- more correlation leading to longer run-times.

Similarly, a bounding box around $\B\beta$ directs the MIO to look for solutions in the neighborhood of a reference point $\B\beta_{0}$.
In our experiments, we chose the reference $\B\beta_{0}$ as the solution obtained by MIO (warm-started with a first order solution)  
and allowing it to run for a few hundred to a few thousand seconds. We observed that the MIO solver in presence of bounding boxes 
in the $\B\beta$-space certified optimality and in the process finding better solutions; much faster  than the $\B\zeta$-bounding box method.

Note that the $\B\beta$-bounding box constraint leads to $O(p)$ and the $\B\zeta$-box leads to 
$O(n)$ constraints. Thus, when $p\gg n$ the additional $\B\zeta$ constraints add a fewer number of extra variables when compared 
to the $\B\beta$ constraints.

\paragraph{Experiments} In the first set of experiments, we  consider the Leukemia dataset with $n = 72, p = 1000$.
We took two different values of $\mmk \in \{ 5, 10 \}$ and for each case we 
ran Algorithm~2 with several random restarts. The best solution thus obtained was used
to warm start the MIO formulation~\eqref{eq-card-form2-1}, which we ran for an additional 3600 seconds. The solution thus obtained is denoted by 
$\B\beta_{0}$. We then consider formulation~\eqref{eq-card-form2-1-xbet} with ${\mathcal L}^{\zeta}_{\ell, \text{loc}} = \infty$ and different
values of ${\mathcal L}^{\beta}_{\ell, \text{loc}} = \text{Frac}$ (as annotated in Figure~\ref{fig-leuk-data-bounding1}) --- the results are displayed in Figure~\ref{fig-leuk-data-bounding1}.

We consider another set of experiments to demonstrate the performance of the MIO in certifying global optimality for different synthetic datasets
with varying $n,p,\mmk$ as well as with different structures on the bounding box.
In the first case, we generated data as per Example~1 with $\rho= 0.9$, $k_{0} = 5$.
We consider the case with $\B\zeta_{0} = \M{X}\B\beta_{0}$, 
${\mathcal L}^{\beta}_{\ell, \text{loc}} = \infty$ and ${\mathcal L}^{\zeta}_{\ell, \text{loc}} = 0.5\| \M{X} \B\beta_{0} \|_{1}$,
where $\B\beta_{0}$ is a $\mmk$-sparse solution obtained from the
 MIO formulation~\eqref{eq-card-form2-1} run with a time limit of 1000 seconds, after being warm-started with Algorithm~2.
 The results are displayed in Figure~\ref{fig-synth-data-bounding1}[Left Panel].
In the second case (with data same as before) we obtained $\B\beta_{0}$ in the same fashion as described before---we took a bounding box around $\B\beta_{0}$, and left  the box constraint around $\M{X}\B\beta_{0}$ inactive, i.e., we set
  ${\mathcal L}^{\zeta}_{\ell, \text{loc}} = \infty $  and ${\mathcal L}^{\beta}_{\ell, \text{loc}} =\| \B\beta_{0}\|_{1}/\mmk$. 
  We performed two sets of experiments, where the data were generated 
  based on different SNR values---the results are displayed in Figure~\ref{fig-synth-data-bounding1} with SNR=1 [Middle Panel] and 
  SNR = 3[Right Panel].

In the same vein,  we have Figure~\ref{fig-synth-data-bounding-p1k} studying the effect of formulations~\eqref{eq-card-form2-1-xbet} for  
 synthetic datasets generated as per Example 1 with $n=50, p = 1000, \rho = 0.9$ and $k_{0} = 5$.

\subsubsection{Statistical Performance}

%%%%%%%%%%%%%%%%%%%%%
%% n << p synthetic examples
%%%%%%%%%%%%%%%%%%%%%

To understand the statistical behavior of MIO when compared to other approaches for learning sparse models, 
we considered synthetic datasets for values of $n$ ranging from $30-50$ and values of $p$ ranging from $1000-2000$.
%In the underdetermined case, we considered synthetic datasets for values of $n$ ranging from $30-50$ and values of $p$ ranging from $1000-2000$. 
%We performed the MIO best subset algorithm on our datasets, as well as {Lasso}, {Sparsenet}, and Algorithm~2 alone. 
The following methods were used for comparison purposes
%\begin{enumerate}
%\item[{\bf (a)}] Algorithm~2. Here we used fifty different random initializations around $\M{0}$, of the form 
%$\min(i-1,1)N(\M{0}_{p \times 1}, 4\M{I}), i = 1, \ldots, 50$
%and took the solution corresponding to the best objective value.
%\item[{\bf (b)}] The MIO approach with warm starts from part (a). 
%\item[{\bf (c)}] The {\texttt {Lasso}} solution.
%\item[{\bf (d)}] The {\texttt {Sparsenet}} solution.
%\end{enumerate}
{\bf (a)} Algorithm~2. Here we used fifty different random initializations around $\M{0}$, of the form 
$\min(i-1,1)N(\M{0}_{p \times 1}, 4\M{I}), i = 1, \ldots, 50$
and took the solution corresponding to the best objective value;
{\bf (b)} The MIO approach with warm starts from part (a); 
{\bf (c)} The {\texttt {Lasso}} solution and 
{\bf (d)} The {\texttt {Sparsenet}} solution.

For methods (a), (b) we considered ten equi-spaced values of  $\mmk$ in the range $[3, 2 k_0]$ (including the optimal value of $k_0$).
For each of the methods, the best model was selected in the same fashion as described in Section~\ref{stat-perf-n>p} using separate validation sets.

In addition, for some examples, we also study the performance of the \emph{debiased} version of the {\texttt{Lasso}}, as described in Section~\ref{stat-perf-n>p}.

%For method (c) we considered eighty different values of $\lambda$ of the form $\lambda_{\max}0.95^{i}, i = 1, \ldots, 80$.
%We obtained the ``optimal'' tuning parameter for each of the above cases by using a separate validation set. Once the model was selected, we again obtained  the  
% prediction error on a held out test set:
%$$\text{Prediction Error} = \| \M{X} \widehat{\B\beta} - \M{X} \B{\beta}^0\|_2^2 /\|\M{X} \B{\beta}^0\|_2^2.$$
%The results were averaged over ten random instances.

\begin{figure}[!h]
\centering
\scalebox{.9}[.5]{\begin{tabular}{cc}
\includegraphics[height = .35\textheight,width = 0.56\textwidth, , trim = 0mm 7mm 1mm 5mm, clip]{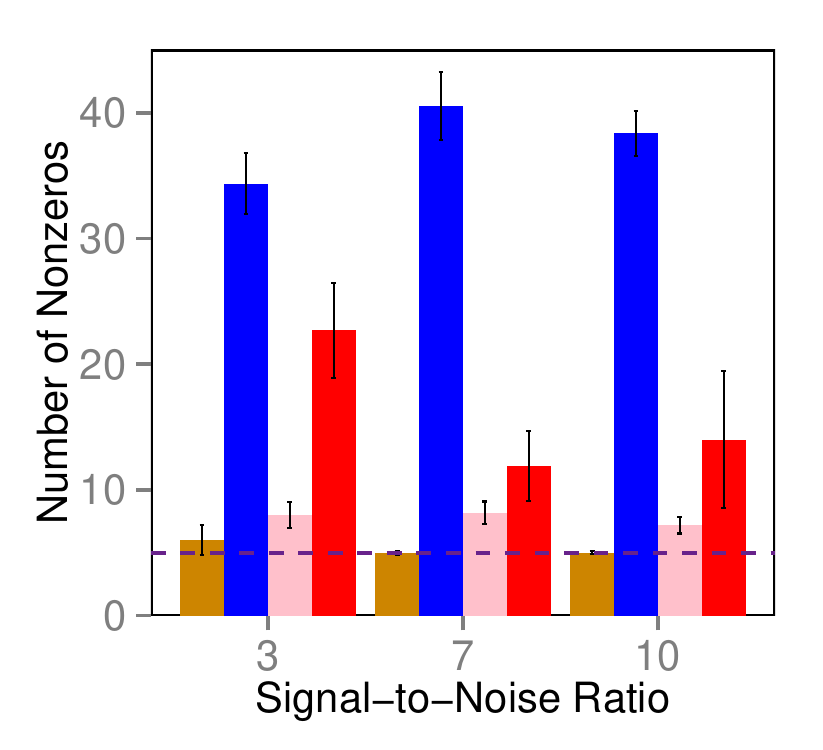}&
\includegraphics[height = .35\textheight, width = 0.45\textwidth, trim = 15mm 7mm 3mm 5mm, clip]{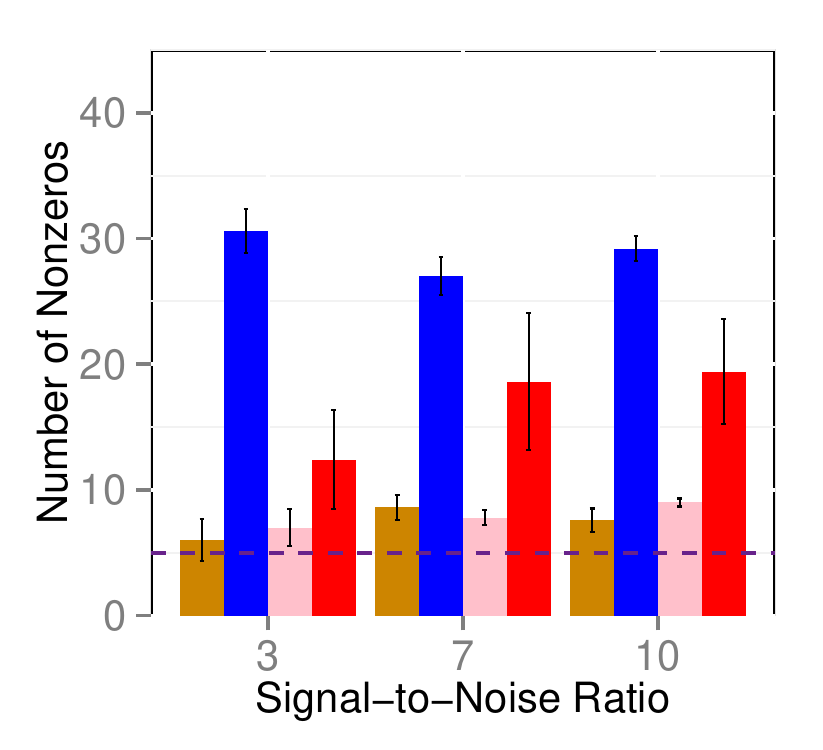}\\
\includegraphics[height = .35\textheight, width = 0.56\textwidth, trim = 2mm 0mm 1mm 5mm, clip]{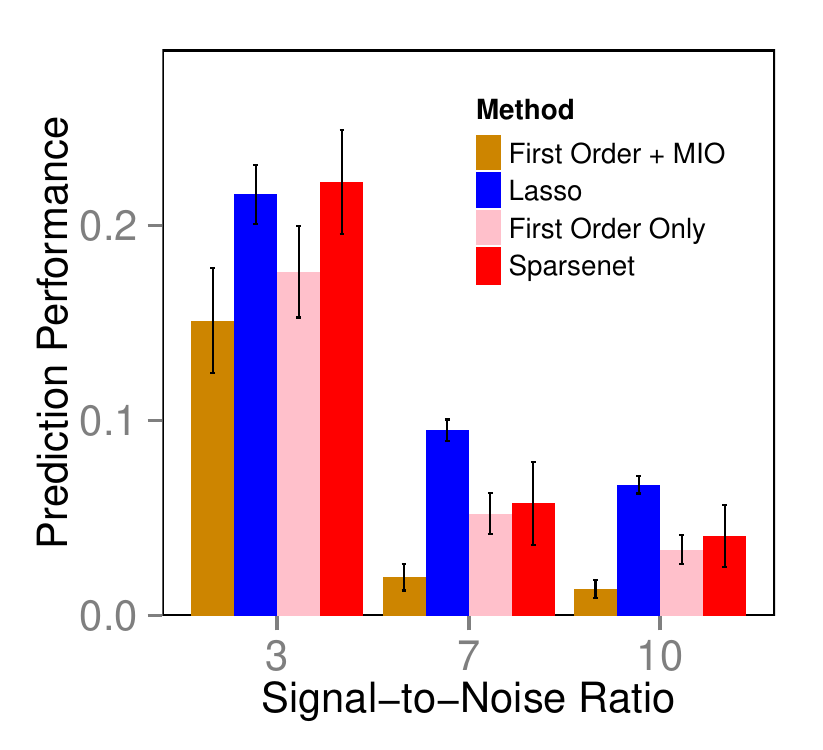}&
\includegraphics[height = .35\textheight,width = 0.45\textwidth, ,trim = 16mm 0mm 3mm 5mm, clip]{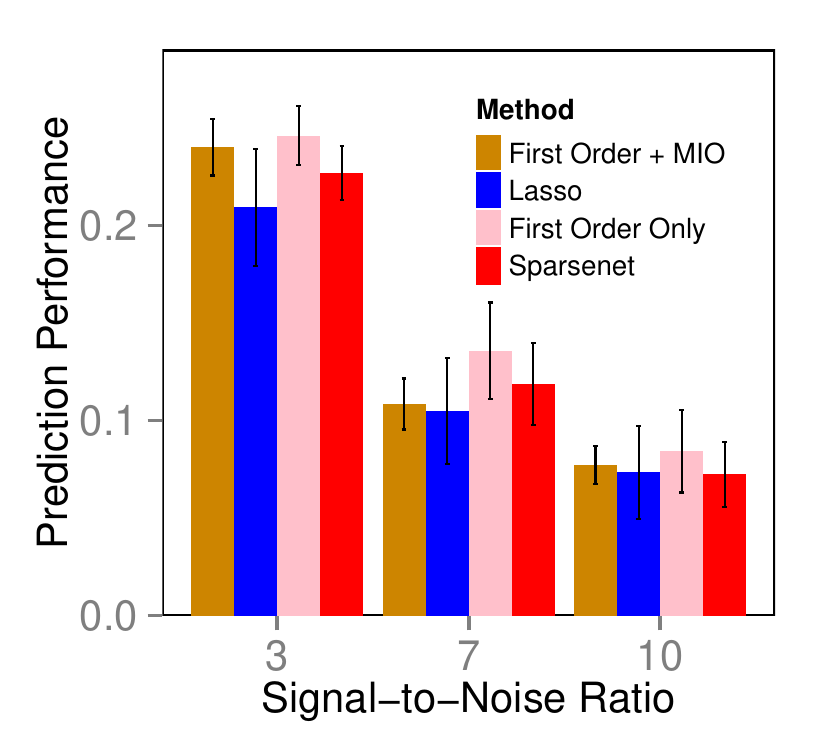}
\end{tabular}}
\caption{ {\small {The sparsity and predictive performance for different procedures:
[Left Panel]  shows Example~1 with $n = 50, p = 1000, \rho = 0.8, k_0 = 5$ and [Right Panel] shows Example~2 with $n = 30, p = 1000$---for each instance
several SNR values have been shown. } }}
\label{fig-n<p1}
\end{figure}

\begin{figure}[!h]
\centering
\scalebox{.9}[.5]{\begin{tabular}{cc}
\includegraphics[height = .35\textheight,width = 0.52\textwidth, , trim = 0mm 7mm 3mm 3mm, clip]{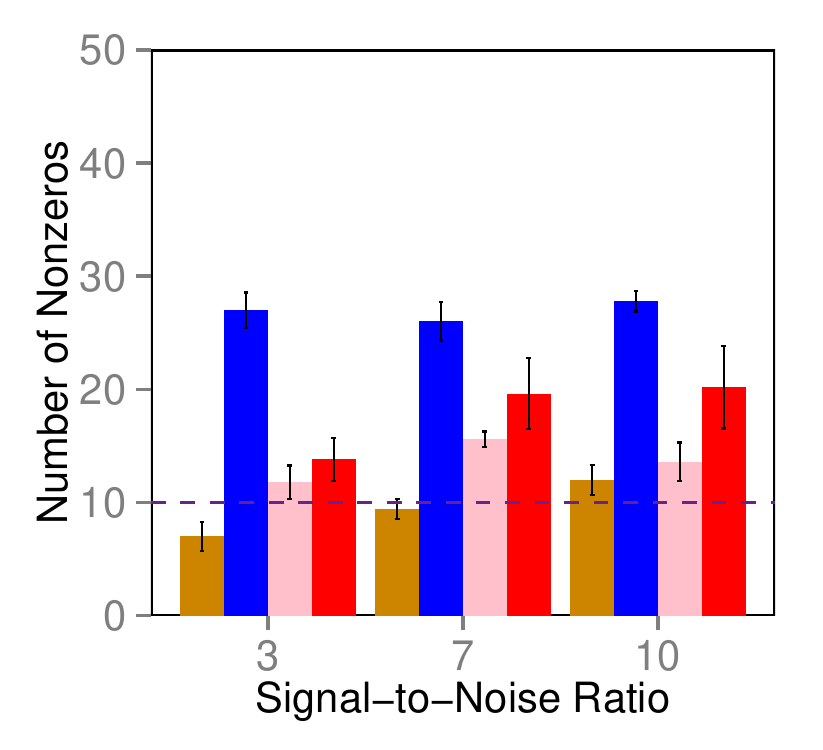}&
\includegraphics[height = .35\textheight, width = 0.43\textwidth, trim = 15mm 7mm 3mm 3mm, clip]{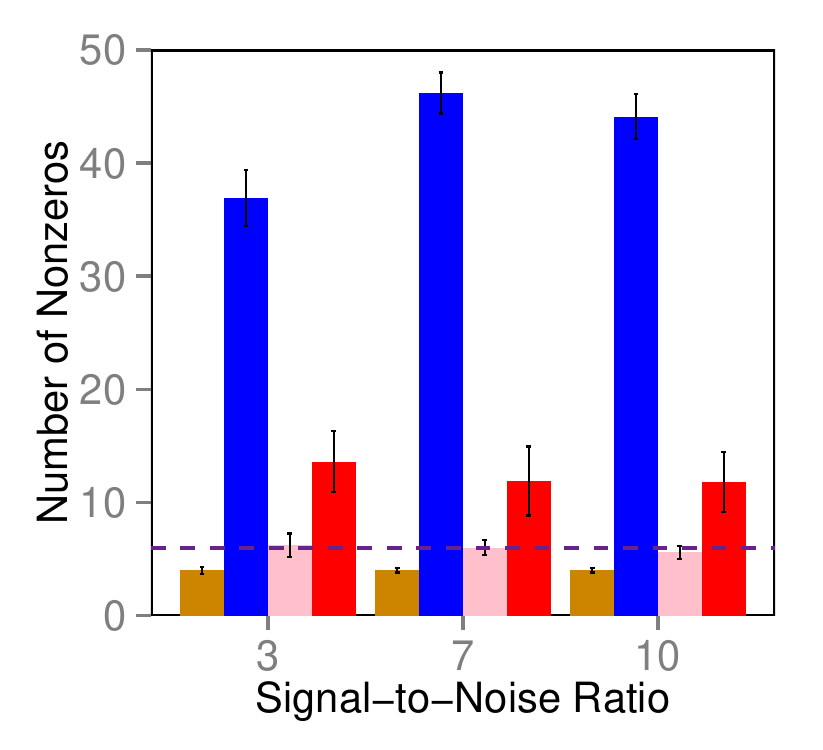}\\

\includegraphics[height = .35\textheight, width = 0.52\textwidth, trim = 2mm 0mm 3mm 3mm, clip]{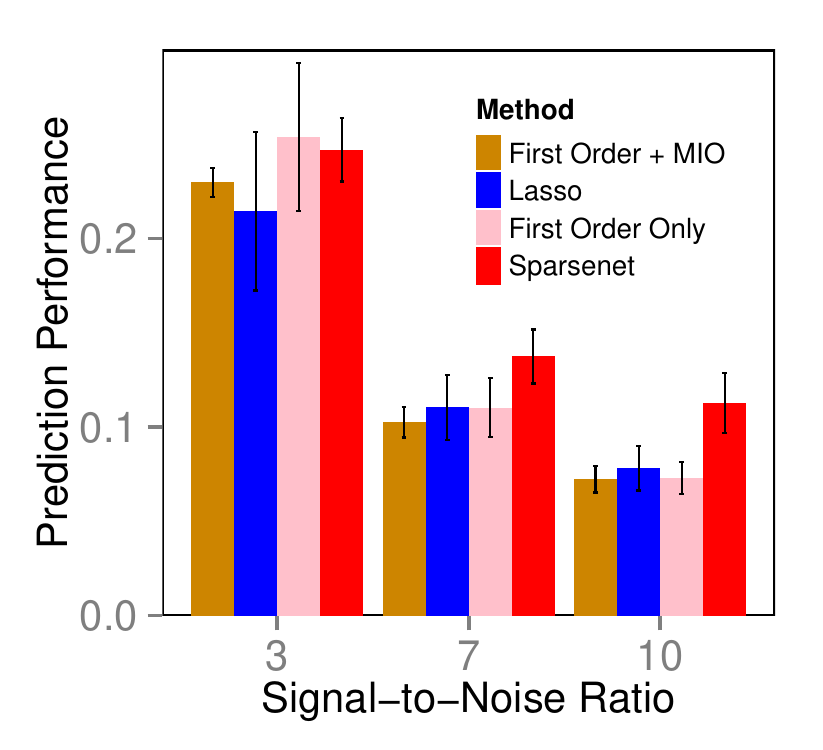}&
\includegraphics[height = .35\textheight,width = 0.43\textwidth, ,trim = 16mm 0mm 3mm 3mm, clip]{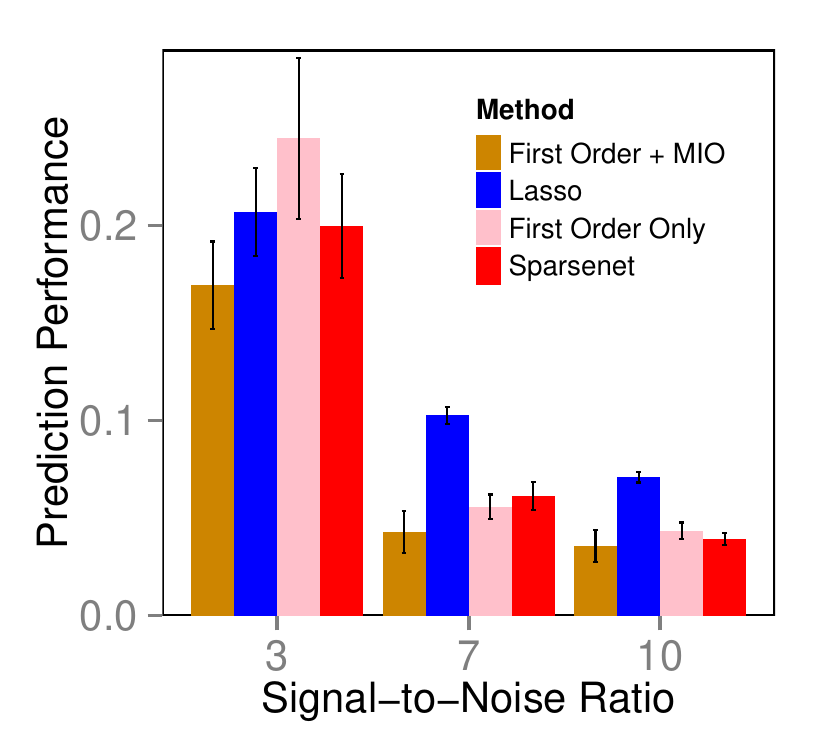}
\end{tabular}}
\caption{{ \small { [Left Panel] Shows performance for data generated according to Example 3 with $n = 30, p = 1000$ 
and [Right Panel]  shows Example~4 with $n = 50, p = 2000$.   } }  }
\label{fig-n<p2}
\end{figure}

In Figure~\ref{fig-n<p1} and Figure~\ref{fig-n<p2} we present selected  representative results from four different examples described in Section~\ref{sec:expt-data-desc}.

%\paragraph{Example 1}
%We took $\sigma_{ij} = \rho^{| i - j |}$ for $i, j \in \{1, \ldots, p \} \times \{ 1, \ldots, p\}$. 
%
%Here, $k_0 = 5$ and $\beta^0_{i} = 1$ for $i \in \{ \kappa_{1}, \ldots, \kappa_{5}\}$ for five equi-spaced
%\footnote{In the case where exactly equi-spaced values are not possible we rounded the indices to the nearest large integer value} values in the interval 
%$\{ 1, 2 , \ldots, p\}$. 
%
%Different values of $n,p,\rho$ and $\text{SNR}$ were considered.
%
%\paragraph{Example 2}
%In this case, we took $\B\Sigma = \M{I}_{p \times p}$ i.e., the entries $\M{X}_{n \times p}$ formed an iid standard Gaussian ensemble.
% Here, $k_0 = 5$ and $\B\beta^{0} = (\M{1}'_{ 5\times 1}, \M{0}'_{ p - 5 \times 1})' \in  \mathbb{R}^{p}$
%
%%[ {\texttt iidGN case 2} ]
%
%\paragraph{Example 3}
%In this case, we took $\B\Sigma = \M{I}_{p \times p}$.
%
%Here, $k_0 = 10$ and  $\beta_{i}^{0} = \frac12 + (10 - \frac12)\frac{(i-1)}{k_0}, i = 1, \ldots, 10$ and $\beta^0_i= 0, \forall i > 10$ --- i.e., a vector with ten nonzero entries, with the nonzero values being 
%equally spaced in the interval $[\frac12, 10]$. 
%
%%[ {\texttt iidGN case 3} ]
%
%
%%beta0=0*randn(pp,1); beta0(1:kpop) = linspace(10,.5,kpop); 
%%kpop = 10;
%
%
%\paragraph{Example 4}
%In this case, we took $\B\Sigma = \M{I}_{p \times p}$.
%Here, $k_0 = 6$ and $\B\beta^{0} = (-10,    -6 ,   -2,     2,     6,    10, \M{0}_{p - 6})$, i.e., a vector with size nonzero entries, equally spaced in the interval $[-10,10]$. 
%
%[ {\texttt iidGN case 4} ]

%% trim is LBRT

%% trim is LBRT

In Figure~\ref{fig-n<p1} the left panel shows the performance of different methods for Example~1 with $n = 50, p = 1000, \rho = 0.8,k_0=5$.
In this example, there are five non-zero coefficients: the features corresponding to the non-zero coefficients are weakly correlated and
a feature having a non-zero coefficient is highly correlated with a  feature having a zero coefficient. In this situation, the {\texttt {Lasso}} selects a very dense model
since it fails to distinguish between a zero and a non-zero coefficient when the variables are correlated---it brings both the coefficients in the model (with shrinkage).
MIO (with warm-start) performs the best---both in terms of predictive accuracy and in selecting a sparse set of coefficients. MIO obtains the sparsest model among the 
four methods and seems to find better solutions in terms of statistical properties than the models obtained by the first order methods alone. 
Interestingly, the ``optimal model'' selected by the first order methods is more dense than that selected by the MIO.  
The number of non-zero coefficients selected by MIO  remains fairly stable across different SNR values, unlike the other three methods.
For this example, we also experimented with the different versions of debiased {\texttt{Lasso}}.
In summary: the best debiased {\texttt{Lasso}} models had performance marginally better than {\texttt{Lasso}} but quite inferior to MIO. See the results in Appendix, Section~\ref{sec:debiased-lasso} for further details.

In Figure~\ref{fig-n<p1} the right panel shows Example~2, with $n=30,p=1000,k_0=5$ and all non-zero coefficients equal one.  In this example, all the methods
perform similarly in terms of predictive accuracy. This is because all non-zero coefficients in $\B\beta^0$ have the same value.
In fact for the smallest value of SNR, the {\texttt {Lasso}} achieves the best predictive model.
In all the cases however, the MIO achieves the sparsest model with favorable predictive accuracy.

In Figure~\ref{fig-n<p2}, for both the examples, the model matrix is an iid Gaussian ensemble. The underlying regression coefficient $\B\beta^0$ however, is 
structurally different than Example~2 (as in Figure~\ref{fig-n<p1}, right-panel). The structure in $\B\beta^0$ is responsible for different statistical behaviors of the four methods
across Figures~\ref{fig-n<p1} (right-panel) and Figure~\ref{fig-n<p2} (both panels). The alternating signs and varying amplitudes of $\B\beta^0$ are responsible for the poor behavior of 
{\texttt {Lasso}}. The MIO (with warm-starts) seems to be the best among all the methods. For Example~3 (Figure~\ref{fig-n<p2}, left panel) the 
predictive performances of  {\texttt {Lasso}}  and  MIO are comparable---the MIO however delivers much sparser models than the  {\texttt {Lasso}}.

The key conclusions are as follows:
\begin{enumerate}
\item The MIO best subset  algorithm has a significant edge in detecting the correct sparsity structure for all examples compared to {\texttt {Lasso}}, {\texttt {Sparsenet}} and the 
stand-alone discrete first order method.

\item For data generated as per Example~1 with large values of $\rho$, the MIO best subset  algorithm gives better predictive performance compared to its competitors.

\item For data generated as per Examples~2 and 3, MIO delivers similar predictive models  like the {\texttt {Lasso}}, but produces much sparser models. In fact, {\texttt {Lasso}} seems to perform marginally better than MIO, as a predictive model for small values of SNR.

\item For Example~4, MIO performs the best both in terms of predictive accuracy and  delivering sparse models.

\end{enumerate}

%\paragraph{Results on the Leukemia dataset}
%We applied the approaches in Items~{\bf (a)}, {\bf (b)} and {\bf (c)} to the semi-synthetic Leukemia data-set.
%The best {\texttt {Lasso}} based model delivered a model with prediction error $0.00458$ (s.e. 0.0067) and model-size $27$.
%The first order method (Algorithm~2) delivered a solution with prediction error $0.0084$ (s.e. 0.0016) and model-size   5.
%The MIO solution, when warm-started with the solution obtained from Algorithm~2 delivered a model with 
%prediction error $0.0663$ (s.e. 0.0141) and model-size  3. The best-subset selection procedure gets a sparser model possibly because of high correlations 
%in the model matrix. 

%The properties of the estimator (after doing CV etc)
%            pred-accuracy | stand-err | nnz 
%{Lasso}_soln =[ 0.0458    0.0067   27.0000 ]
%FO_soln : [0.0084    0.0016    5.0000]
%BS_soln = [0.0663    0.0141    3.0000 ]

%%%%%%%%%%%%%%%%%%%%%
%% LAD case
%%%%%%%%%%%%%%%%%%%%%

\section{Computational Results for Subset Selection with  Least Absolute Deviation  Loss}\label{sec:cLAD}

In this section, we demonstrate how our method can be used for the best subset selection problem with LAD objective~\eqref{lad-l0}.

Since the main focus of this paper is the least squares loss function, we consider  only a few representative examples for the LAD case.
The LAD loss is appropriate when the  error follows a heavy tailed distribution.
The datasets used for the experiments parallel those described in Section~\ref{sec:expt-data-desc}, the difference being in the distribution of 
$\epsilon$. We took $\epsilon_{i}$ iid from a double exponential distribution with variance $\sigma^2$. 
The value of $\sigma^2$ was adjusted to get different values of SNR.

\paragraph{Datasets analysed} We consider a set-up similar to Example~1 (Section~\ref{sec:expt-data-desc}) with $k_{0} = 5$ and $\rho = 0.9$.
Different choices of $(n,p)$ were taken to cover both the overdetermined ($n = 500, p = 100$) 
and high-dimensional   cases  ($n = 50, p = 1000$ and $n = 500, p = 1000$).

The other competing methods used for comparison were (a) discrete first order method~(Section~\eqref{sec:appli-ls-bs-2}) (b) MIO warm-started with the first order solutions and 
(c) the LAD loss with $\ell_{1}$ regularization:
\begin{equation*}\label{lad-L1}
\min \;\; \| \M{y} - \M{X} \B\beta \|_{1}  + \lambda \| \B\beta \|_{1}, 
\end{equation*}
which we denote by {\texttt {LAD-{Lasso}}}.
The training, validation and testing were done in the same fashion as in the least squares case. For each method, we report the 
number of non-zeros in the optimal model and associated prediction accuracy~\eqref{def-pred-error-1}.

\begin{figure}[!h]
\centering
\scalebox{.95}[.55]{\begin{tabular}{ccc}
\includegraphics[height = .35\textheight,width = 0.45\textwidth, ,trim = 0mm 0mm 0mm 0mm, clip]{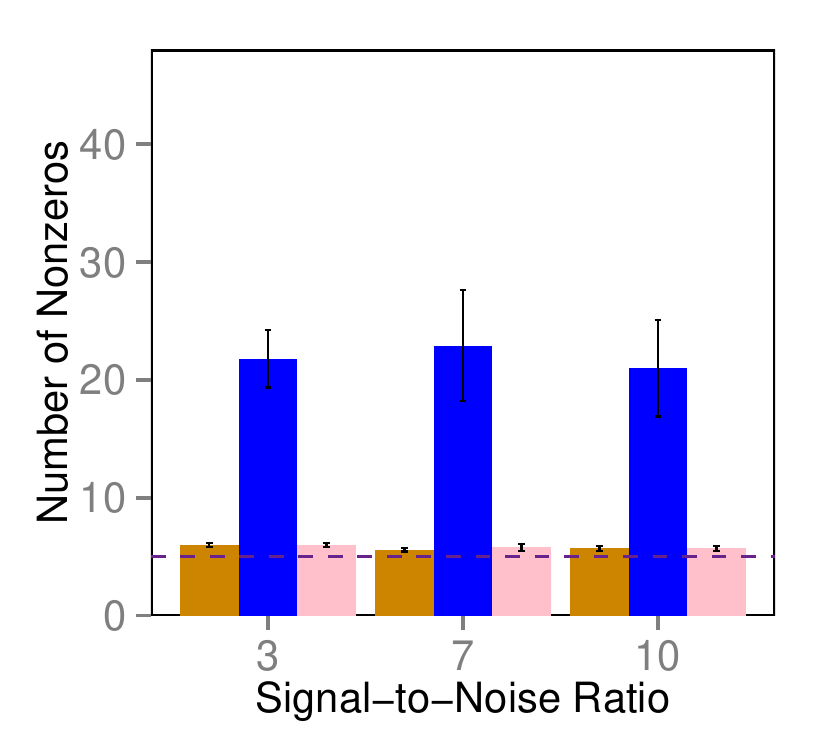}&&
\includegraphics[height = .35\textheight, width = 0.45\textwidth, trim = 0mm 0mm 0mm 0mm, clip]{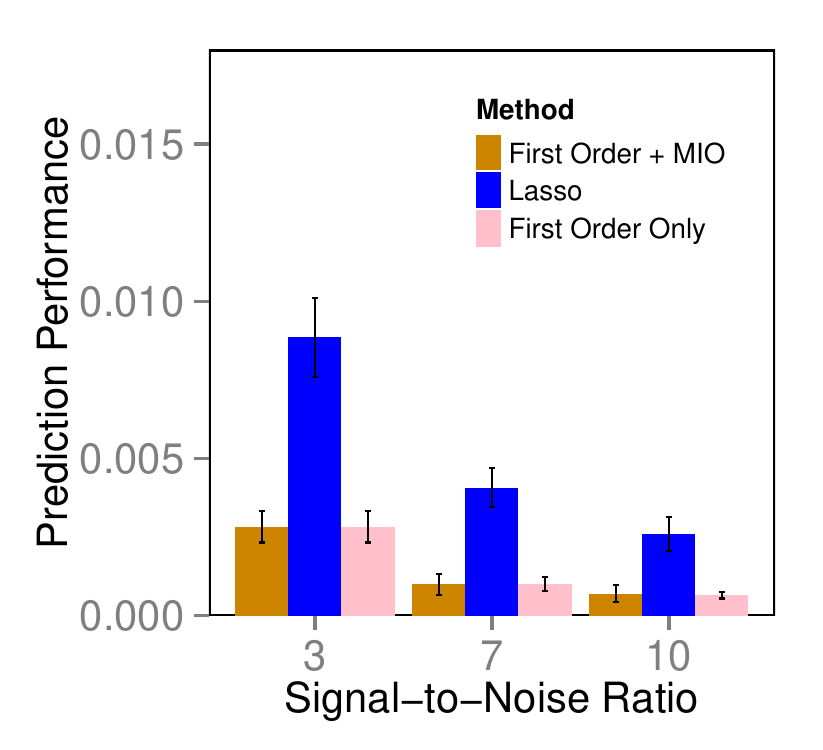}
\end{tabular}}
\caption{{ \small {The sparsity and predictive performance for different procedures for $n = 500, p = 100$ for  Problem~\eqref{lad-l0}.
The data is generated as per Example~1 with $\rho= 0.9, k_0=5$ and double exponential errors---further details are available in the text. 
The acronym ``{Lasso}''  refers to {\texttt LAD-{Lasso}}~\eqref{lad-L1}. The MIO is seen to deliver sparser models with better predictive accuracy when compared to the {\texttt LAD-{Lasso}}. } } }
\label{fig-LAD1}
\end{figure}

\begin{figure}[thb]
\centering
\scalebox{.94}[.4]{\begin{tabular}{ccc}
\includegraphics[height = .35\textheight, width = 0.45\textwidth, trim =2.5mm 8mm 1mm 3mm, clip]{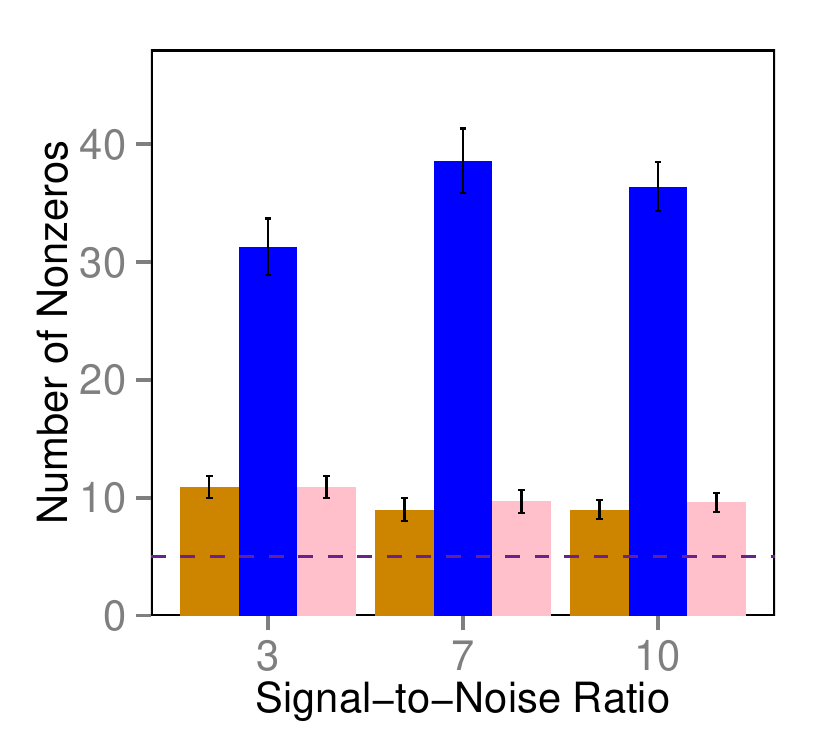}&&
\includegraphics[height = .35\textheight,width = 0.45\textwidth, trim = 7mm 8mm 3mm 3mm, clip]{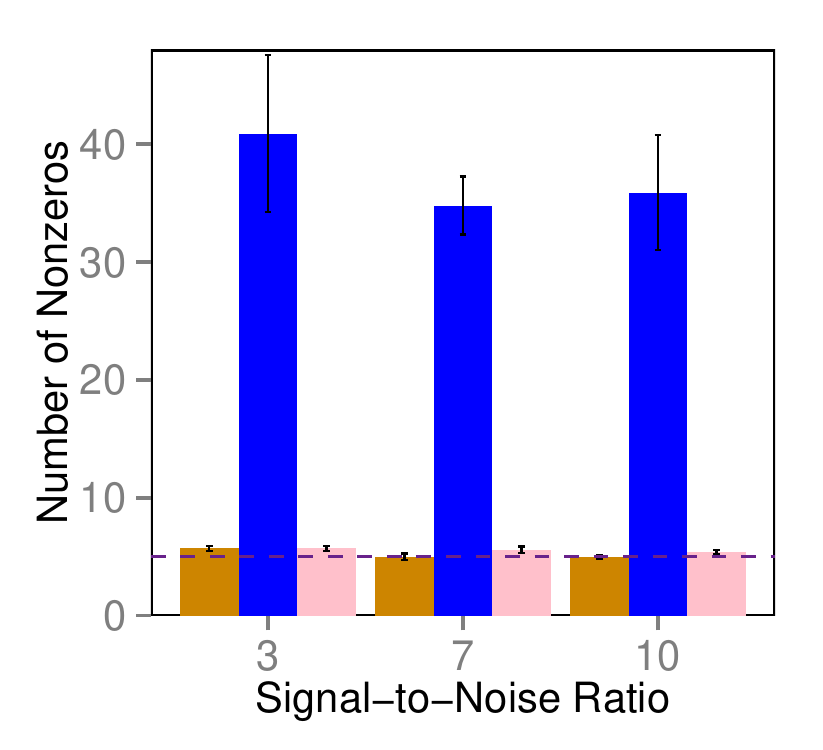}\\\\
\includegraphics[height = .35\textheight,width = 0.45\textwidth,  trim = 3.5mm 2mm 1mm 3mm, clip]{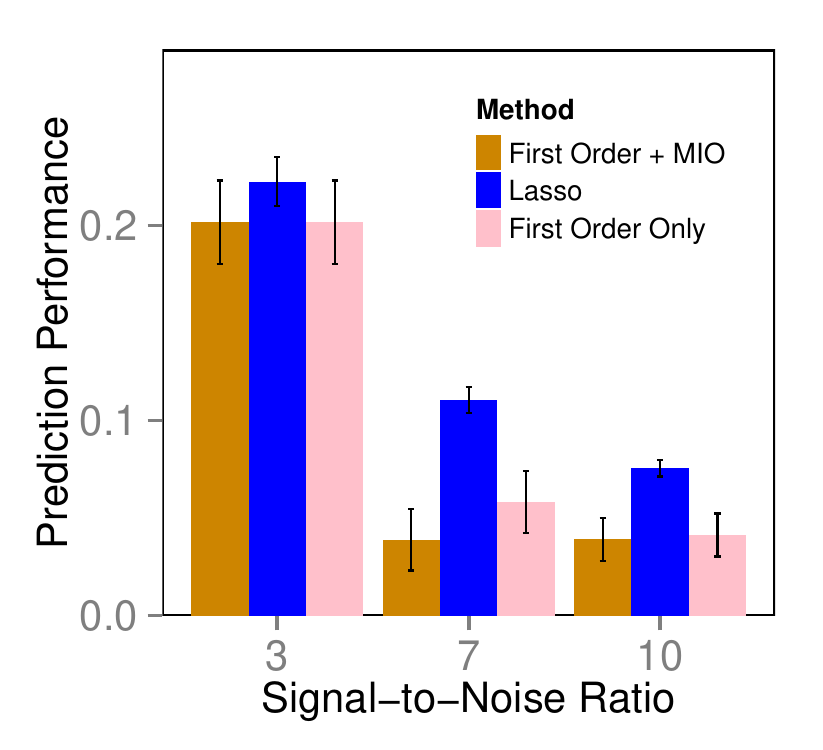}&&
\includegraphics[height = .35\textheight,width = 0.45\textwidth, trim = 7mm 2mm 3mm 3mm, clip]{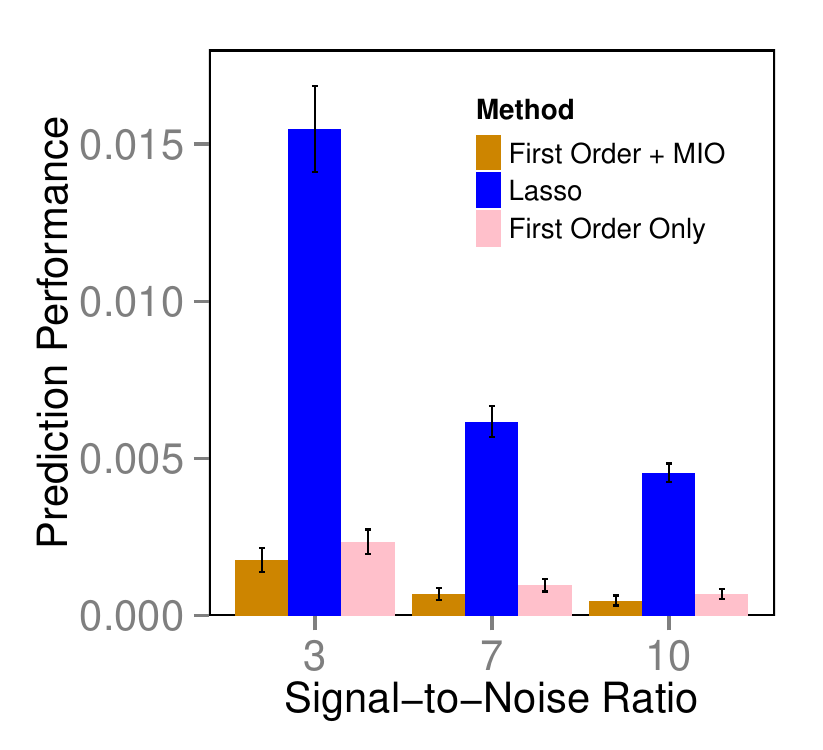}
\end{tabular}}
\caption{{ \small {Figure showing the number of nonzero values and predictive performance for different values of $n$ and $p$ for Problem~\eqref{lad-l0} (as in Figure~\ref{fig-LAD1}).
[Left panel] has $n = 50, p = 1000$
 and [Right panel] has $n = 500, p = 1000$.} } }
\label{fig-LAD2}
\end{figure}

Figure~\ref{fig-LAD1} compares the MIO approach with others for LAD in the overdetermined case ($n > p$).
Figure~\ref{fig-LAD2} does the same for the high-dimensional case ($p\gg  n $).
The conclusions parallel those for the least squares case. Since, in the example considered, the features corresponding to the non-zero coefficients are weakly correlated and
a feature having a non-zero coefficient is highly correlated with a  feature having a zero coefficient---the {\texttt {LAD-{Lasso}} } selects an overly dense model and misses out in terms of prediction error.
Both the MIO (with warm-starts) and the discrete first order methods behave similarly---much better than $\ell_{1}$ regularization schemes.
As expected, we observed that subset selection with least squares loss leads to inferior models for these examples, due to a heavy-tailed distribution of the errors. 

The results in this section are similar to the least squares case. The MIO approach provides an edge both in terms of sparsity and predictive accuracy compared to {Lasso} both for the 
overdetermined and the high-dimensional   case.  

\section{Conclusions}\label{sec:conc}

In this paper, we have revisited the classical   best subset selection problem of choosing $\mmk$  out of $p$ features in linear regression given $n$ observations
using  a modern optimization lens, i.e., 
MIO and a discrete extension of first order methods from continuous optimization. Exploiting the astonishing progress of MIO solvers in the last twenty-five  years,
we have shown that this approach   solves problems with $n$ in the 1000s and $p$ in the 100s in minutes to provable optimality,
 and finds near optimal   solutions  for $n$ in the 100s and $p$ in the 1000s in minutes. Importantly, the solutions provided by the MIO approach significantly outperform
 other state of the art methods like  {\texttt {Lasso}} in achieving  sparse models with good predictive power.  
 Unlike all other methods, the MIO approach always provides a guarantee on its sub-optimality even if the algorithm is terminated early. 
 Moreover,  it  can 
 accommodate side constraints on  the coefficients of the  linear regression and also 
 extends   to finding best subset solutions for the least absolute deviation loss function.
 
 While continuous optimization methods have played and continue to play an important role in statistics over the years, discrete optimization methods 
 have not. The evidence in this paper as well as in \cite{bertsimas2014least} suggests that MIO methods are tractable and lead to desirable properties 
 (improved accuracy and  sparsity among others) at the expense of higher, but still reasonable, computational times.

 \section*{Acknowledgements}
 We would like to thank the Associate editor and two reviewers for their comments that helped us improve the paper. 
 A major part of the work was performed when R.M. was at Columbia University.
% R.M. would like to thank Trevor Hastie, Arian Maleki, Robert Tibshirani and Cun-Hui Zhang for helpful discussions.
%  
%\clearpage

%%%%%%%%%%%%%%%%%%%%%
%% Appendix
%%%%%%%%%%%%%%%%%%%%%

%%%%%%%%%%%%%%%%%%%%%
%% N >> P
%%%%%%%%%%%%%%%%%%%%%
%\input{SectionCompTestsn>p.tex}

%%%%%%%%%%%%%%%%%%%%%
%% N << P
%%%%%%%%%%%%%%%%%%%%%
\clearpage
%\subsection{The Underdetermined Case - Rahul}
%\input{SectionCompTestsn<p.tex}

%%%%%%%%%%%%%%%%%%%%%
%% References
%%%%%%%%%%%%%%%%%%%%%
%\newpage
%\nocite{*} %make all of the references show up

\bibliographystyle{plainnat_my} 

\bibliography{rahul_dbm3,sparsityPaper,agst_new}

\begin{thebibliography}{62}
\providecommand{\natexlab}[1]{#1}
\providecommand{\url}[1]{\texttt{#1}}
\expandafter\ifx\csname urlstyle\endcsname\relax
  \providecommand{\doi}[1]{doi: #1}\else
  \providecommand{\doi}{doi: \begingroup \urlstyle{rm}\Url}\fi

\bibitem[sup()]{supercomputer}
{Top500 Supercomputer Sites, Directory page for Top500 lists. Result for each
  list since June 1993}.
\newblock \url {http://www.top500.org/statistics/sublist/}.
\newblock Accessed: 2013-12-04.

\bibitem[Bertsimas and Mazumder(2014)]{bertsimas2014least}
D.~Bertsimas and R.~Mazumder.
\newblock Least quantile regression via modern optimization.
\newblock \emph{The Annals of Statistics}, 42\penalty0 (6):\penalty0
  2494--2525, 2014.

\bibitem[Bertsimas and Shioda(2009)]{bertsimas2009algorithm}
D.~Bertsimas and R.~Shioda.
\newblock Algorithm for cardinality-constrained quadratic optimization.
\newblock \emph{Computational Optimization and Applications}, 43\penalty0
  (1):\penalty0 1--22, 2009.

\bibitem[Bertsimas and Weismantel(2005)]{bertsimas2005optimization_new}
D.~Bertsimas and R.~Weismantel.
\newblock \emph{Optimization over integers}.
\newblock Dynamic Ideas Belmont, 2005.

\bibitem[Bickel et~al.(2009)Bickel, Ritov, and
  Tsybakov]{bickel2009simultaneous}
P.~Bickel, Y.~Ritov, and A.~Tsybakov.
\newblock Simultaneous analysis of lasso and dantzig selector.
\newblock \emph{The Annals of Statistics}, pages 1705--1732, 2009.

\bibitem[Bienstock(1996)]{bienstock1996computational}
D.~Bienstock.
\newblock Computational study of a family of mixed-integer quadratic
  programming problems.
\newblock \emph{Mathematical programming}, 74\penalty0 (2):\penalty0 121--140,
  1996.

\bibitem[Bixby(2012)]{bixby}
R.~E. Bixby.
\newblock A brief history of linear and mixed-integer programming computation.
\newblock \emph{Documenta Mathematica, Extra Volume: Optimization Stories},
  pages 107--121, 2012.

\bibitem[Blumensath and Davies(2008)]{blumensath2008iterative}
T.~Blumensath and M.~Davies.
\newblock Iterative thresholding for sparse approximations.
\newblock \emph{Journal of Fourier Analysis and Applications}, 14\penalty0
  (5-6):\penalty0 629--654, 2008.

\bibitem[Blumensath and Davies(2009)]{blumensath2009-acha}
T.~Blumensath and M.~Davies.
\newblock Iterative hard thresholding for compressed sensing.
\newblock \emph{Applied and Computational Harmonic Analysis}, 27\penalty0
  (3):\penalty0 265--274, 2009.

\bibitem[Boyd and Vandenberghe(2004)]{BV2004}
S.~Boyd and L.~Vandenberghe.
\newblock \emph{Convex Optimization}.
\newblock Cambridge University Press, Cambridge, 2004.

\bibitem[B{\"u}hlmann and {van-de-Geer}(2011)]{buhlmann2011statistics}
P.~B{\"u}hlmann and S.~{van-de-Geer}.
\newblock \emph{Statistics for high-dimensional data}.
\newblock Springer, 2011.

\bibitem[Bunea et~al.(2007)Bunea, Tsybakov, Wegkamp,
  et~al.]{bunea2007aggregation}
F.~Bunea, A.~B. Tsybakov, M.~H. Wegkamp, et~al.
\newblock Aggregation for gaussian regression.
\newblock \emph{The Annals of Statistics}, 35\penalty0 (4):\penalty0
  1674--1697, 2007.

\bibitem[Candes et~al.(2008)Candes, Wakin, and Boyd]{boyd08-new}
E.~Candes, M.~Wakin, and S.~Boyd.
\newblock Enhancing sparsity by reweighted $\ell_1$ minimization.
\newblock \emph{Journal of Fourier Analysis and Applications}, 14\penalty0
  (5):\penalty0 877--905, 2008.

\bibitem[Candes(2008)]{candes2008restricted}
E.~Candes.
\newblock The restricted isometry property and its implications for compressed
  sensing.
\newblock \emph{Comptes Rendus Mathematique}, 346\penalty0 (9):\penalty0
  589--592, 2008.

\bibitem[Cand{\`e}s and Plan(2009)]{candes2009near}
E.~Cand{\`e}s and Y.~Plan.
\newblock Near-ideal model selection by $\ell_{1}$ minimization.
\newblock \emph{The Annals of Statistics}, 37\penalty0 (5A):\penalty0
  2145--2177, 2009.

\bibitem[Candes and Tao(2006)]{candes2006near}
E.~Candes and T.~Tao.
\newblock Near-optimal signal recovery from random projections: Universal
  encoding strategies?
\newblock \emph{Information Theory, IEEE Transactions on}, 52\penalty0
  (12):\penalty0 5406--5425, 2006.

\bibitem[Chen et~al.(1998)Chen, Donoho, and Saunders]{CDS1998}
S.~Chen, D.~Donoho, and M.~Saunders.
\newblock Atomic decomposition by basis pursuit.
\newblock \emph{SIAM Journal on Scientific Computing}, 20\penalty0
  (1):\penalty0 33--61, 1998.

\bibitem[Dettling(2004)]{dettling2004bagboosting}
M.~Dettling.
\newblock Bagboosting for tumor classification with gene expression data.
\newblock \emph{Bioinformatics}, 20\penalty0 (18):\penalty0 3583--3593, 2004.

\bibitem[Donoho(2006)]{donoho2006}
D.~Donoho.
\newblock For most large underdetermined systems of equations, the minimal
  $\ell^1$-norm solution is the sparsest solution.
\newblock \emph{Communications on Pure and Applied Mathematics}, 59:\penalty0
  797--829, 2006.

\bibitem[Donoho and Johnstone(1993)]{Donoho93idealspatial}
D.~Donoho and I.~Johnstone.
\newblock Ideal spatial adaptation by wavelet shrinkage.
\newblock \emph{Biometrika}, 81:\penalty0 425--455, 1993.

\bibitem[Donoho and Elad(2003)]{donoho2003optimally}
D.~Donoho and M.~Elad.
\newblock Optimally sparse representation in general (nonorthogonal)
  dictionaries via $\ell_{1}$ minimization.
\newblock \emph{Proceedings of the National Academy of Sciences}, 100\penalty0
  (5):\penalty0 2197--2202, 2003.

\bibitem[Donoho and Huo(2001)]{donoho2001uncertainty}
D.~Donoho and X.~Huo.
\newblock Uncertainty principles and ideal atomic decomposition.
\newblock \emph{Information Theory, IEEE Transactions on}, 47\penalty0
  (7):\penalty0 2845--2862, 2001.

\bibitem[Efron et~al.(2004)Efron, Hastie, Johnstone, and Tibshirani]{LARS}
B.~Efron, T.~Hastie, I.~Johnstone, and R.~Tibshirani.
\newblock Least angle regression (with discussion).
\newblock \emph{Annals of Statistics}, 32\penalty0 (2):\penalty0 407--499,
  2004.
\newblock ISSN 0090-5364.

\bibitem[Fan and Li(2001)]{Fan01}
J.~Fan and R.~Li.
\newblock Variable selection via nonconcave penalized likelihood and its oracle
  properties.
\newblock \emph{Journal of the American Statistical Association}, 96\penalty0
  (456):\penalty0 1348--1360(13), 2001.

\bibitem[Fan and Lv(2011)]{fan2011nonconcave}
J.~Fan and J.~Lv.
\newblock Nonconcave penalized likelihood with {NP}-dimensionality.
\newblock \emph{Information Theory, IEEE Transactions on}, 57\penalty0
  (8):\penalty0 5467--5484, 2011.

\bibitem[Fan and Lv(2013)]{fan2013asymptotic}
Y.~Fan and J.~Lv.
\newblock Asymptotic equivalence of regularization methods in thresholded
  parameter space.
\newblock \emph{Journal of the American Statistical Association}, 108\penalty0
  (503):\penalty0 1044--1061, 2013.

\bibitem[Frank and Friedman(1993)]{FF93}
I.~Frank and J.~Friedman.
\newblock A statistical view of some chemometrics regression tools (with
  discussion).
\newblock \emph{Technometrics}, 35\penalty0 (2):\penalty0 109--148, 1993.

\bibitem[Friedman(2008)]{FJ08}
J.~Friedman.
\newblock Fast sparse regression and classification.
\newblock Technical report, Department of Statistics, Stanford University,
  2008.

\bibitem[Friedman et~al.(2007)Friedman, Hastie, Hoefling, and
  Tibshirani]{FHT2007}
J.~Friedman, T.~Hastie, H.~Hoefling, and R.~Tibshirani.
\newblock Pathwise coordinate optimization.
\newblock \emph{Annals of Applied Statistics}, 2\penalty0 (1):\penalty0
  302--332, 2007.

\bibitem[Furnival and Wilson(1974)]{FW74}
G.~Furnival and R.~Wilson.
\newblock Regression by leaps and bounds.
\newblock \emph{Technometrics}, 16:\penalty0 499--511, 1974.

\bibitem[Greenshtein(2006)]{greenshtein2006best}
E.~Greenshtein.
\newblock Best subset selection, persistence in high-dimensional statistical
  learning and optimization under $\ell_{1}$ constraint.
\newblock \emph{The Annals of Statistics}, 34\penalty0 (5):\penalty0
  2367--2386, 2006.

\bibitem[Greenshtein and Ritov(2004)]{greenshtein2004persistence}
E.~Greenshtein and Y.~Ritov.
\newblock Persistence in high-dimensional linear predictor selection and the
  virtue of overparametrization.
\newblock \emph{Bernoulli}, 10\penalty0 (6):\penalty0 971--988, 2004.

\bibitem[Gurobi~Optimization(2013)]{gurobi}
I.~Gurobi~Optimization.
\newblock Gurobi optimizer reference manual, 2013.
\newblock URL \url{http://www.gurobi.com}.

\bibitem[Hastie et~al.(2009)Hastie, Tibshirani, and Friedman]{FHT-09-new}
T.~Hastie, R.~Tibshirani, and J.~Friedman.
\newblock \emph{The Elements of Statistical Learning, Second Edition: Data
  Mining, Inference, and Prediction (Springer Series in Statistics)}.
\newblock Springer New York, 2 edition, 2009.
\newblock ISBN 0387848576.

\bibitem[Knight and Fu(2000)]{KF2000}
K.~Knight and W.~Fu.
\newblock Asymptotics for lasso-type estimators.
\newblock \emph{Annals of Statistics}, 28\penalty0 (5):\penalty0 1356--1378,
  2000.

\bibitem[Loh and Wainwright(2013)]{loh2013regularized}
P.-L. Loh and M.~Wainwright.
\newblock Regularized m-estimators with nonconvexity: Statistical and
  algorithmic theory for local optima.
\newblock In \emph{Advances in Neural Information Processing Systems}, pages
  476--484, 2013.

\bibitem[Lv and Fan(2009)]{lv2009unified}
J.~Lv and Y.~Fan.
\newblock A unified approach to model selection and sparse recovery using
  regularized least squares.
\newblock \emph{The Annals of Statistics}, pages 3498--3528, 2009.

\bibitem[Mazumder et~al.(2011)Mazumder, Friedman, and Hastie]{mhf-09-jasa}
R.~Mazumder, J.~Friedman, and T.~Hastie.
\newblock Sparsenet: Coordinate descent with non-convex penalties.
\newblock \emph{Journal of the American Statistical Association},
  117(495):\penalty0 1125--1138, 2011.

\bibitem[Meinshausen and B\"{u}hlmann(2006)]{MB2006}
N.~Meinshausen and P.~B\"{u}hlmann.
\newblock High-dimensional graphs and variable selection with the lasso.
\newblock \emph{Annals of Statistics}, 34:\penalty0 1436--1462, 2006.

\bibitem[Miller(2002)]{miller2002subset}
A.~Miller.
\newblock \emph{Subset selection in regression}.
\newblock CRC Press Washington, 2002.

\bibitem[Natarajan(1995)]{natarajan1995sparse}
B.~Natarajan.
\newblock Sparse approximate solutions to linear systems.
\newblock \emph{SIAM journal on computing}, 24\penalty0 (2):\penalty0 227--234,
  1995.

\bibitem[Nemhauser(2013)]{nem}
G.~Nemhauser.
\newblock Integer programming: the global impact.
\newblock Presented at EURO, INFORMS, Rome, Italy, 2013.
\newblock \url {http://euro2013.org/wp-content/uploads/Nemhauser_EuroXXVI.pdf}.
  Accessed: 2013-12-04.

\bibitem[Nesterov(2005)]{nest_05}
Y.~Nesterov.
\newblock Smooth minimization of non-smooth functions.
\newblock \emph{Mathematical Programming, Series A}, 103:\penalty0 127--152,
  2005.

\bibitem[Nesterov(2007)]{nest-07}
Y.~Nesterov.
\newblock Gradient methods for minimizing composite objective function.
\newblock Technical report, Center for Operations Research and Econometrics
  (CORE), Catholic University of Louvain, 2007.
\newblock Technical Report number 76.

\bibitem[Nesterov(2004)]{nesterov2004introductory}
Y.~Nesterov.
\newblock \emph{Introductory Lectures on Convex Optimization: A Basic Course}.
\newblock Kluwer, Norwell, 2004.

\bibitem[Raskutti et~al.(2011)Raskutti, Wainwright, and
  Yu]{raskutti2011minimax}
G.~Raskutti, M.~Wainwright, and B.~Yu.
\newblock Minimax rates of estimation for high-dimensional linear regression
  over-balls.
\newblock \emph{Information Theory, IEEE Transactions on}, 57\penalty0
  (10):\penalty0 6976--6994, 2011.

\bibitem[Rockafellar(1996)]{rock-conv-96}
R.~Rockafellar.
\newblock \emph{Convex Analysis}.
\newblock Princeton University Press, Princeton, 1996.
\newblock ISBN 0691015864.
\newblock URL
  \url{http://www.amazon.com/exec/obidos/redirect?tag=citeulike07-20\&path=ASIN/0691015864}.

\bibitem[Shen et~al.(2013)Shen, Pan, Zhu, and Zhou]{shen2013constrained}
X.~Shen, W.~Pan, Y.~Zhu, and H.~Zhou.
\newblock On constrained and regularized high-dimensional regression.
\newblock \emph{Annals of the Institute of Statistical Mathematics},
  65\penalty0 (5):\penalty0 807--832, 2013.

\bibitem[Tibshirani(1996)]{Ti96}
R.~Tibshirani.
\newblock Regression shrinkage and selection via the lasso.
\newblock \emph{Journal of the Royal Statistical Society, Series B},
  58:\penalty0 267--288, 1996.

\bibitem[Tibshirani(2011)]{tibshirani2011regression}
R.~Tibshirani.
\newblock Regression shrinkage and selection via the lasso: a retrospective.
\newblock \emph{Journal of the Royal Statistical Society: Series B (Statistical
  Methodology)}, 73\penalty0 (3):\penalty0 273--282, 2011.

\bibitem[Tropp(2006)]{tropp2006just}
J.~Tropp.
\newblock Just relax: Convex programming methods for identifying sparse signals
  in noise.
\newblock \emph{Information Theory, IEEE Transactions on}, 52\penalty0
  (3):\penalty0 1030--1051, 2006.

\bibitem[van~de Geer et~al.(2011)van~de Geer, B{\"u}hlmann, and
  Zhou]{van2011adaptive}
S.~Geer, P.~B{\"u}hlmann, and S.~Zhou.
\newblock The adaptive and the thresholded lasso for potentially misspecified
  models (and a lower bound for the lasso).
\newblock \emph{Electronic Journal of Statistics}, 5:\penalty0 688--749, 2011.

\bibitem[Wainwright(2009)]{wainwright2009sharp}
M.~Wainwright.
\newblock Sharp thresholds for high-dimensional and noisy sparsity recovery
  using-constrained quadratic programming (lasso).
\newblock \emph{Information Theory, IEEE Transactions on}, 55\penalty0
  (5):\penalty0 2183--2202, 2009.

\bibitem[Zhang(2010{\natexlab{a}})]{zhang2010nearly}
C.-H. Zhang.
\newblock Nearly unbiased variable selection under minimax concave penalty.
\newblock \emph{The Annals of Statistics}, 38\penalty0 (2):\penalty0 894--942,
  2010{\natexlab{a}}.

\bibitem[Zhang and Huang(2008)]{ZH08}
C.-H. Zhang and J.~Huang.
\newblock The sparsity and bias of the lasso selection in high-dimensional
  linear regression.
\newblock \emph{Annals of Statistics}, 36\penalty0 (4):\penalty0 1567--1594,
  2008.

\bibitem[Zhang and Zhang(2012)]{zhang2012general}
C.-H. Zhang and T.~Zhang.
\newblock A general theory of concave regularization for high-dimensional
  sparse estimation problems.
\newblock \emph{Statistical Science}, 27\penalty0 (4):\penalty0 576--593, 2012.

\bibitem[Zhang(2010{\natexlab{b}})]{tzhang-09N}
T.~Zhang.
\newblock Analysis of multi-stage convex relaxation for sparse regularization.
\newblock \emph{The Journal of Machine Learning Research}, 11:\penalty0
  1081--1107, 2010{\natexlab{b}}.

\bibitem[Zhang et~al.(2014)Zhang, Wainwright, and Jordan]{zhang2014lower}
Y.~Zhang, M.~Wainwright, and M.~I. Jordan.
\newblock Lower bounds on the performance of polynomial-time algorithms for
  sparse linear regression.
\newblock \emph{arXiv preprint arXiv:1402.1918}, 2014.

\bibitem[Zhao and Yu(2006)]{ZY2006}
P.~Zhao and B.~Yu.
\newblock On model selection consistency of lasso.
\newblock \emph{Journal of Machine Learning Research}, 7:\penalty0 2541--2563,
  2006.

\bibitem[Zheng et~al.(2014)Zheng, Fan, and Lv]{Lv-2014}
Z.~Zheng, Y.~Fan, and J.~Lv.
\newblock High dimensional thresholded regression and shrinkage effect.
\newblock \emph{Journal of the Royal Statistical Society: Series B (Statistical
  Methodology)}, 76\penalty0 (3):\penalty0 627--649, 2014.
\newblock ISSN 1467-9868.

\bibitem[Zou(2006)]{zou2006a}
H.~Zou.
\newblock The adaptive lasso and its oracle properties.
\newblock \emph{Journal of the American Statistical Association}, 101:\penalty0
  1418--1429, 2006.

\bibitem[Zou and Li(2008)]{zouli08}
H.~Zou and R.~Li.
\newblock One - step sparse estimates in nonconcave penalized likelihood
  problems.
\newblock \emph{The Annals of Statistics}, 36\penalty0 (4):\penalty0
  1509--1533, 2008.

\end{thebibliography}

%\bibliography{rahul_dbm2,mht_refs,sparsityPaper,new_agst_new}
%\bibliography{mht_refs,rahul_dbm,sparsityPaper,rahul_dbm2,new_agst_new}
%\bibliography{mht_refs,sparsityPaper,rahul_dbm,rahul_dbm2,new_agst_new}

%%%%\printbibliography

\clearpage

\begin{appendix}

\section*{Appendix and Supplementary Material}

\section{Additional  Details for Section~2}\label{sec:intro-append-1}

%\subsection{Speed Improvements in Computer Hardware}
%See Figure~\ref{super}, presenting the speed improvements in computer hardware over the past 25 years.

%\medskip
%\medskip

%\begin{figure}[H]
%\centering
%\includegraphics[scale=0.6]{supercomputer.pdf}
%\includegraphics[scale=0.6]{supercomputer_year.pdf}
%\caption{Log of Peak Supercomputer Speed from 1993--2013.}
%\label{super}
%\end{figure}
%ubs-data-1,ubs-data-2,subsub-opt-cvx-1

\subsection{Solving the convex quadratic optimization Problems in Section~\ref{subsub-opt-cvx-1}}\label{app:subsub-opt-cvx-1}
We show here that the convex quadratic optimization problems appearing in Section~\ref{subsub-opt-cvx-1} are indeed quite simple and can be solved with
small computational cost.

We first consider Problem~\eqref{ubs-data-1}, the computation of $u_{i}^{-}$ which is a minimization problem.
We assume without loss of generality that the feasible set of problem~\eqref{ubs-data-1} is non-empty. 
Thus by standard results in quadratic optimization~\cite{BV2004}, it follows that, there exists a $\tau$ such that:
$$ \nabla \left(\frac{1}{2} \| \M{y} - \M{X} \B\beta \|_{2}^2 + \tau \beta_{i} \right) =  0,$$
where, $\nabla$ denotes derivative wrt $\B\beta$ and a $\B\beta$ that satisfies the above gradient condition must also be feasible for Problem~\eqref{ubs-data-1}. 
Simplifying the above equation, we get:
$$ \M{X}'\M{X} \B\beta  = \M{X}'\M{y} - \tau e_{i},$$
where, $e_{i}$ is a vector in $\Re^p$ such that its $i$th coordinate is one with the remaining equal to zero. Simplifying the above expression, we have 
$$ \|\M{y} - \M{X} \B\beta\|^2_2  = \| (\mathbb{I} - P_{X} ) \M{y}  + \tau q_{i} \|_{2}^2.$$ 
Above, $\mathbb{I}$ is the identity matrix of size $p \times p$ and $P_{X}$ is the familiar projection matrix given by $\M{X} (\M{X}'\M{X})^{-1} \M{X}'$ \footnote{Note that we assume here that $p >n$ which typically guarantees that $\M{X}'\M{X}$ is invertible with probability one, provided the entries of 
$\M{X}$ are drawn from a continuous ditribution.} and $q_{i} =  \M{X} (\M{X}'\M{X})^{-1} e_{i}$. Observing that 
$\|\M{y} - \M{X} \B\beta\|^2_2  = \text{UB}$, one can readily estimate $\tau$ that satisfies the above simple quadratic equation. This leads to the solution of $\tau$, which 
subsequently leads to the optimal value $\widetilde{\B\beta}$ that solves Problem~\eqref{ubs-data-1}. This readily leads to the optimum of Problem~\eqref{ubs-data-1}.

The above argument readily applies to Problem~\eqref{ubs-data-1}, for the computation of $u_{i}^{+}$ by writing it as an equivalent minimization problem and observing that:
$$-u_{i}^{+} = \min_{\B\beta}  \;\; -\beta_{i}\;\; \;\; s.t. \;\;\;\; \frac12 \| \M{y} - \M{X} \B\beta \|^2_{2} \leq \text{UB}.$$

The above derivation can also be adapted to the case of Problem~\eqref{ubs-data-2}. 
Towards this end, notice that for estimating $v_{i}^-$ the above steps (for computing $u_{i}^-$)
will be modified: $e_{i}$  gets replaced by $\M{x}_{i} \in \Re^p$ (the $i$th row of $\M{X}$); and 
$P_{X}$ denotes the projection matrix onto the column space of $\M{X}$, even if the matrix $\M{X}'\M{X}$ is not invertible (since here, we consider arbitrary $n,p$). 

In addition, the Problems~\eqref{ubs-data-1} for the different variables and~\eqref{ubs-data-2} for the different samples; can be solved \emph{completely} independently, in parallel.

\subsection{Details for Section~\ref{general-1-variants} }\label{app:general-1-variants}

Note that in Problem~\eqref{eq-card-form2-1} we consider a uniform bound on $\beta_{i}$'s:
$-{\mathcal M}_{U}  \leq \beta_{i}  \leq {\mathcal M}_{U},$ for all $i = 1, \ldots, p$. 
Note that some of the variables $\beta_{i}$ may have larger 
amplitude than the others, thus it may be reasonable to have bounds depending upon the variable index $i$.
Thusly motivated, for added flexibility, one can consider the following (adaptive)
bounds on $\beta_{i}$'s: $-{\mathcal M}^i_{U}  \leq \beta_{i}  \leq {\mathcal M}^i_{U}$ for $i = 1, \ldots, p$. 
The  parameters ${\mathcal M}^i_{U} $ can be taken as $\max \{ |u^+_{i}| , |  u^{-}_{i} | \},$ as defined in~\eqref{ubs-data-1}.

More generally, one can also consider asymmetric bounds on $\beta_{i}$ as: $u_{i}^-  \leq \beta_{i}  \leq u_{i}^+$ for all $i$.

Note that the above ideas for bounding $\beta_{i}$'s can also be extended to obtain sample-specific bounds on $\langle \M{x}_{i}, \B\beta \rangle$ for 
$i = 1, \ldots, n$.

The bounds on $\|\widehat{\B\beta}\|_{1}$ and $\|\M{X}\widehat{\B\beta}\|_{1},\|\M{X}\widehat{\B\beta}\|_{\infty}$ can also be adapted to the above 
variable dependent bounds on $\beta_{i}$'s.

While the above modifications may lead to marginally improved performances, we do not dwell much on these improvements mainly for the sake of a clear exposition.

\subsection{Proof of Proposition~\ref{prop-oper-norm-inverse-1}}\label{proof-prop-oper-norm-1}~\\

{\it Proof}
\begin{enumerate}
\item[{\bf (a)} ] Given a set $I$, we define   $\M{G}:=\M{X}'_{I}\M{X}_{I}  -\M{I},$ and let $g_{ij}$ denote the $(i,j)$th entry of $\M{G}$. 
For any $\M{u} \in  \mathbb{R}^{k}$ we have
\begin{align}
 \max_{\| \M{u} \|_{1} = 1} \|\M{G} \M{u} \|_1 =~ & \max_{\| \M{u} \|_{1} = 1} \left( \sum_{i = 1}^{k}  \left | \sum_{j = 1}^{k} g_{ij} u_{j}\right| \right)&  \nonumber \\
  \leq~& \max_{\| \M{u} \|_{1} = 1} \left(  \sum_{i= 1 }^{k}  \sum_{j= 1 }^{k}   | u_{j}| |g_{ij}| \right)& \nonumber \\
  = ~&  \max_{\| \M{u} \|_{1} = 1} \left(  \sum_{j = 1 }^{k}   | u_{j} |  \sum_{i \neq j } |g_{ij}|  \right) &(g_{jj}=0 )\nonumber \\
\leq ~&  \max_{\| \M{u} \|_{1} = 1} \left(  \mu [k-1] \| \M{u}\|_{1} \right)  & \left( \sum_{i \neq j } |g_{ij}| \leq \mu [k-1] \right) \nonumber \\
=~ & \mu [k-1] . & \nonumber 
\end{align}
\item[{\bf (b)}] Using    $ \M{X}'_{I} \M{X}_{I}= \M{I} + \M{G} $ and  standard power-series convergence (which is valid since $ \| \M{G} \|_{1,1} < 1$)
we obtain 
$$~~~~~~~ \| (\M{X}'_{I} \M{X}_{I})^{-1} \|_{1,1} = \| \left(\M{I} + \M{G}\right)^{-1} \|_{1,1} = \sum_{i=0}^{\infty} \|  \M{G} \|^i_{1,1} 
\leq \frac1{1 - \| \M{G} \|_{1,1} } \leq \frac1{ 1- \mu [k-1]}.~~~~~~~\hfill \Box$$
\end{enumerate}

\subsection{Proof of Theorem~\ref{prop-beta-bound1}}\label{proof-prop-beta-bound1}~\\

{\it Proof}
\begin{enumerate}
\item[{\bf (a)}] 
Since $\widehat{\B\beta}_{I} = (\M{X}'_{I} \M{X}_{I})^{-1}\M{X}'_{I}\M{y}$
we have 
\begin{equation}
\label{eq:temp34}
  \| \widehat{\B\beta}\|_{1}= \|\widehat{\B\beta}_{I}\|_{1} \leq \|(\M{X}'_{I} \M{X}_{I})^{-1}\|_{1,1} \|\M{X}'_{I}\M{y}\|_{1} .
  \end{equation}
Note that 
\begin{equation}
\label{eq:temp35}
\|\M{X}'_{I}\M{y}\|_{1} =\sum_{j \in I} | \langle \M{X}_{j} , \M{y} \rangle | \leq \max_{I, | I | = k} \sum_{j \in I} | \langle \M{X}_{j} , \M{y} \rangle | \leq 
\sum_{j=1}^{k} | \langle \M{X}_{(j)} , \M{y} \rangle | .  \end{equation}
Applying Part (b) of Proposition~\ref{prop-oper-norm-inverse-1}  and \eqref{eq:temp35} to  \eqref{eq:temp34}, we obtain 
\eqref{l1-norm-bd-beta} .

\item[{\bf (b)}] 
We write 
$\widehat{\B\beta}_{I} = \M{A} \M{y}$ for $\M{A }= (\M{X}'_{I} \M{X}_{I})^{-1}\M{X}'_{I}$. If $\M{a}_{i}, i= 1, \ldots, k$ denote the rows of $\M{A}$ we have: 
\begin{equation}
\label{eq:a11}
 \|\widehat{\B\beta}_{I} \|_{\infty} = \max_{i = 1, \ldots, k } |\langle \M{a}_{i}, \M{y} \rangle | \leq \left(\max_{i = 1, \ldots, k } \| \M{a}_{i}\|_2 \right)\|\M{y}\|_{2}.
 \end{equation}
For every $i=1,\ldots ,k$ we have
\begin{align}
 \|\M{a}_{i}\|_{2} \leq & \max_{\| \M{u} \|_2 =  1} \| \M{A}\M{u} \|_2  \nonumber  \\
  =& \max_{\|\M{u}\|_2 = 1} \|(\M{X}'_{I} \M{X}_{I})^{-1}\M{X}'_{I} \M{u}\|_2 \nonumber  \\
%\leq& \max_{\|u\|_2 = 1} \|(\M{X}'_{I} \M{X}_{I})^{-1}\|_2 \|\M{X}'_{I}u\|_2 \\
%\leq & \lambda_{\max}(\M{X}'_{I} \M{X}_{I})^{-1}\lambda_{\max}(\M{X}_{I}) \\
%=& \lambda_{\max}(\M{X}_{I})/\lambda^2_{\min}(\M{X}_{I})
\leq & \lambda_{\max}\left((\M{X}'_{I} \M{X}_{I})^{-1}\M{X}'_{I}\right) \nonumber \\
= & \max\;\; \left\{ \frac{1}{d_1},  \ldots, \frac{1}{d_k}\right \}, \label{eq:aa2}
\end{align}
where $d_{1}, \ldots, d_{k} $ are the (nonzero) singular values of the matrix $\M{X}_{I}$.  
To see how one arrives at~\eqref{eq:aa2} let us denote the singular value decomposition  of   $\M{X}_{I}= \M{U D V}'$ with $ \M{D} = \diag \left(d_{1},d_{2}, \ldots,d_{k} \right).$ 
We  then have $$ (\M{X}'_{I} \M{X}_{I})^{-1}\M{X}'_{I} = (\M{VD}^{-2}\M{V}') (\M{U D V}')' = \M{V D}^{-1} \M{U'}$$ and the singular values of 
$(\M{X}'_{I} \M{X}_{I})^{-1}\M{X}'_{I} $ are thus  $1/d_i$, $i=1,\ldots ,k$.  

The eigenvalues of $\M{X}'_{I}\M{X}_{I}$ are $d_{i}^2 $ and from \eqref{eq:ab1}
we obtain that 
$d_i^2 \geq \eta_k$.  Using  \eqref{eq:aa2} we thus obtain 
\begin{equation}
\label{eq:a13}
\max_{i = 1, \ldots, k } \| \M{a}_{i}\|_2  \leq \frac{1}{ \sqrt{\eta_k}}.
\end{equation}
Substituting the bound \eqref{eq:a13} to \eqref{eq:a11} we obtain 
\begin{equation}
\label{lb-linftybeta-1} 
\| \widehat{\B\beta}_{I} \|_{\infty} \leq \frac{1}{\sqrt{\eta_{k}}}\|\M{y}\|_{2}.
\end{equation}
Using the notation   $\tilde{\M{A}} = (\M{X}'_{I} \M{X}_{I})^{-1}$,  we have 
\begin{align}
 \|\widehat{\B\beta}_{I} \|_{\infty} = &  \max_{i = 1, \ldots, k } |\langle \tilde{\M{a}}_{i}, \M{X}'_{I}\M{y} \rangle |  \nonumber \\
 \leq  & \left(\max_{i = 1, \ldots, k|  } \| \tilde{\M{a}}_{i}\|_2 \right) \| \M{X}'_{I}\M{y}\|_2 \nonumber  \\
 \leq & 
\lambda_{\max}\left( (\M{X}'_{I}\M{X}_{I})^{-1} \right)  \| \M{X}'_{I}\M{y}\|_2  \nonumber  \\
= &  \left(\max_{i=1,\ldots, k} \frac{1}{d_i^2} \right) \cdot    \sqrt{ \sum_{j \in I} | \langle \M{X}_{j}, \M{y} \rangle |^2 } \nonumber  \\
\leq & \frac{1}{\eta_{k}}\sqrt{\sum_{j=1}^{k} | \langle \M{X}_{(j)}, \M{y} \rangle |^2}.  \label{eq:a14}
\end{align}
 Combining \eqref{lb-linftybeta-1}  and \eqref{eq:a14} we obtain~\eqref{upper-bd-linbet011}. 
 
\item[{\bf (c)}]  
We  have 
\begin{equation}\label{eqn-1-0-0}
 \|\M{X}_{I} \widehat{\B\beta}_{ I}\|_{1} \leq  \sum_{i=1}^{n}  | \langle \M{x}_{i}, \widehat{\B\beta}_{I}  \rangle | \leq \sum_{i=1}^{n} \|\M{x}_{i}\|_{\infty}\|\widehat{\B\beta}_{{I}}\|_{1} 
 =  \sum_{i=1}^{n} \|\M{x}_{i}\|_{\infty}\|\widehat{\B\beta}_{I}\|_{1} .
 \end{equation}
Let $\M{P}_{I} := \M{X}_{I} (\M{X}'_{I}\M{X}_{I} )^{-1}\M{X}'_{I}$ denote the  projection onto the columns of 
$\M{X}_{I}$. We have 
 $\|\M{P}_{I}\M{y}\|_{2} \leq \|\M{y}\|_{2}$, leading to:
\begin{equation}\label{eqn-1-0-1}
\|\M{X}_I \widehat{\B\beta}_{I}\|_{1}  = \|\M{P}_{I}\M{y}\|_{1}  \leq \sqrt{k}  \|\M{P}_{I}\M{y}\|_{2}   \leq \sqrt{k} \| \M{y}\|_{2},
\end{equation}
where we used  that for any $\M{a}\in  \mathbb{R}^{m}$, we have $\sqrt{m} \| \M{a} \|_{2} \geq \|\M{a}\|_{1}$.
Combining~\eqref{eqn-1-0-0} and~\eqref{eqn-1-0-1} we obtain~\eqref{l1-norm-bd-xbeta}.

\item[{\bf (d)}] 
For any vector $\B\beta_{I}$ which has zero entries in the coordinates outside $I$,
we have:
$$ \|\M{X} \B\beta_{I}\|_{\infty} \leq  \max_{i =1 , \ldots, n } | \langle \M{x}_{i}, \B\beta_{I}  \rangle | \leq 
\max_{i = 1, \ldots, n }  \| \M{x}_{i}\|_{1:k} \| \B\beta_{I} \|_{\infty}, $$
leading to~\eqref{upper-bd-linxbet011}. $ \hfill \Box$
\end{enumerate}

\section{Proofs and Technical Details for Section~3}

\subsection{Proof of Proposition~\ref{prop-of-suff-dec1}}\label{proof-prop-of-suff-dec1}~\\
{\it Proof}
\begin{enumerate}
\item[{\bf (a)}] Let $\B\beta$ be a vector satisfying  $\|\B\beta\|_{0} \leq \mmk $.  Using 
 the notation $\widehat{\B\eta} \in \M{H}_{{\mmk} } \left(\B\beta - \frac1L \nabla g(\B\beta) \right)$ we have the following chain of inequalities:
\begin{align}
g(\B\beta)&= Q_{L}(\B\beta, \B\beta) &  \nonumber \\
%\label{line-1-1} \\
 &\geq \inf_{\| \B\eta \|_{0} \leq {\mmk}  } \;\; Q_{L}(\B\eta, \B\beta) & \nonumber \\
 &= \inf_{\| \B\eta \|_{0} \leq {\mmk} } \;\; \left ( \frac{L}{2} \| \B\eta - \B\beta \|_2^2 + \langle \nabla g(\B\beta), \B\eta - \B\beta  \rangle + g(\B\beta) \right )& \nonumber \\
&=  \inf_{\| \B\eta \|_{0} \leq {\mmk} } \;\; \left (  \frac{L}{2} \left\| \B\eta - \left ( \B\beta - \frac{1}{L} \nabla g(\B\beta) \right) \right\|_2^2  
- \frac{1}{2L} \|\nabla g(\B\beta) \|_2^2  + g(\B\beta) \right) & \nonumber  \\
&=  \;\; \left (  \frac{L}{2} \| \widehat{\B\eta} - \left ( \B\beta - \frac{1}{L} \nabla g(\B\beta) \right) \|_2^2  
- \frac{1}{2L} \|\nabla g(\B\beta) \|_2^2  + g(\B\beta) \right) &  ~{\rm (From}~\eqref{eta-hat-HT1}) \nonumber \\
&=  \left ( \frac{L}{2} \left \|\widehat{\B\eta} - \B\beta \right\|_2^2 +
 \langle \nabla g(\B\beta),\widehat{\B\eta} - \B\beta  \rangle + g(\B\beta) \right ) & \nonumber  
\end{align}
\begin{align}
&=  \left ( \frac{L - \ell}{2} \left \|\widehat{\B\eta} - \B\beta \right\|_2^2 + 
\frac{\ell}{2} \left \|\widehat{\B\eta} - \B\beta \right\|_2^2 +
 \langle \nabla g(\B\beta),\widehat{\B\eta} - \B\beta  \rangle + g(\B\beta) \right ) & \nonumber  \\
&\geq  \frac{L - \ell}{2} \left \|\widehat{\B\eta} - \B\beta \right\|_2^2 + 
\underbrace{\left ( \frac{\ell}{2} \left \|\widehat{\B\eta} - \B\beta \right\|_2^2 +
 \langle \nabla g(\B\beta),\widehat{\B\eta} - \B\beta  \rangle + g(\B\beta) \right )}_{Q_{\ell}(\widehat{\B\eta}, \B\beta)} & \nonumber  \\
  & \geq   \frac{L - \ell}{2} \left \|\widehat{\B\eta} - \B\beta \right\|_2^2 + g(\widehat{\B\eta}). &  ~{\rm (From}~\eqref{major-1})   \nonumber 
\end{align}
This  chain of inequalities leads to: 
\begin{equation}\label{eqn-suff-decrease-1-gen}
g(\B\beta) -  g(\widehat{\B\eta})   \geq \frac{L - \ell}{2} \left \| \widehat{\B\eta}- \B\beta \right\|_2^2.
\end{equation}
Applying \eqref{eqn-suff-decrease-1-gen} for   $\B\beta = \B\beta_{m}$ and 
$ \widehat{\B\eta}=\B\beta_{m+1}$, the vectors generated by Algorithm 1, we obtain \eqref{eqn-suff-decrease-1}.
This implies that the objective values
$g ( \B\beta_{m}  )$ are decreasing and since the sequence is bounded below
$(g(\B\beta)\geq 0)$, we obtain that  $g(\B\beta_{m} )$ converges as $m \rightarrow \infty$.

\item[{\bf (b)}]  If $ L > \ell$ and from part (a), the result follows. 

\item[{\bf (c)}] 
%We begin by observing that the condition $\| \liminf_{m \rightarrow \infty} \B\beta_{m} \|_{0} = \mmk$ 
%is equivalent to  $\liminf_{m \rightarrow \infty}  \min_{i: \beta_{mi} \neq 0 } |\beta_{mi}| > 0$.
The condition $\underline{\alpha}_{k} >0$ means that for all $m$ sufficiently large, the entry 
$|\beta_{(k),m}|$ will remain (uniformly) bounded away from zero. 
We will use this to prove that the support of $\B\beta_{m}$ converges. For the purpose of establishing contradiction
suppose that  the support does not converge.  Then,  there are infinitely many values of 
$m'$ such that $ \M{1}_{m'} \neq \M{1}_{m'+1}$. Using the fact that
$\| \B\beta_{m}\|_{0} = \mmk$ for all large $m$ we have
\begin{equation}\label{contra-1-1}
\;\;  \|\B\beta_{m'} - \B\beta_{m'+1}\|_{2} \geq 
\;\; 
\sqrt{ \beta_{m', i }^2+\beta_{m'+1, j}^2} \geq
\;\;   \frac{| \beta_{m', i }| + | \beta_{m'+1, j}| }{\sqrt{2}},
\end{equation} 
where  $i,j$ are such that  $ \beta_{m'+1,i}= \beta_{m', j}=0$.
As $m' \rightarrow \infty$, the quantity in the rhs of~\eqref{contra-1-1} remains bounded away from zero
since $\underline{\alpha}_{k} >0$.
%$\liminf_{m \rightarrow \infty}  \min_{i: \beta_{mi} \neq 0 } |\beta_{mi}| > 0$. 
This contradicts the fact that $\B\beta_{m+1} - \B\beta_{m} \rightarrow \M{0},$ as established in part (b). 
Thus, $\M{1}_{m}$ converges, and since $\M{1}_{m}$ is a discrete sequence, 
it converges after finitely many iterations, that is $\M{1}_{m}= \M{1}_{m+1}$  for all $m \geq M^*$.
Algorithm~1 becomes a 
vanilla gradient descent algorithm, restricted to the space $\M{1}_{m}$ for $m \geq M^*$. Since a gradient descent algorithm 
for minimizing a convex function over a closed convex set leads to a sequence of iterates that converge~\cite{rock-conv-96,nesterov2004introductory}, we conclude that
Algorithm~1 converges.
Therefore, 
 the sequence $ \B\beta_{m} $  converges to  $\B\beta^*$,  a  first order stationarity point:
$$ \B\beta^*  \in  \M{H}_{{\mmk}} \left(\B\beta^*  - \frac1L\nabla g( \B\beta^* ) \right) .$$

\item[{\bf (d)}]  Let ${\mathcal I}_{m} \subset \{1, \ldots, p \}$  denote the set of  $\mmk$ largest values of  the vector 
$\left(\B\beta_{m} - \frac1L \nabla g(\B\beta_{m}) \right)$ in absolute value.  By  
 the definition of  $\M{H}_{{\mmk} } \left(\B\beta_{m} - \frac1L \nabla g(\B\beta_{m}) \right)$, we have 
$$
\left| \left(\B\beta_{m} - \frac1L \nabla g(\B\beta_{m}) \right)_{i} \right| \geq  \left| \left(\B\beta_{m} - \frac1L \nabla g(\B\beta_{m}) \right)_{j}  \right|,
$$
for  all $i,j$ with  $i \in {\mathcal I}_{m}$ and $ j \notin {\mathcal I}_{m}$.
Thus, 
\begin{equation}\label{diff-1-1-11}
\liminf_{m\rightarrow \infty} \;\; \min_{i \in {\mathcal I}_{m}} \;\; \left| \left(\B\beta_{m} - \frac1L \nabla g(\B\beta_{m}) \right)_{i}  \right| \geq  
\liminf_{m\rightarrow \infty} \;\; \max_{j \notin {\mathcal I}_{m}} \;\; \left| \left(\B\beta_{m} - \frac1L \nabla g(\B\beta_{m}) \right)_{j}  \right| .
\end{equation}
Moreover, 
$$
 \left(\B\beta_{m} -  \M{H}_{{\mmk} } \left(\B\beta_{m} - \frac1L \nabla g(\B\beta_{m}) \right)\right)_{i} = \begin{cases}
\displaystyle  \frac1L (\nabla g(\B\beta_{m}))_{i} ,  & i \in  {\mathcal I}_{m},  \vspace{3pt}\\
\beta_{m,i} , & \text{otherwise} . 
 \end{cases}
$$
 Using  the fact that $\B\beta_{m+1} - \B\beta_{m} \rightarrow \M{0}$ we have 
$$
 (\nabla g(\B\beta_{m}))_{i}  \rightarrow 0,  i \in  {\mathcal I}_{m}\;\; \text{and} \;\; \beta_{m,j}  \rightarrow 0 , j  \notin  {\mathcal I}_{m}
$$
as $m \rightarrow \infty$.
Combining  with~\eqref{diff-1-1-11} we have that:
$$ \liminf_{m \rightarrow \infty} \;\; \min_{i \in {\mathcal I}_{m}} \;\; | \B\beta_{mi} | \geq   \liminf_{m \rightarrow \infty}\;\; \max_{j \notin {\mathcal I}_{m}} \;\;  \frac1L \left|  \left( \nabla g(\B\beta_{m})\right)_j \right| =   \frac1L
 \liminf_{m \rightarrow \infty} \;\; \| \nabla g ( \B\beta_{m} ) \|_{\infty}  .$$
Since, $\liminf_{m \rightarrow \infty} \;\; \min_{i \in {\mathcal I}_{m}} \;\; | \B\beta_{mi} | = \underline{\alpha}_{k} = 0 $ (by hypothesis), 
the lhs of the above inequality equals zero, which leads to  
$\liminf_{m \rightarrow \infty} \;\; \| \nabla g ( \B\beta_{m} ) \|_{\infty} = 0 $. 

\item[{\bf (e)}]
We build on the proof of Part (d).

It follows from equation~\eqref{diff-1-1-11} (by suitably modifying `$\liminf$' to `$\limsup$') that:
$$\underbrace{ \limsup_{m \rightarrow \infty} \;\; \min_{i \in {\mathcal I}_{m}} \;\; | \B\beta_{mi} |}_{\overline{\alpha}_{k}} \geq   \limsup_{m \rightarrow \infty}\;\; \max_{j \notin {\mathcal I}_{m}} \;\;  \frac1L \left|  \left( \nabla g(\B\beta_{m})\right)_j \right| =   \frac1L
 \limsup_{m \rightarrow \infty} \;\; \| \nabla g ( \B\beta_{m} ) \|_{\infty} .$$
Note that the lhs of the above inequality is $\overline{\alpha}_{k}$ which is zero (by hypothesis), thus 
$ \| \nabla g ( \B\beta_{m} ) \|_{\infty}  \rightarrow  0$ as $m \rightarrow \infty$.

Suppose $\B\beta_{\infty}$ is a limit point of the sequence $\B\beta_{m}$. 
Thus there is a subsequence 
$ m'  \subset \{ 1, 2, \ldots, \}$ such that $\B\beta_{m'} \rightarrow \B\beta_{\infty}$ and 
$g(\B\beta_{m'}) \rightarrow g(\B\beta_{\infty})$. Using the continuity of the gradient and hence the function $\cdot \mapsto \| \nabla g(\cdot)\|_{\infty}$ we have that
$ \| \nabla g ( \B\beta_{m'} ) \|_{\infty}  \rightarrow  \| \nabla g(\B\beta_{\infty})\|_{\infty} =  0$ as $m' \rightarrow \infty$.
Thus $\B\beta_{\infty}$ is a solution to the unconstrained (without cardinality constraints) optimization problem  $\min~g(\B\beta)$. Since $g(\B\beta_{m})$ is a 
decreasing sequence, $g(\B\beta_{m})$ converges to the minimum of $g(\B\beta)$. $\hfill \Box$

%$\nabla g( \B\beta_{m'}) \rightarrow \M{0}$, i.e., 
%$\liminf_{m \rightarrow \infty} \;\; \nabla g( \B\beta_{m}) \rightarrow \M{0}$.
%Since, $g(\B\beta_{m})$ is a decreasing sequence, this implies that $g(\B\beta_{m}) \rightarrow g(\B\beta^*),$
%where, $\B\beta^*$ is an unconstrained (without cardinality constraints)  solution   to  $\min~g(\B\beta)$. $\hfill \Box$

%If $L > \ell$, we  have
%$$ g ( \B\beta_{m}  ) - g( \B\beta_{m+1} ) \rightarrow 0 \;\; \text{and} \;\;  \|  \B\beta_{m+1}  -  \B\beta_{m}  \|_2^2  \rightarrow 0,$$
%which implies that $\M{1}_{m}-\M{1}_{m+1}=0$ for all $m\geq M^*$,  as there are finitely many possible supports.
%
%\item[{\bf (c)}]  Once the support stabilizes, then Algorithm 1 is exactly like a steepest decent method, which converges
%(see \cite{nesterov2004introductory}). Therefore, 
% the sequence $ \B\beta_{m} $ i  converges to  $\widehat{\B\beta}$ that  satisfies:
%$$  \M{H}_{{\mmk}} \left( \widehat{\B\beta}  - \frac1L\nabla g( \widehat{\B\beta} ) \right) = \widehat{\B\beta} , $$
%that is   $\widehat{\B\beta}$ is a  first order stationarity point. $\hfill \Box$
\end{enumerate}

\subsection{Proof of Proposition~\ref{hard-thresh-char-1}}\label{app:proof-hard-thresh-char-1}~\\
{\it Proof:}\\
We provide a proof of Proposition~\ref{hard-thresh-char-1}, for the sake of completeness.

It suffices to consider $ |c_{i}| > 0$  for all $i$. Let $\B\beta$ be an optimal solution to Problem~\eqref{defn:HT1}
and let $S :=\{ i: \beta_{i} \neq 0 \}$. The 
objective function is given by 
 $\sum_{i\not\in S } |c_{i}|^2 + \sum_{i\in S} (\beta_{i} - c_{i})^2$. 
 Note that by selecting $\beta_i=c_i$ for $i\in S$, we can make the objective function 
  $\sum_{i\not\in S } |c_{i}|^2 $.
Thus, to minimize the objective function,  $S$ must correspond to the indices of the 
largest $k$ values of $|c_{i}|, i \geq 1. \hfill \Box$

\subsection{Proof of Proposition~\ref{singleton-set-1}}\label{proof-singleton-set-1}~\\
{\it Proof}\\
This follows from Proposition~\ref{prop-of-suff-dec1}, Part (a), which implies that:
$$g({\B\eta}) -  g(\hat{\B\eta})   \geq \frac{L - \ell}{2} \left \|\hat{\B\eta} - {\B\eta} \right\|_2^2,$$
for any $\hat{\B\eta} \in \M{H}_{{k} } \left(\B\eta - \frac1L \nabla g(\B\eta) \right)$. Now by the definition of 
$\M{H}_{\mmk}(\cdot),$ we have $g(\B\eta) = g(\hat{\B\eta})$ which along with $L > \ell$ implies that the rhs of the above inequality is zero:
thus $\left \|\hat{\B\eta} - {\B\eta} \right\|_2 =0$, i.e., $ \B\eta = \hat{\B\eta}$. Since the choice of $\hat{\B\eta}$ was arbitrary, it follows that 
$\B\eta$ is the only element in the set $\M{H}_{{k} } \left(\B\eta - \frac1L \nabla g(\B\eta) \right)$. $\hfill \Box$

\subsection{Proof of Proposition~\ref{singleton-set-2}}\label{proof-singleton-set-2}~\\
{\it Proof}\\
The proof follows by noting that $\widehat{\B\beta}$ is $k$-sparse along with Proposition~\ref{prop-of-suff-dec1}, Part (a), which implies that:
$$g(\widehat{\B\beta}) -  g(\hat{\B\eta})   \geq \frac{L - \ell}{2} \left \|\widehat{\B\beta} - \hat{\B\eta}\right\|_2^2,$$
for any $\hat{\B\eta} \in \M{H}_{{k} } \left( \widehat{\B\beta} - \frac1L \nabla g(\widehat{\B\beta}) \right)$. Now, by the definition of $\widehat{\B\beta}$
we have $g(\widehat{\B\beta}) = g(\hat{\B\eta})$ which along with $L > \ell$ implies that the rhs of the above inequality is zero:
thus $\widehat{\B\beta}$ is a first order stationary point. $\hfill \Box$

\subsection{Proof of Theorem~\ref{FO - complexity-bound1}}\label{proof-FO-complexity-bound1}~\\
{\it Proof}\\
Summing inequalities ~\eqref{eqn-suff-decrease-1} for $ 1 \leq m \leq M.$
we obtain 
\begin{equation}
\sum_{m=1}^{M} \left ( g ( \B\beta_{m}  ) - g( \B\beta_{m+1} )  \right )  \geq 
 \frac{L - \ell}{2}\sum_{m=1}^{M} \| \B\beta_{m+1} - \B\beta_{m} \|_2^2, 
 \end{equation}
leading to 
$$ g ( \B\beta_{1}  ) - g( \B\beta_{M+1} )  \geq  \frac{M (L - \ell) }{2}  \min_{m = 1, \ldots, M}  \| \B\beta_{m+1} - \B\beta_{m} \|_2^2.$$
Since the decreasing sequence $g ( \B\beta_{m+1}  )$ converges to  $g (\B\beta^*)$  by Proposition~\ref{prop-of-suff-dec1} we have:
$$~~~~~~~~~~~~~~~~ \frac{g ( \B\beta_{1} ) -  g(\B\beta^*)}{M} \geq \frac{g ( \B\beta_{1}  ) - g( \B\beta_{M+1} )}{M} \geq 
 \frac{  (L - \ell) }{2}  \min_{m = 1, \ldots, M}  \|\B\beta_{m+1} - \B\beta_{m} \|_2^2.  ~~~~~~~~~~~~~~~~~~~~~ \hfill   \Box $$

\subsection{Proof of Proposition~\ref{prop-n-1-1}}\label{proof-prop-n-1-1}~\\
{\it Proof} \\
If $\B\eta$ is a first order stationary point with $\|\B\eta \|_0 \leq k$,
it follows from the argument following Definition~\ref{defn-FO-stat-1}, that there is a set $I \subset \{ 1, \ldots, p \}$ with $|I^c| = k$ such 
that $\nabla_{i} g(\B\eta) = 0$ for all $i \notin I$ and $\eta_{i} = 0$ for all $i \notin I$.
Let $\mu_{i}:=\eta_{i} - \frac 1 L \nabla_{i} g(\B\eta)$ for $i = 1, \ldots, p$. Suppose $I_{k}$ denotes the set of indices corresponding to the 
top $k$ ordered values of $|\mu_{i}|$. Note that:
\begin{equation}
\mu_{i} = \eta_{i}, \;\;  i \in I_{k}  \quad \text{and} \quad |\mu_{j}| = |  \frac 1 L \nabla_{j} g(\B\eta)|,\;\; j  \notin I_k.
\end{equation}
For $i \in I_{k}$ and $j \notin I_k$ we have $ | \mu_{i} | \geq | \mu_{j} |$.
This implies that 
$ |\eta_{i}| \geq |  \frac 1 L \nabla_{j} g(\B\eta)|$. Since $\B\eta \in \M{H}_{{\mmk} } \left(\B\eta - \frac1L \nabla g(\B\eta) \right)$ and $\|\B\eta\|_{0} <k$, it follows that 
$0=\min_{i \in I_{k}}  | \eta_{i} | = \min_{i \in I_{k}}  | \mu_{i}|$. We thus have that
$\nabla_{j} g(\B\eta) = 0$ for all $j \notin I_{k}$. In addition, note that $ \nabla_{i} g(\B\eta) = 0$ for all $i \in I_{k}$. 
Thus it follows that $\nabla g(\B\eta) = \M{0}$ and hence $\B\eta \in \argmin_{\B\eta} \; g(\B\eta)$. $\hfill \Box$

\section{Brief Review of Statistical Properties for the subset selection problem}\label{app:sec:stat-prop1}
In this section, for the sake of completeness we briefly review some of the properties of solutions to Problem~\eqref{eq-card-k}. 

Suppose the linear 
model assumption is true, i.e., $\M{y} = \M{X} \B\beta^0 + \B\epsilon$, with $\epsilon_{i} \stackrel{\text{iid}}{\sim} \text{N} (0, \sigma^2)$. 
Let $\widehat{\B\beta}$ denote a solution to~\eqref{eq-card-k}.
\cite{raskutti2011minimax} showed that, with probability greater than 
$1 - \exp (-c_{1} k  \log(p/k))$, the worst case (over $\B\beta^0$) predictive performance has the following upper bound: 
\begin{equation}\label{ub-1-raskutti}
\max_{\B\beta^0: \|\B\beta^0\|_{0}  \leq k } \; \; \frac{1}{n} \left \| \M{X} \B\beta^0 - \M{X}\widehat{\B\beta}   \right \|_{2}^2 \leq c_{2}  \sigma^2 \frac{k \log(p/k)}{n}, 
\end{equation}
where, $c_{1}, c_{2}$ are universal constants.  Similar results also appear in~\cite{bunea2007aggregation,zhang2014lower}. 
Interestingly, the upper bound~\eqref{ub-1-raskutti} does \emph{not} depend upon 
$\M{X}$. 
%This property is different than  \texttt{{Lasso}} based solutions, where the quality of the upper bound 
%depends upon $\M{X}$, as discussed below. 
Unless $p/k = O(1)$, the upper bound appearing in~\eqref{ub-1-raskutti} is of the order $O(\sigma^2 \frac{k \log(p)}{n})$ where the constants are universal.
In terms of the expected (worst case) predictive risk, an upper bound is given by~\cite{zhang2014lower}:
\begin{equation}\label{ub-1-raskutti-e}
\max_{\B\beta^0: \|\B\beta^0\|_{0}  \leq k } \; \; \frac{1}{n}  \mathbb E \left( \left \| \M{X} \B\beta^0 - \M{X}\widehat{\B\beta}   \right \|_{2}^2 \right)  \lesssim  \sigma^2 \frac{k \log(p)}{n}, 
\end{equation}
where, the symbol ``$\lesssim$'' means ``$\leq$'' upto some universal constants.

A natural question is how do the bounds for {\texttt{Lasso}}-based solutions compare with~\eqref{ub-1-raskutti-e}?
%begs a few interesting questions:
%how does the upper bound for {\texttt{Lasso}}-based solutions compare to that established in~\eqref{ub-1-raskutti-e}?
%Are there theoretical lower bounds on the performance of  {\texttt{Lasso}}-based solutions? 
%Fortunately answers to both the above questions are available in the rich body of work in statistical literature---see for example the nice work of~\cite{zhang2014lower}, and references therein. 
In a recent paper~\cite{zhang2014lower}, the authors derive upper and lower bounds of the prediction performance of the thresholded version of the 
{\texttt {{Lasso}}} solution, which we present briefly.  
Suppose
$$\hat{\B\beta}_{\ell_{1}} \in \argmin_{\B\beta} \; \frac{1}{2n} \| \M{y} - \M{X} \B\beta\|_{2}^2 + \lambda_{n} \| \B\beta\|_{1}$$ denotes a {\texttt{{Lasso}}} solution 
for $\lambda_{n}= 4 \sigma \sqrt{\frac{\log p}{n}}$. Let $\hat{\B\beta}_{\text{TL}}$ denote the thresholded version of the {\texttt{{Lasso}}} solution, which retains the 
top $k$ entries of $\hat{\B\beta}_{\ell_{1}} $ in absolute value and sets the remaining to zero. The bounds on the predictive performances of {\texttt{{Lasso}}} based solutions 
depend upon a restricted eigen-value type condition. Following~\cite{zhang2014lower}, we define, for any subset $S \in \{ 1, 2, \ldots, p \}$,  the quantity:
$C(S) := \{ \B\beta | \|\B\beta_{S^c}\|_{1} \leq 2 \| \B\beta_{S}\|_{1} \},$ where, $\| \B\beta_{S}\|_{1} = \sum_{j \in S} |\beta_{j}|$ and 
$\| \B\beta_{S^c}\|_{1} = \sum_{j \in S^c} |\beta_{j}|$. 
We say that the matrix $\M{X}$ satisfies a restricted eigen-value type condition with parameter $\gamma(X)$ if it satisfies the following:
$$ \frac{1}{n} \| \M{X} \B\beta\|_{2}^2 \geq \gamma(\M{X}) \| \B\beta\|_{2}^2 \;\;\;\quad \text{for}\quad \B\beta \in \cup_{S: |S|=k} C(S).$$
Note that $\gamma(\M{X})\leq 1$ and $\gamma(\M{X})$ is also related to the so called compatibility condition~\cite{buhlmann2011statistics}.
In an insightful paper, \cite{zhang2014lower} show that under such restricted eigenvalue type conditions the following holds:
\begin{equation}\label{zhang-ub-1}
\frac{\sigma^2}{\gamma(\M{X}_{\text{bad}})^2} \frac{k^{1-\delta} \log(p)}{n} \lesssim \max_{\B\beta^0: \|\B\beta^0\|_{0}  \leq k } \frac{1}{n} 
\mathbb E \left( \left \| \M{X} \B\beta^0 - \M{X}\widehat{\B\beta}_{\text{TL}}   \right \|_{2}^2 \right) \lesssim \frac{\sigma^2}{\gamma(\M{X})^2} \frac{k \log(p)}{n}
\end{equation}
%\begin{equation}\label{zhang-ub-1}
%\max_{\B\beta^0: \|\B\beta^0\|_{0}  \leq k } \frac{1}{n} 
%\mathbb E \left( \left \| \M{X} \B\beta^0 - \M{X}\widehat{\B\beta}_{\text{TL}}   \right \|_{2}^2 \right) \lesssim \frac{\sigma^2}{\gamma(\M{X})} \frac{k \log(p)}{n}
%\end{equation}
%%with probability $1 - \exp(-c_{4} k \log(p))$; where, $c_{3}, c_{4}$ are universal constants.
%\cite{zhang2014lower} also derive lower bounds on the predictive performance of the thresholded {Lasso} solutions. 
In particular, the lower bounds apply to \emph{bad} design matrices $\M{X}_{\text{bad}}$ 
%such that
%\begin{equation}\label{lb-zhang-1}
%\frac{\sigma^2}{\gamma(\M{X}_{\text{bad}})} \frac{k^{1-\delta} \log(p)}{n} \lesssim \max_{\B\beta^0: \|\B\beta^0\|_{0}  \leq k } \frac{1}{n} \mathbb E \left( \left \| \M{X} \B\beta^0 -
% \M{X}\widehat{\B\beta}_{\text{TL}}   \right \|_{2}^2 \right) 
%\end{equation}
for some arbitrarily small scalar $\delta>0$. In fact~\cite{zhang2014lower} establish a result stronger than~\eqref{zhang-ub-1}, where, 
$\widehat{\B\beta}_{\text{TL}}$ can be replaced by a $k$-sparse estimate delivered by a polynomial time method. 
 The bounds displayed in~\eqref{zhang-ub-1} show that there is a significant \emph{gap} between the predictive performance of 
subset selection procedures (see bound~\eqref{ub-1-raskutti-e}) and {\texttt{Lasso}} based $k$-sparse solutions---the magnitude of the gap depends upon 
how small $\gamma(\M{X})$ is. $\gamma(\M{X})$ can be small if the pairwise correlations between the features of the model matrix is quite high. These results complement
our experimental findings in Section~\ref{sec:computation-ls}.

An in-depth analysis of the properties of solutions to the Lagrangian version of Problem~\eqref{eq-card-k}, namely,
Problem~\eqref{L0-lag-1} is presented in~\cite{zhang2012general}.
\cite{raskutti2011minimax,zhang2012general} also analyze the errors in the regression coefficients: $\| \B\beta^0 - \widehat{\B\beta} \|_{2}$, 
under further minor assumptions on the model matrix $\M{X}$.
\cite{zhang2012general,shen2013constrained} provide interesting theoretical analysis of the variable selection properties of~\eqref{eq-card-k} and~\eqref{L0-lag-1}, showing that
subset selection procedures have superior  variable selection properties over {\texttt{Lasso}} based methods.

In passing, we remark that~\cite{zhang2012general} develop statistical properties of \emph{inexact} solutions to Problem~\eqref{L0-lag-1}. This may serve as 
interesting theoretical support for \emph{near global} solutions to Problem~\eqref{eq-card-k}, where the certificates of sub-optimality are delivered by our MIO framework in terms of 
global lower bounds. A precise and thorough understanding of the statistical properties of sub-optimal solutions to Problem~\eqref{eq-card-k} is left for an interesting piece of future work.

\section{Additional Details on Experiments and Computations} \label{more-expts-1}

\subsection{Some additional figures related to the radii of bounding boxes}\label{sec:append-bounding-box}
\begin{figure}[!h]
\begin{center}
 { \sf Evolution of the  MIO gap for~\eqref{eq-card-form2-1-xbet}, effect of bounding box radii ($n=50, p = 1000$).} \\

   % SNR = 1   \hspace{4cm} SNR = 3 

 %\medskip
 %\medskip
 
 ${\mathcal L}^{\zeta}_{\ell, \text{loc}} = \infty $  and ${\mathcal L}^{\beta}_{\ell, \text{loc}} =2 \| \B\beta_{0}\|_{1}/\mmk$ \\
 
 \scalebox{1.}{\includegraphics[height = .3\textheight, width = 0.45\textwidth, trim = 0mm 12mm 3mm 15mm, clip]{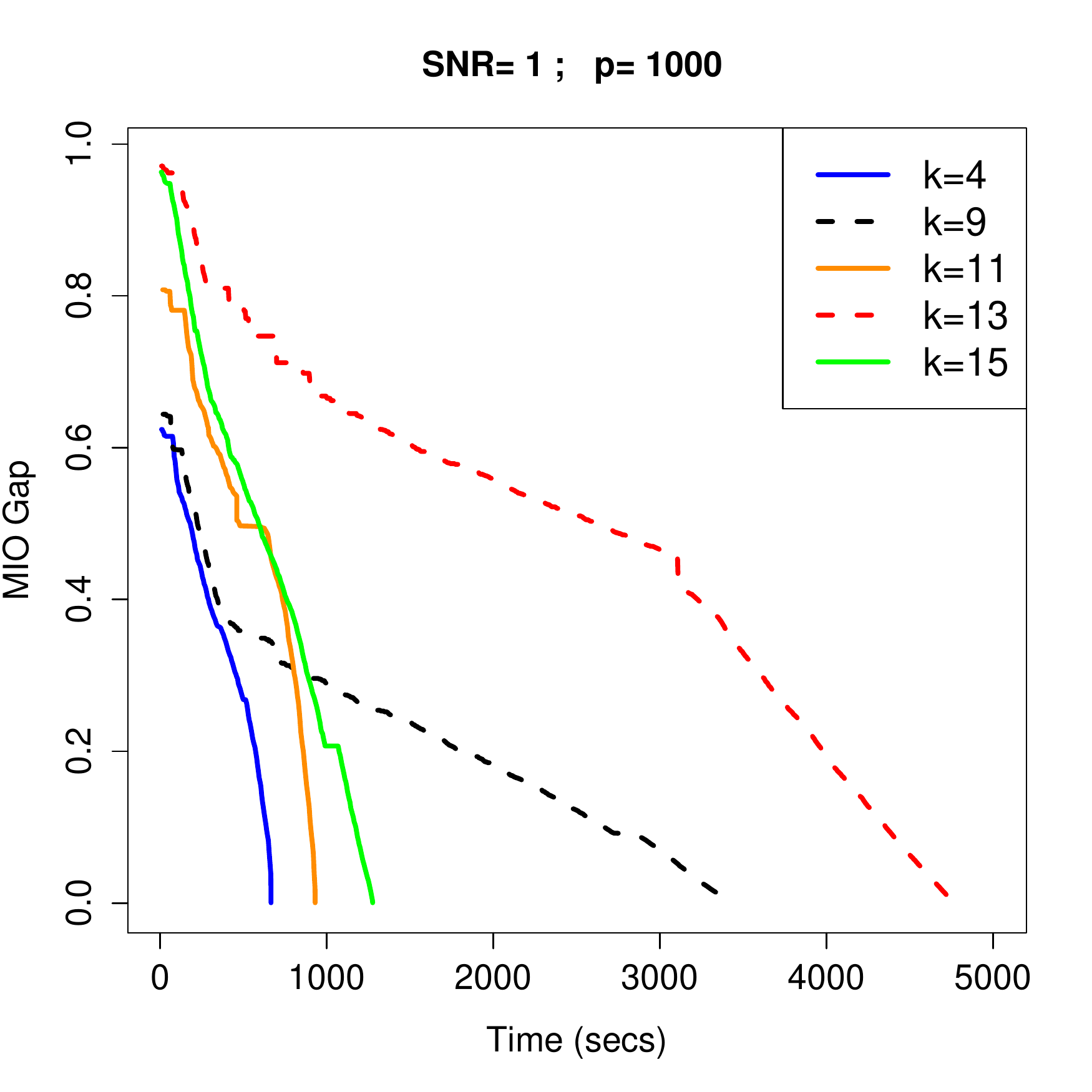}}
 \scalebox{1.}{\includegraphics[height = .3\textheight, width = 0.45\textwidth, trim = 6mm 12mm 3mm 15mm, clip]{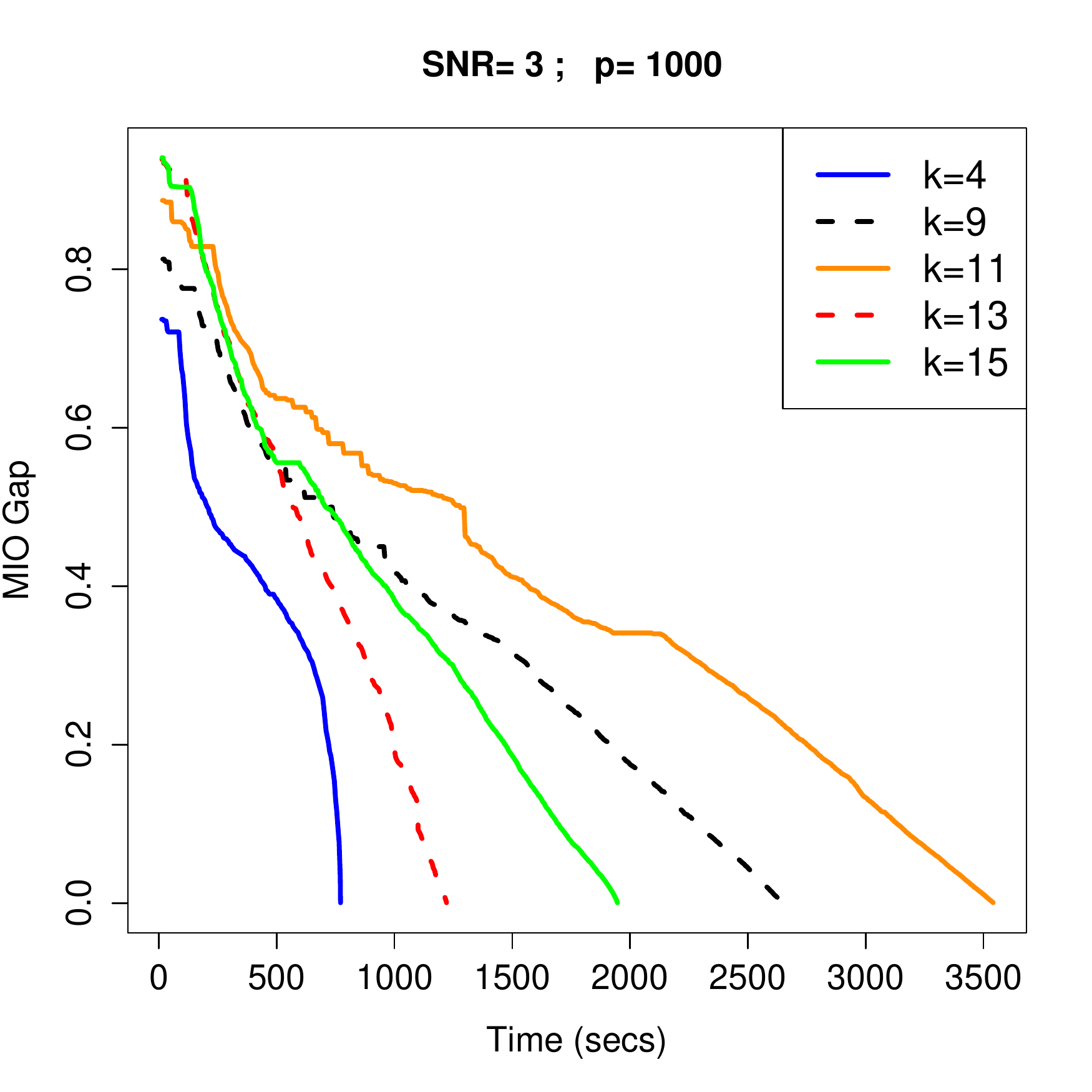}} 
 
 %\medskip
 
${\mathcal L}^{\zeta}_{\ell, \text{loc}} = \infty $  and ${\mathcal L}^{\beta}_{\ell, \text{loc}} =\| \B\beta_{0}\|_{1}/\mmk$\\
 
 \scalebox{1.}{\includegraphics[height = .3\textheight, width = 0.45\textwidth, trim = 0mm 6mm 3mm 15mm, clip]{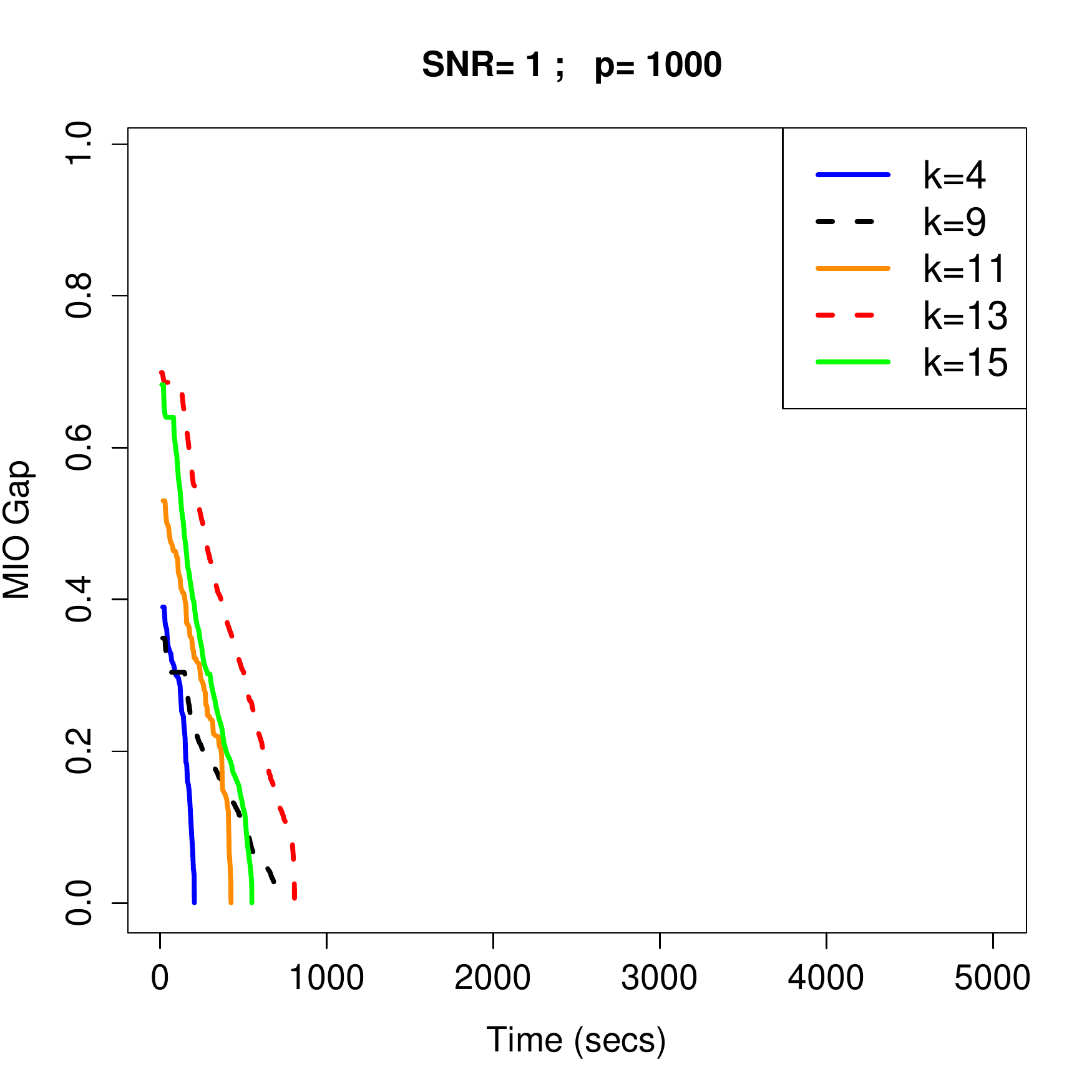}}
 \scalebox{1.}{\includegraphics[height = .3\textheight, width = 0.45\textwidth, trim = 6mm 6mm 3mm 15mm, clip]{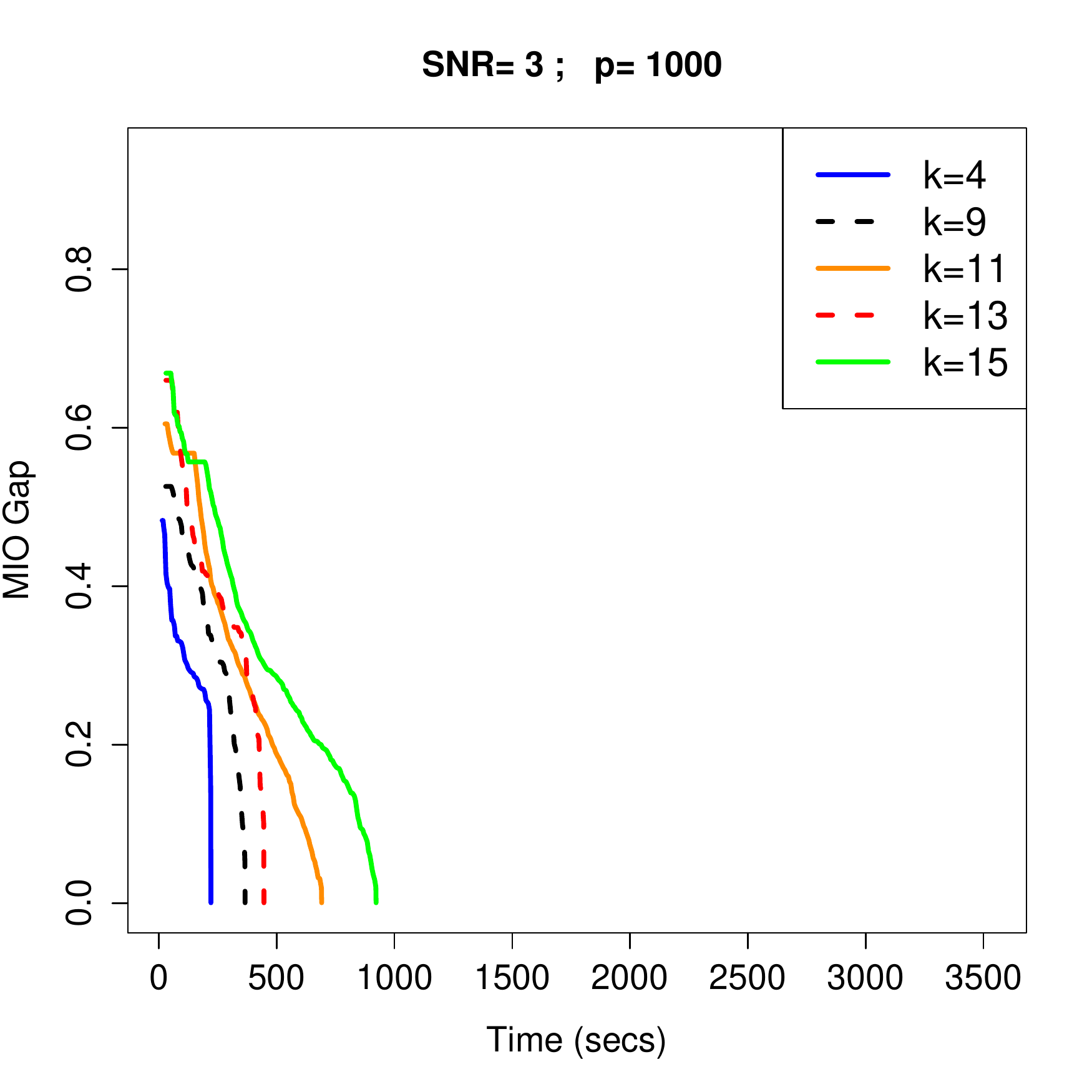}}
 \end{center}
\caption{{ \small {
The evolution of the MIO gap with varying radii of bounding boxes for MIO formulation~\eqref{eq-card-form2-1-xbet}.
The top panel has radii twice the size of the bottom panel.}
The dataset considered is generated as per Example 1 with $n=50, p = 1000, \rho = 0.9$ and $k_{0} = 5$
for different values of SNR: [Left Panel] SNR = 1, [Right Panel] SNR = 3.  For each case, different values of $\mmk$ have been considered.
The top panel has a bounding box radii which is twice the corresponding case in the lower panel. 
As expected, the times for the MIO gaps to close depends upon the radii of the boxes. The optimal 
solutions obtained were found to be insensitive to the choice of the bounding box radius.}  }
 \label{fig-synth-data-bounding-p1k}
\end{figure}

Some figures illustrating the effect of the bounding box radii are presented in Figure~\ref{fig-synth-data-bounding-p1k}.

\subsection{{\texttt{Lasso}}, Debiased {\texttt{Lasso}} and MIO} \label{sec:debiased-lasso}

We present here comparisons of the debiased {\texttt{Lasso}} with MIO and {\texttt{Lasso}}. 

Debiasing is often used to mitigate the shrinkage imparted by the {\texttt{Lasso}} regularization parameter. This is done by performing an unrestricted least squares
on the support selected by the {\texttt{Lasso}}. Of course the results will depend upon the tuning parameter used for the problem. We use two methods towards this end.
In the first method we find the best {\texttt{Lasso}} solution (by obtaining an optimal tuning parameter based on minimizing predictive error on a held out validation set); we then obtain the 
un-regularized least squares solution for that {\texttt{Lasso}} solution. 
This typically performed worse than {\texttt{Lasso}} in all the experiments we tried---see Tables~\ref{tab-deb-1-1} and~\ref{tab-deb-1-2}.
The unrestricted least squares solution on the optimal 
model selected by the {\texttt{Lasso}} (as shown in Figure~\ref{fig-n500p100}) had worse predictive performance than the {\texttt{Lasso}}, 
with the same sparsity pattern, as shown in Table~\ref{tab-deb-1-1}.
 This is probably due to overfitting since the model selected by the {\texttt{Lasso}} is quite dense compared to $n,p$.
Table~\ref{tab-deb-1-2} presents the results for $50=n \ll p=1000$. We consider the same example presented in Figure 9, Example 1. 
First of all, Table~\ref{tab-deb-1-2} presents the prediction performance of {\texttt{Lasso}} after debiasing---we considered the same tuning parameter considered 
optimal for the {\texttt{Lasso}} problem. We see that as in the case of Table~\ref{tab-deb-1-1}, the debiasing does not lead to improved performance in terms of prediction error.

We thus experimented with another variant of the debiased {\texttt{Lasso}}, where for every $\lambda$ we computed the {\texttt{Lasso}} solution~\eqref{lass-lag}
and obtained $\hat{\B\beta}_{\text{Deb}, \lambda}$ by performing an unrestricted least squares fit on the support selected by the {\texttt{Lasso}} solution at $\lambda$. 
This method can be thought of delivering feasible solutions for Problem~\eqref{eq-card-k}, for a value of $k:=k(\lambda)$ determined by the {\texttt{Lasso}} solution at $\lambda$.
The success of this method makes a case in support of using criterion~\eqref{eq-card-k}.  The tuning parameter was then selected by minimizing predictive performance on a held out test validation set. This method in general performed better than {\texttt{Lasso}} in delivering a sparser model with better predictive accuracy than the {\texttt{Lasso}}. 
The performance of the debiased {\texttt{Lasso}} was similar to {\texttt{Sparsenet}} and was in general inferior to MIO by orders of magnitude, especially for the problems 
where the pairwise correlations between the variables was large and SNR was low and $n \ll p$.
The results are presented in Table~\ref{tab-deb-2-1},\ref{tab-deb-2-2} (for the case $n>p$) and~\ref{tab-deb-3-1} and~\ref{tab-deb-3-2} (for the case $n\ll p$).

\begin{table}[!ht]

\centering

{\bf{Debiasing at optimal {\texttt{Lasso}} model, $n>p$}} 

\medskip

\scalebox{1.1}{
\begin{tabular}{|c|c|c|}
    \hline
  SNR & $\rho$ &Ratio: {\texttt{Lasso}}/ Debiased {\texttt{Lasso}}  \\ \hline \hline 
    6.33& 0.5&	0.33 \\ \hline
3.17	&0.5& 0.54	\\ \hline
1.58	&0.5& 0.53	\\ \hline \hline
6.97		&0.8& 0.67	\\ \hline
3.48	&0.8& 0.64	\\ \hline
1.74 &0.8& 0.63	\\ \hline \hline
8.73	&0.9	& 1\\ \hline
4.37	&0.9& 0.58 \\ \hline
2.18	&0.9	& 0.61  \\ \hline
       \end{tabular}}
        \caption{\small{{\texttt{Lasso}}  and Debiased {\texttt{Lasso}} corresponding to the numerical experiments of Figure 4, for Example 1 with $n = 500, p = 100, \rho \in \{0.5, 0.8, 0.9\} $ and $k_0 = 10$. 
        Here, ``Ratio'' equals the ratio of the prediction error of the {\texttt{Lasso}} and the debiased {\texttt{Lasso}} at the optimal tuning parameter selected by the {\texttt{Lasso}}.}}
         \label{tab-deb-1-1}
\end{table}

\begin{table}[!ht]

\centering

{\bf{Debiasing at optimal {\texttt{Lasso}} model, $n\ll p$}} 

\medskip

\scalebox{1.1}{  \begin{tabular}{|c|c|c|}
    \hline
 SNR & $\rho$ & Ratio: {\texttt{Lasso}}/Debiased {\texttt{Lasso}}  \\ \hline \hline 
10	&0.8& 0.90	\\ \hline
7	&0.8& 1.0	\\ \hline
3 &0.8& 0.91	\\ \hline 
       \end{tabular}}
        \caption{\small{ {\texttt{Lasso}} and Debiased {\texttt{Lasso}} corresponding to the numerical experiments of Figure 9, for Example 1 with $n = 50, p = 1000, \rho = 0.8 $ and $k_0 = 5 $.  Here, ``Ratio'' equals the ratio of the prediction error of the {\texttt{Lasso}} and the debiased {\texttt{Lasso}} at the optimal tuning parameter selected by the {\texttt{Lasso}}.}} 
        \label{tab-deb-1-2}
\end{table}

The performance of this model 
was comparable with {\texttt{Sparsenet}}---it was better than {\texttt{Lasso}} in terms of obtaining a sparser model with better predictive accuracy. 
However, the performance of MIO was significantly better than the debiased version of the {\texttt{Lasso}}, especially for larger values of $\rho$ and smaller SNR values.

\begin{table}[!ht]

\centering

{\bf{Sparsity of Selected Models}}, $n>p$ 

\medskip

    \scalebox{1.1}{\begin{tabular}{|c|c|c|c|c|}  \hline
    SNR& $\rho$		& {\texttt{Lasso}} 			&Debiased {\texttt{Lasso}} & MIO  \\   \hline  
6.33 & 0.5 & 27.6 (2.122) & 10.9 (0.65) & 10.8 (0.51)                   \\ \hline 
3.17 & 0.5 & 27.7 (2.045) & 10.9 (0.65) & 10.1 (0.1)\\ \hline 
1.58 & 0.5 & 28.0 (2.276) & 10.9 (0.65) & 10.2 (0.2)\\ \hline \hline
6.97 & 0.8 & 34.1 (3.60) & 10.4 (0.15) & 10 (0.0) \\ \hline 
3.48 & 0.8 & 34.0 (3.54) & 10.9 (0.55) & 10.2 (0.2)\\ \hline 
1.74 & 0.8 & 33.7  (3.49) & 13.7 (1.50) & 10  (0.0)\\ \hline \hline
8.73 & 0.9 & 25.9  (0.94) & 13.9 (0.68) & 10.5 (0.17) \\ \hline 
4.37 & 0.9 & 34.6 (3.23) & 18.1 (1.30) & 10.2 (0.25) \\ \hline 
2.18 & 0.9 & 34.7 (3.28) & 20.5 (1.85) & 10.1 (0.10) \\ \hline 
    \end{tabular}}
    \caption{\small{Number of non-zeros in the selected model by {\texttt{Lasso}}, Debiased {\texttt{Lasso}}, and MIO corresponding to the numerical experiments of Figure 4, for Example 1 with $n = 500, p = 100, \rho \in \{0.5, 0.8, 0.9\} $ and $k_0 = 10$. The tuning parameters for all three models were selected separately 
    based on the best predictive model on a held out validation set.
     Numbers within brackets denote standard-errors. Debiased {\texttt{Lasso}} leads to less dense models than {\texttt{Lasso}}. When $\rho$ is small and SNR is large, the model size of debiased {\texttt{Lasso}} performance is similar to MIO.   However, for larger values of $\rho$ and smaller values of SNR
subset selection leads to orders of magnitude sparser solutions than debiased {\texttt{Lasso}}.  }}
        \label{tab-deb-2-1}

\end{table}

\begin{table}[!ht]

\centering

{\bf{Predictive Performance of Selected Models}},  $n>p$

\medskip
   \scalebox{.9}{\begin{tabular}{|c|c|c|c|c|c|}
    \hline
SNR& $\rho$		& {\texttt{Lasso}} 			&Debiased {\texttt{Lasso}} & MIO    & Ratio: \\  
  &&&&&  Debiased {\texttt{Lasso}}/MIO \\ \hline  
6.33 & 0.5 & 0.0384 (0.001) & 0.0255 (0.002) & 0.0266 (0.001) & 1.0  \\ \hline 
3.17 & 0.5 & 0.0768 (0.003) & 0.0511 (0.004) & 0.0478 (0.002) &1.0 \\ \hline 
1.58 & 0.5 & 0.1540 (0.007) & 0.1021 (0.009) & 0.0901 (0.009)& 1.1 \\ \hline \hline
6.97 & 0.8 & 0.0389 (0.002) & 0.0223 (0.001) & 0.0231 (0.002) & 1.0 \\ \hline 
3.48 & 0.8 & 0.0778 (0.004) & 0.0464 (0.003) & 0.0484 (0.004)& 1.0 \\ \hline 
1.74 & 0.8 & 0.1557 (0.007) & 0.1156 (0.008) & 0.0795 (0.008) & 1.5 \\ \hline\hline  
8.73 & 0.9 & 0.0325 (0.001) & 0.0220 (0.002) & 0.0197 (0.002) & 1.2 \\ \hline 
4.37 & 0.9 & 0.0632 (0.002) & 0.0532 (0.003) & 0.0427 (0.008)&  1.3 \\ \hline 
2.18 & 0.9 & 0.1265 (0.005) & 0.1254 (0.006) & 0.0703 (0.011)&  1.8  \\ \hline 
   \end{tabular}}
        \caption{\small{ Predictive Performance for tests of {\texttt{Lasso}}, Debiased {\texttt{Lasso}}, and MIO corresponding to the numerical experiments of Figure 4, for Example 1 with $n = 500, p = 100, \rho \in \{0.5, 0.8, 0.9\} $ and $k_0 = 10$. Numbers within brackets denote standard-errors. 
        The tuning parameters for all three models were selected separately 
    based on the best predictive model on a held out validation set. When $\rho$ is small and SNR is large, debiased {\texttt{Lasso}} performance is similar to MIO. However, for larger values of $\rho$ and smaller values of SNR
subset selection performs better than debiased {\texttt{Lasso}} based solutions.  }}\label{tab-deb-2-2}
\end{table}

% a1<-c(0.0255,0.0511,.1021,0.0223,.0464,.1156,0.022,.0532,.1254) [deb - {\texttt{Lasso}}]
% b1<-c(0.0266,0.0478,0.0901,0.0231,0.0484,0.0795,0.0197,0.0427,0.0703)  [mio ]
% [1,] 0.9586466
% [2,] 1.0690377
% [3,] 1.1331853
% [4,] 0.9653680
% [5,] 0.9586777
% [6,] 1.4540881
% [7,] 1.1167513
% [8,] 1.2459016
% [9,] 1.7837838

We then follow the method described above (for the $n>p$ case), where we consider a sequence of models $\hat{\B\beta}_{\text{Deb}, \lambda}$ and find the $\lambda$ that delivers the best predictive model 
on a held out validation set. 
\begin{table}[!ht]

\centering

{\bf{Sparsity of Selected Models}}, $n\ll p$ 

\medskip

    \scalebox{1.1}{\begin{tabular}{|c|c|c|c|c|}  \hline
    SNR& $\rho$		& {\texttt{Lasso}} 			&Debiased {\texttt{Lasso}} & MIO  \\   \hline  
10 & 0.8 & 25.7 (1.73) & 7.9 (0.43) & 5 (0.12) 	\\ \hline
7 & 0.8 & 27.8 (2.69) & 8.1 (0.43) & 5 (0.16)	\\ \hline
3 & 0.8 & 28.0 (2.72) & 10.0 (0.88) & 6 (1.18)	\\ \hline
    \end{tabular}}
    \caption{\small{Number of non-zeros in the selected model by {\texttt{Lasso}}, Debiased {\texttt{Lasso}}, and MIO corresponding to the numerical experiments of Figure 9, for Example 1with $n = 50, p = 1000, \rho = 0.8 $ and $k_0 = 5 $. Numbers within brackets denote standard-errors. The tuning parameters for all three models were selected separately 
    based on the best predictive model on a held out validation set.
    Debiased {\texttt{Lasso}} leads to less dense models than {\texttt{Lasso}} but more dense models than MIO.  The performance gap between MIO and debiased {\texttt{Lasso}} 
    becomes larger with lower values of SNR.}}\label{tab-deb-3-1}
\end{table}

\begin{table}[!ht]

\centering

{\bf{Predictive Performance of Selected Models}}, $n\ll p$ 

\medskip
   \scalebox{1.1}{ \begin{tabular}{|c|c|c|c|c|c|}
    \hline
SNR& $\rho$		& {\texttt{Lasso}} 			&Debiased  & MIO    & Ratio: \\  
  &&& {\texttt{Lasso}} &&  Debiased {\texttt{Lasso}}/ MIO\\ \hline  
10 & 0.8 & 0.084 (0.004) & 0.046 (0.003) & 0.014 (0.005) &  3.3   \\ \hline
7 & 0.8 & 0.122 (0.005) & 0.070 (0.004) & 0.020 (0.007)  &     3.5  \\ \hline
3 & 0.8 & 0.257 (0.012) & 0.185 (0.016) & 0.151 (0.027)  &      1.2 \\ \hline
   \end{tabular}}
\caption{\small {Predictive performances of {\texttt{Lasso}}, Debiased {\texttt{Lasso}}, and MIO corresponding to the numerical experiments of Figure 9, for Example 1with $n = 50, p = 1000, \rho = 0.8 $ and $k_0 = 5$. 
Numbers within brackets denote standard-errors. The tuning parameters for all three models were selected separately 
    based on the best predictive model on a held out validation set. MIO consistently leads to better predictive models than Debiased {\texttt{Lasso}} and ordinary {\texttt{Lasso}}. Debiased {\texttt{Lasso}} performs better than ordinary {\texttt{Lasso}}}.}\label{tab-deb-3-2}
\end{table}

%\subsection{Additional results: Subset selection for LAD}\label{app:sec:l0-lad}
%Figures showing the performances of subset selection for the LAD problem versus the Lasso penalized version are presented in Figures~\ref{fig-LAD1} and~\ref{fig-LAD2}.

\end{appendix}

\clearpage

\end{document}